%% file: census.tex
\documentclass[trackchanges,twocolumn]{aastex7}
\usepackage{hyperref}
\usepackage{float}

\graphicspath{{./}{figures/}}
\shorttitle{Pulsar Census: SKA-Low prototype station}
\shortauthors{Kumar et. al.}

\begin{document}

\title{50-250 MHz Pulsar Census with an SKA-Low prototype station: Spectra and Polarization}

\author[]{Pratik Kumar}
\affiliation{International Centre for Radio Astronomy Research, Curtin University, Bentley, WA 6102, Australia}
\email[show]{pratik.kumar@curtin.edu.au}  

\author{Marcin Sokolowski} 
\affiliation{International Centre for Radio Astronomy Research, Curtin University, Bentley, WA 6102, Australia}
\email{marcin.sokolowski@curtin.edu.au}

\author{Randall Wayth}
\affiliation{SKAO, Science Operations Centre, CSIRO ARRC, 26 Dick Perry Avenue, Kensington WA 6151, Australia}
\affiliation{International Centre for Radio Astronomy Research, Curtin University, Bentley, WA 6102, Australia}
\email{r.wayth@curtin.edu.au}

\begin{abstract}
Low-frequency pulsar observations are crucial for understanding pulsar emission spectra and population physics, as well as for probing the interstellar medium (ISM) and Earth's ionosphere. We report the largest low-frequency pulsar census conducted in the southern hemisphere, covering 50–250 MHz, using the EDA2, an SKA‑Low prototype station. In this survey, we detected 120 pulsars, including 23 first-time detections below 150 MHz and 5 below 100 MHz. For each source, we provide integrated pulse profiles and flux‑density measurements across five sub-bands spanning 50–250 MHz. We also obtained improved dispersion measure (DM) values for 110 pulsars, with a median absolute DM correction of about 0.1 pc cm$^{-3}$. We measured significant Faraday rotation for 40 pulsars with improved rotation measure (RM) values for 4 pulsars, as well as phase-resolved RM variation in J1453-6413. Full-polarimetric pulse profiles are provided for all these pulsars, with multi-frequency polarimetric data for 20 of them. These results will enhance future SKA‑Low science: refining pulsar population models, informing survey strategies, and advancing characterization of both the ISM and the ionosphere through low-frequency pulsar monitoring. 

\end{abstract}




\section{Introduction} \label{sec:intro} 

Pulsars are highly magnetized, rotating neutron stars that emit radiation by converting their rotational energy into dipole radiation across the electromagnetic (EM) spectrum, but mainly at radio frequencies. However,   
even after their discovery at low radio frequencies \citep[81.5 MHz,][]{hewish} almost six decades ago, their observations have remained elusive at the lowest frequencies, with most discoveries and studies reported around $\sim$ 1 GHz. Even among those discovered, a large fraction remains undetected and poorly understood at low frequencies below $\sim$ 300 MHz. Many factors contribute to this, both technological and observational limitations. While many advancements have been made in the past decade in technology and instrumentation with the addition of new radio frequency observatories like the Low Frequency Array \citep[LOFAR;][]{Vanhaarlem}; the Murchison Widefield Array \citep[MWA;][]{Tingay}; the Long Wavelength Array \citep[LWA;][]{gtaylor}; and NenuFAR \citep[][]{zarka}, most pulsars have not been detected at low frequencies, partly also due to observational constraints. 

The ability to detect a source with high signal-to-noise (S/N) ratio also depends on the background noise, given the high Galactic background emission at low frequencies, which follows a power law with an index of -2.6 \citep{Haslam}. Therefore, higher sensitivity is required to detect fainter sources. In general, pulsars are classified as steep spectrum sources, S$_{\nu}$ $\propto$ $\nu^{\alpha}$, where S$_{\nu}$ is the flux density at frequency $\nu$ and $\alpha$ is the spectral index with a mean value of -1.6 \citep{jankowski441psr}. However, their radio spectra have been extensively studied for decades in order to understand their population as well as their complex emission mechanism \citep[see, e.g.,][]{Sieber,Malov,malofeev,Lorimer1995}. These studies suggest a turnover in pulsar spectra at low frequencies attributed to synchrotron self-absorption \citep{Sieber} in the pulsar magnetosphere or free-free absorption \citep{Malov} in the interstellar medium (ISM). Studies of pulsars at higher frequencies show that flux density spectra follow similar trends among normal and millisecond pulsars (MSPs) \citep{jankowski441psr}. At the same time, some studies of MSPs have shown that the power law behavior continues to the lowest frequencies \citep{Kuzmin}, unlike normal pulsars, although more recent studies indicate a spectral turnover in MSP spectra as well \citep{rahulsharan}. This suggests the need for improving the frequency coverage of MSPs at frequencies below 300 MHz. The combined effect of higher system noise and low flux density due to spectral turnover makes it difficult to detect more pulsars at low frequencies.

In addition to these, the effects of dispersion ($\propto \nu^{-2}$) and scattering ($\propto \nu^{-4}$) on pulsar emission, which have a very strong frequency dependence, can not be ignored \citep{bhat2004}. With improvements in computing, dispersion effects can be minimized with the application of coherent de-dispersion techniques, whereas multi-path scattering introduces asymmetric broadening, which is more difficult to correct. Nevertheless, the strong frequency dependence of these effects presents an opportunity to make high-precision measurements of pulsar dispersion and scattering. An application is to improve high-precision pulsar timing array (PTA) experiments, geared towards detecting gravitational waves, whose sensitivity is affected due to variability in the ISM \citep{Hemberger2008}. Recent studies have shown that low-frequency observations can provide accurate measurements of pulsar scattering index \citep{Bansal,Kirsten2019}, measurements of solar wind variability \citep{Tiburzi,Kumar}, and even time-dependent scattering variability \citep{Donner,Kumar2025}, which could all contribute towards improving the noise modeling, thereby increasing the sensitivity of PTAs. Pulsars also scintillate due to turbulence in the interstellar medium. This causes the measured signal to fluctuate, leading to variations in measured flux density of pulsars. This effect is more pronounced at low frequencies \citep{Bhat_2023}.

As pulsars are also highly polarized sources, propagation through the ionized and magnetized ISM also imparts a frequency-dependent distortion to the linearly polarized signal via Faraday Rotation (FR) on the received signal. Again, since this effect is proportional to $\lambda^2$, observations at low frequencies allows for more robust measurements of pulsar Rotation Measure (RM), which is the product of line-of-sight (LOS) magnetic field and the intervening electron density.  
We can probe the magneto-ionic structure of the ISM, large-scale magnetic field structure of the galaxy \citep{Sobey}. It is possible to determine the average LOS magnetic field strength of the ISM, using pulsar observations as,
\begin{equation} \label{eq:1}
   <B_{\parallel}>= 1.232\, \frac{RM}{DM}\, \mu G
\end{equation}

where RM in rad m$^{-2}$ and DM (in pc cm$^{-3}$) is the pulsar dispersion measure (electron density integrated along the LOS). Long-term studies of pulsar RM and its variation also allow us to understand the feasibility of different ionospheric electron density models and their usefulness on different time scales \citep[see, e.g.,][]{Kumar2025}. Finally, pulsar polarizations are also important to understand their complex emission characteristics \citep{vonhoensbroech}. 

\subsection{Motivations and science goals} \label{sec:motivations}
Detecting, characterizing and monitoring pulsars is a key science goal for the future Square Kilometre Array(SKA)-Low radio telescope. In addition to studying pulsars themselves, high precision DMs are required for PTAs to detect low frequency gravitational waves \citep{janssen}. Pulsars are typically detected at higher frequencies, hence observations of known pulsars at lower frequencies provides high precision DM and RM measurements as well as measurements of any spectral turnover at low frequencies. Additionally, pulsar observations and detections at low frequencies are also affected propagation effects such as scattering and dispersion\citep{bhat2004}. This wide bandwidth information is required for estimating the number of detections of future pulsar surveys. Low frequency observations between 50-250 MHz also provide very large fractional bandwidth to study the evolution of pulse profiles and the identification of interesting sources. Low-frequency data from the presented survey will provide these information and improve our understanding of pulsar spectra.


\section{Observations}\label{sec:obs}
The Engineering Development Array2 (EDA2) is a low-frequency aperture array comprising 256 MWA dual polarization bowtie dipoles, with modified low-noise amplifiers (LNAs) that improve the sensitivity down to $\sim$50 MHz. The dipoles are distributed in a pseudo-random layout with a diameter of 35 m. This is a second-generation prototype station built as a verification system for low-frequency Square Kilometer Array (SKA-Low) \citep{caiazzo2017ska}, operating between 50-350 MHz. Details of EDA2 system verification, design, calibration, and performance are presented in \citet{Wayth2022}.
The observing band is split into 512 coarse channels, separated by $\sim$ 0.781 MHz. These coarse channels are oversampled using a polyphase filterbank (PFB) to get a channel width of $\sim$ 0.926 MHz. The station can operate in stand-alone mode, both as an interferometer and as a beamformed array. 

Pulsar observations are carried out using the beamformed mode, by coherently summing the complex dipole voltages to form a phase-array beam. The sensitivity of the phase array varies as a function of frequency, pointing direction, and local sidereal time (LST), with the highest sensitivity at zenith. The station beams are electronically steered in a pointing direction, using a digital beamforming method implemented in the firmware of the Tile Processing Modules \citep[TPMs;][]{naldi2017}, by applying a phase delay to each dipole in the signal chain.
A census of pulsars was carried out with an initial system at intermediate frequencies \citep{lee2022} deployed with a maximum time resolution of 1.08 $\mu$s, which could capture one coarse channel. EDA2 was also used to search for Fast Radio Bursts (FRBs) with an improved bandwidth and high time resolution imaging capability providing revised rates at low frequencies \citep{marcintransient}. With improvements to hardware and software, the current work makes use of an upgraded system that can capture up to $\sim$ 100 coarse channels. This improved system was also used to analyze the largest sample of Crab giant pulses at low frequencies\citep{crabgiant}. Compared to the MWA at MRO, which is a precursor to SKA-Low, EDA2 has real-time beamforming capability, with more manageable data volumes and turnaround times for processing which includes time required for calibration and beamforming of raw data. Whereas, MWA-Voltage capture system (VCS) performs tied-array offline beamforming which required storing original complex voltages at a very high data rate of 28 TB h$^{-1}$ \citep{tremblay2015}, which makes EDA2 more feasible for data-intensive applications such as pulsar observations across the usable frequency range and high cadence pulsar monitoring campaigns. 

\subsection{Target selection}\label{sec:obs2}

\begin{figure*}[htbp!]
\includegraphics[width=\textwidth,scale=0.9]{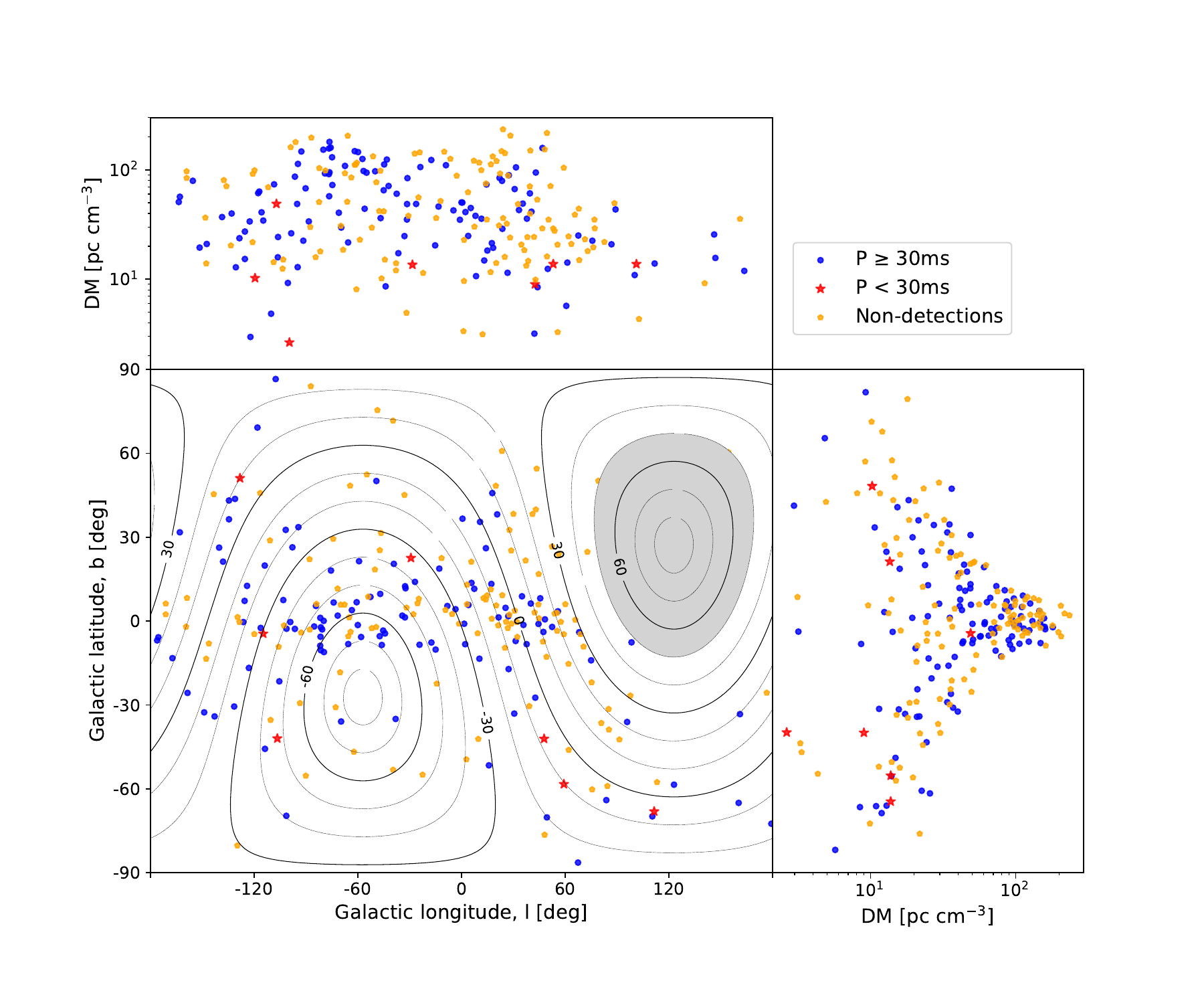}
\caption{Distribution of census targets in Galactic coordinates. Red and blue indicate the detected targets, with periods below and above 30 ms, respectively, and non-detected targets are shown in yellow. Contours show the line of equal declination, and the shaded region shows the region inaccessible due to low elevation.}
\label{fig:distribution}
\end{figure*}

The observing capability of EDA2 spans three octaves in frequency, and the sensitivity of the instrument varies with conditions and observing setup. As such, the direction-dependent response of these dipoles results in a system equivalent flux density (SEFD), which is a measure of sensitivity, varying by a factor of $\sim$ 2 between $\sim$ 1.3-2.2 kJy as a function of elevation(at frequency 160 MHz where the array is most sensitive). Furthermore, the frequency-dependent response of the instrument degrades the sensitivity by a factor of $\sim$ 4-8 at 70 MHz and similarly by $\sim$ 3-5 at 300 MHz. These values are for the zenith response of EDA2; at lower elevations, the sensitivity is lower. The dependence of EDA2 SEFD on frequency, LST, and elevation is detailed in \citep{sokolowski2022}, and a python code\footnote{\url{https://github.com/marcinsokolowski/station_beam}} along with a webpage\footnote{\url{http://skalowsensitivitybackup-env.eba-daehsrjt.ap-southeast-2.elasticbeanstalk.com}} has been provided for easy viewing.
Since a large fraction of known pulsars are in the galactic plane region, we also note that, due to the high system temperature, the sensitivity in those parts of the sky is lower than stated above. Based on the sensitivity and using available measurements of pulsar flux density we estimated the expected flux density at 160 MHz using \textit{PULSAR SPECTRA}\footnote{\url{https://github.com/NickSwainston/pulsar_spectra}} \citep{swainston2022}, and also by scaling literature values by a spectral index of -1.4, we limited our target sample to $>$ 20 mJy at 160 MHz, to avoid very long on source integration time. Additionally, we put the following constraints: (1) declination limit of $\delta <$ +30$^{\circ}$, except for a few known bright sources, to avoid loss in sensitivity due to low elevation, and (2) a DM cut-off of 200 pc cm$^{-3}$ to avoid non detections due to smearing caused by scattering at low frequencies. Figure \ref{fig:distribution} shows the distribution of all observed targets on the sky plane in Galactic coordinates. 

\subsection{Observations and data reduction}\label{sec:obs3}
Given the peak sensitivity of the instrument at 160 MHz and system limitations, all observations to confirm pulsar detections were carried out with 80 coarse channels centered around 169 MHz. Observations were carried out for an integration time between 0.5 and 2 hours based on expected flux density, and to limit the size of the raw data files ($\sim$ 0.45 TB hr$^{-1}$ for 28 MHz bandwidth).
Post detection in this frequency range, we did follow up observations over the EDA2 observing band. Since the goal was to study the evolution of pulsar spectra, their profile, and other features, the 50-250 MHz band was split into 5 separate bands. Due to the presence of radio frequency interference (RFI) in the FM band (88-108 MHz), coarse channels below 88 MHz were combined together to create a single frequency band pulse profile. For frequencies above 108 MHz, and up to 250 MHz, the band was split into 4 frequency chunks of $\sim$ 45 MHz centered around 131 MHz, and $\sim$ 28 MHz centered around 169, 197, and 225 MHz. The lower frequency was given a wider bandwidth due to the presence of some known satellite downlinks in this band \citep{dylan}, increased system temperature, and the potential turnover in pulsar spectra at these frequencies.

All observations were scheduled at respective target transit times to maximize available EDA2 sensitivity. The corresponding observing epochs are indicated under DM$_{\mathrm{Epoch}}^{\mathrm{EDA2}}$ column in Table \ref{tab:dmperiod}. The raw beamformed voltage data (dada file) was recorded for each coarse channel on the EDA2 server and was subsequently reduced to folded archives. Each dada file was reduced to 512 phase bins, 256 fine channels coherently de-dispersed, and 30 s averaged archives, using the \textit{DSPSR} package\footnote{\url{https://dspsr.sourceforge.net}} \citep{dspsr}. Since we could not store dada files for more than $\sim$3.5 hours of observations, data reduction time was estimated, and observations were spaced to get the previous data reduced before the next observation. All of this scheduling and data reduction was automated using a combination of python and bash scripts, which calculates the source transit time and checks the subsystem of the station so that there are no overlapping observations. After data reduction, 36 coarse channels around 169 MHz were combined into a single archive after removing the overlapping frequencies due to oversampled PFB, using a combination of \textit{psrsh} and \textit{psradd} tools from \textit{PSRCHIVE} software package\footnote{\url{https://psrchive.sourceforge.net/}} \citep{vanstraten,2004PASA...21..302H}. Bright RFI was excised using python in the combined file, in time and frequency, using median statistics. Detection was made using a combination of visual inspection, comparing against available pulsar profiles in EPN pulsar catalog\footnote{\url{https://psrweb.jb.man.ac.uk/epndb/}} and \textit{pdmp} tool from \textit{PSRCHIVE}. In case of marginal detection, the pulsar was reobserved with a larger observing bandwidth to increase the sensitivity and S/N. Among the 240 observed targets, we detected 120 pulsars in our observations. The final list of detected pulsars was then re-observed in the 50-250 MHz range to study the frequency dependence of the spectrum and profiles.

\section{Results and Analysis}
\subsection{Average pulse profile}\label{sec:pulse_prof}
As stated in Section \ref{sec:obs3}, observations were performed in five frequency bands, post detection around 160 MHz. Raw voltage data was processed using \textit{DSPSR}, applying pulsar ephemeris and other metadata to create coarse channel archives. These were subsequently combined to create a broadband profile, which was excised for RFI. After the RFI removal, the combined archives were averaged in time and frequency to get the integrated pulse profiles. The same process was applied for each of the five frequency bands, as described in Section \ref{sec:obs3}. In some cases, nearby frequency bands were combined to get an improved detection. The average pulse profiles for 120 pulsars with some example pulsars across multiple frequency bands are shown in Figure \ref{fig:psrprof}.

\subsection{DM correction}\label{sec:dm}
We applied the \textit{PSRCHIVE} program \textit{pdmp} to find optimal period and DM to improve the pulsar ephemeris, by maximizing the signal-to-noise of the frequency and time averaged profile. The search was performed in each frequency band of census, and the final period and DM correspond to maximum signal-to-noise (S/N) ratio. Additionally, the diagnostic plots from \textit{pdmp} were also used as a further verification of all detections reported in Section \ref{sec:pulse_prof}. We used a DM step size of 0.01 pc cm$^{-3}$ for our search, and pulsars with significant corrections were reprocessed with the updated DM to improve average pulse profiles. 

From the current sample, 110 pulsars showed a significant change in DM, with a median absolute value of 0.1 pc cm$^{-3}$, over a span of 8 to 38 years. Some notable examples which showed a large change in DM are PSR J1711-5350, J1240-4124, J0856-6137, and J1418-3921, where the DM change was -0.85, -0.83, 0.76, and 0.59 pc cm$^{-3}$ respectively. Since pulsar dispersion has an inverse square dependency on observing frequency\citep{handbook}, low frequency observations are generally more sensitive to DM measurements. Hence, we could get a significant change in DM in some cases, compared to those reported in ATNF pulsar catalog \citep{atnf}, generally from high frequency observations. Although it should be pointed out that the DM estimated using S/N maximization may be biased by the profile frequency evolution, which is more pronounced at low frequencies. Table \ref{tab:dmperiod} provides the list of all detected pulsars along with their catalog period, DM, and the corrected DM obtained from this work. Apart from higher sensitivity to pulsar DMs and profile evolution at low-frequencies, Section \ref{sec:dmvar} elaborates on other reasons for observed DM variations in pulsars.

\subsection{Calibration and mean flux density}\label{sec:calib}
The flux density (S$_{\nu}$) of a pulsar can be obtained from the radiometer equation in \cite{handbook} below:
\begin{equation} \label{eq:2}
    S_{\nu} = \gamma \frac{ \text{S/N }  T_\mathrm{sys}}{G \sqrt{n_\mathrm{p} \Delta \nu T}}\sqrt{\frac{W_\mathrm{eq}}{P-W_\mathrm{eq}}}
\end{equation}

\noindent where $\gamma$=1 is a correction factor, S/N is the signal to noise ratio of the pulsar average profile, P is the pulse period, T is the total integration time, $\Delta \nu$ is the effective bandwidth of the data, $n_\mathrm{p}$ is the number of polarizations, $\frac{G}{T_\mathrm{sys}}$=SEFD, and $W_\mathrm{eq}$ is the equivalent width of the pulse. Hence, we need to measure the SEFD and profile width for each observation to estimate the pulsar flux density.
The SEFD for EDA2 has already been measured using data, for a range of frequencies and observing parameters, and an estimate of the EDA2 SEFD is available from the sensitivity calculator in \cite{sokolowski2022}. These results were further compared against full electromagnetic simulation of the station beam in the commercial simulation package \textit{FEKO}\footnote{\url{https://altair.com/feko}}. It is worth pointing out that SEFD estimates for total Stokes I are obtained by taking a quadrature sum of two linear polarizations of the antenna, which is only strictly valid for an unpolarized source and high source elevation, which can lead to mis-estimations at low elevations by $\sim 10-20 \%$ \citep{Sutinjo}.

The pulse width is measured from the average pulse profiles in Section \ref{sec:pulse_prof}, by modeling the profile as a sum of Gaussian components by visually inspecting each profile. The maximum number of Gaussian components used for profile fitting was determined using Bayesian information criteria \citep[BIC,][]{BIC}. Based on this fitting, we determined the S/N of the profile, along with pulse width at half-maximum (\textit{w$_{50}$}) and ten percent of the maximum (w$_\mathrm{10}$). The flux density values are estimated using \textit{w$_{50}$} measurements. While this entire method may not be the most accurate way to calculate fluxes, but it's likely the only possible one for low-frequency arrays. To account for all errors including SEFD measurements, we assume a total error of 50$\%$ in the calculated flux density measurements. Using these measurements along with other catalog measurements, we fit spectra using the \textit{PULSAR SPECTRA} package. Table \ref{tab:fluxdensity} provides the calculated flux density measurements for all detected pulsars at multiple frequencies, along with pulse width measurements and other parameters in equation \ref{eq:2}. 

\subsection{Pulsar Spectra}\label{sec:spec}
Since the variation in flux density measurements and calibration from different telescopes can introduce systematic errors in spectral modeling, we adopt the method developed in \citep{jankowski441psr} and use other literature measurements to get a robust fit of spectra using the \textit{PULSAR SPECTRA} package. Details of the implementation of spectral modeling are given by \cite{swainston2022}. In summary, it uses maximization of a modified Gaussian likelihood function, which downweights outlier data points. We used the five analytical models from \citep{jankowski441psr}, simple power-law (PL), broken power-law (BPL), power-law with a low frequency turnover (PLL), power-law with a high frequency cut-off (PLH), and double turnover spectra (DT), where the best fitting model was determined using the Akaike information criteria (AIC). This measures the information retained in the model without overfitting. The model with the lowest AIC is chosen as the best-fitted model. Additionally, we determine the probability, P$_{best}$, as the inverse sum of likelihoods relative to the best fitting model, that the best fitting model is the true model. Table \ref{tab:spectable} gives the spectral fitting parameters for all 120 pulsars detected in this census. 

Among these pulsars, 20 pulsars show a change in their spectral model, and four pulsars, namely J0038-2501, J0636-4549, J1225-5556, and J1510-4422, had no spectra before the addition of new low-frequency data from EDA2. However, since the model fitting is sensitive to the data, there are cases where the AIC value of best fit model is close to one or more of the other analytical models. 
Figure \ref{fig:spectra} shows the flux density spectra for some of these pulsars which showed a changed in spectra with the addition of census data shown in green stars, along with the spectra of other census detected pulsars. It is worth pointing out that we have used the spectral fitting method as described in \citet{swainston2022}, without any pulsar by pulsar fine tuning, given the large number of pulsars in this census.

Additionally, 13 pulsars show an improvement in their spectral modeling with the addition of EDA2 measurements and are marked with a star in Table \ref{tab:spectable}. Among the EDA2 pulsars, 52 show a turnover in spectra at low frequencies, including five of the seven detected millisecond pulsars. The mean value of low frequency turn-over in these pulsars is 115$\pm$10 MHz, much better constrained than the value of 130$\pm$80 in \cite{Izvekova}. For pulsars with a simple power law behavior, the mean spectral index is -1.78$\pm$0.17, which agrees with the value of -1.6$\pm$0.03 in \cite{jankowski441psr}.

\subsection{RM-synthesis}\label{sec:rmsynth}
Due to the high level of linear polarization in pulsars \citep[see, e.g,][]{kerr} and pulsar RM being proportional to $\lambda^2$, makes low frequency observation appealing for precision RM measurement. At the same time, low spatial resolution at these wavelengths can also cause depolarization. Although the simplest approach to measuring RM is by fitting a linear model to the polarization angle as a function of $\lambda^2$, and can be readily done using \textit{rmfit} routine in \textit{PSRCHIVE}, this assumes that there are no additional depolarizing sources along the LOS. This can be resolved using RM-synthesis techniques by \cite{Burns1966} and further developed by \cite{Brentjens}.
We can describe the complex linear polarization in terms of Stokes parameters (I, Q, U, V) as
\begin{equation}\label{eq:3}
    P = Q+iU = pIe^{i2\psi} = Le^{i2\psi}
\end{equation}
where I is the total intensity, Q and U are the components of linear polarization, V is the circular polarization, L is the linear polarized intensity, and \textit{p} and $\psi$ are the fractional degree and polarization angle of linear polarization. While propagating through a magnetized plasma, polarization vector is rotated as a function of observing wavelength squared($\lambda^2$). For a single point source along the LOS, the rotation is linearly proportional to $\lambda^2$
\begin{equation}\label{eq:4}
    \psi = \psi_0 +RM\lambda^2
\end{equation}
where the proportionality constant RM along the LOS is given by:
\begin{equation}\label{eq:5}
    RM = 1.232 \int n_e(l)B_{\parallel}(l)dl 
\end{equation}
where $n_e(l)$ is the spatial electron number density (cm$^{-3}$) and  $B_{\parallel}(l)$ is the LOS magnetic field ($\mu$G), both function of position(l). Although pulsars are Faraday thin sources, application of RM-synthesis methods over the linear approach allows us to discern weakly chromatic noise due to instrumental leakage, from astrophysical RMs. Additionally, theoretical predictions claim that there should not be intrinsic Faraday rotation within pulsar magnetospheres \citep{Wanghan}. On the other hand, \cite{Noutsos2009} claim to  have detected such effects as variations of RM as a function of pulse phase. Our RM measurements can be used to confirm these results. This can help us improve models of the magnetosphere. Hence, following \cite{Burns1966}, we consider the generalized RM, known as the Faraday depth ($\phi$), defined at each point along the LOS to the source. Using equation \ref{eq:3} and \ref{eq:4}, we can write,
\begin{equation}\label{eq:6}
    P(\lambda^2) = Le^{2i(\psi_0+\phi\lambda^2)}
\end{equation}
Assuming the same $\psi_0$ at all Faraday depths, we can write the Fourier relation:
\begin{equation}\label{eq:7}
    P(\lambda^2) = \int^{+\infty}_{-\infty} F(\phi) e^{2i\phi\lambda^2} d\phi 
\end{equation}
where $F(\phi)$ is the Faraday dispersion function (FDF) representing intrinsic linear polarization, and will show a peak at an RM given by equation \ref{eq:4} for a single compact source along the LOS. Any deviation from this could be astrophysical or instrumental in nature. 

Given that $\lambda^2<0$ solutions are not physical and that we can only sample the complex linear polarization at discrete values, after inversion, the above equation can be written as a discrete Fourier transform(DFT):
\begin{equation}\label{eq:8}
    \tilde{F}(\phi) = \frac{1}{N_c} \sum_{j=1}^{N_c} P(\lambda_j^2)e^{-2i\phi(\lambda_j^2-\lambda_o^2)} 
\end{equation}
where $N_c$ is the number of channels, j is the channel index, $\lambda_j$ is the channel wavelength, and $\lambda_0$ is a reference wavelength following \citep{Brentjens}. Since we are only concerned about the RM value and hence the peak of FDF, the choice of $\lambda_0$ is arbitrary. \cite{Brentjens} also showed that the discrete synthesized FDF in equation \ref{eq:8} is a convolution of the 'real' FDF with the RM spread function (RMSF):
\begin{equation}\label{eq:9}
    \tilde{R}(\phi) = \frac{1}{N_c} \sum_{j=1}^{N_c} e^{-2i\phi(\lambda_j^2-\lambda_o^2)}
\end{equation}

which can be deconvolved from the synthesized FDF in equation \ref{eq:8}, using RM-CLEAN algorithm implemented by \citep{Heald}, to obtain the best S/N ratio. This helps reduce confusion due to instrumental leakage, which peaks at $\phi \sim0$ rad m$^{-2}$. This RMSF peak also limits the resolution of the FDF, and is quantified by full width half maximum (FWHM) \citep{Schnitzeler} as, 
\begin{equation}\label{eq:10}
    \delta\phi(\frac{rad}{m^2}) = \frac{3.8}{\lambda_{max}^2-\lambda_{min}^2}  
\end{equation}
where $\lambda_{max}$ and $\lambda_{min}$ are the span of data wavelength in meters.

\subsection{Ionospheric RM}\label{sec:ionrm}
\begin{figure}[htbp!]
\includegraphics[width=0.47\textwidth]{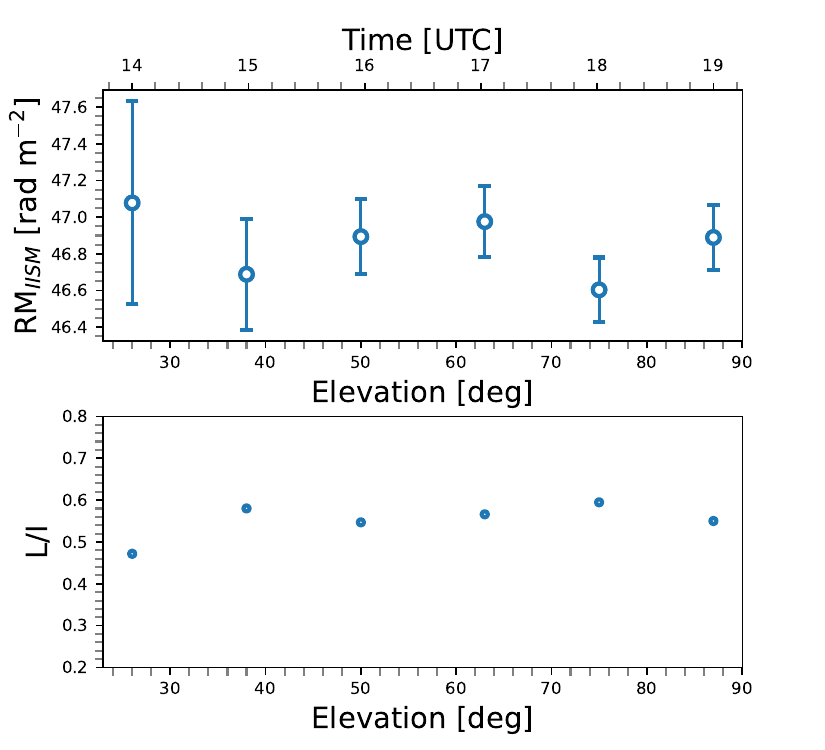}
\caption{(Top) RM$_{ISM}$ for pulsar J0630-2834 as a function of source elevation and observing time in UTC on MJD 60640. (Bottom) Variation of fractional linear polarization with elevation.}
\label{fig:elev}
\end{figure}

Since the Earth's ionosphere imparts a significant RM (RM$_{ion}$) to the observed RM (RM$_{obs}$), it needs to be subtracted to get the 'true' RM and estimate the ISM magnetic field:
\begin{equation}\label{eq:11}
    RM_{ISM} = RM_{obs} - RM_{ion}
\end{equation}
Ionospheric corrections depend on Earth's magnetic field, which has small variations \citep{Alken2021}, and on the total electron content (TEC) of the ionosphere, which changes significantly on all time scales. Previous studies have shown that due to these variations, it is difficult to apply precise correction. Hence, to get most reliable measurements TEC at the telescope site should be measured with a GPS receiver with high time resolution \citep{Kumar2025}. However, since this setup is not readily available, we have to rely on global ionospheric models, which interpolate values from a discrete set of receivers to create a map of the ionosphere on hourly timescales.
Following this, we employed two methods to measure the ionospheric contribution to the RM: 1) using the method described in \cite{Kumar2025} along with IGS and CODE ionospheric maps, which provided the most accurate RM measurements in \cite{Kumar2025} on short and long timescales, 2) using the \textit{IonFR} code \citep{Sotomayor} along with the IGS model. Both models assume a single thin shell of ionospheric plasma for the calculation of RM values. The mean and median of $RM_{ISM}$ from both these methods were compared against literature values, and were found to be the same in each case, hence final values are reported using the \textit{IonFR} method. We also investigated the measurement of ionospheric subtracted RM as a function of observing time and target elevation, to assess the validity of this method and quantify any variation in RM measurements due to instrumental leakages at low elevations.
Figure \ref{fig:elev} shows the RM$_{ISM}$ for pulsar J0630-2834 as a function of source elevation and observing time in UTC on MJD 60640. We see that all values are consistent within error bars, with the error increasing around UTC=15, which is around sunrise for MRO, during which we expect higher variations in the ionospheric conditions. This pulsar has been previously shown to exhibit no RM variations on long timescales \citep{Kumar2025}.

\subsection{Pulsar RM and phase resolved RM}\label{sec:phaserm}
We applied the RM-synthesis method to each of our detected pulsars, in each frequency band, to measure the pulsar's highest S/N RM as well as RM as a function of pulse phase, the profiles were downsampled to 64 or 128 bins and $<1000$ fine frequency channels to increase the S/N per bin. The RM-synthesis method was then applied to measure the pulsar RM as a function of pulse phase in the Faraday depth range of $-200<\phi<200$, which covers the range of expected RM for these pulsars as reported previously in ATNF catalog, with a step size of $0.01$ rad m$^{-2}$, and FDF peaks with S/N $>6$ were reported for each pulsar. For pulsars with detections across multiple EDA2 frequency bands, the final value is reported using the one that has the highest S/N. In total, we have detected significant pulsar RM in 40 pulsars. Table \ref{tab:rmtable} provides our RM measurements and ionospheric corrections for census pulsars, along with corresponding values reported in the ATNF catalog and at frequencies around 1.3 GHz with the MeerKAT telescope \citep{Posselt}. Despite the low RM value for many of these pulsars and their FDF peak falling within the FWHM of RMSF peak generated by instrumental polarization at $\phi \sim 0$ rad m$^{-2}$ given by equation \ref{eq:10}, 28 of these measurements are unaffected by RMSF peak due to the instrumental leakage.

\begin{figure}[htbp!]
\includegraphics[width=0.49\textwidth, height=16cm]{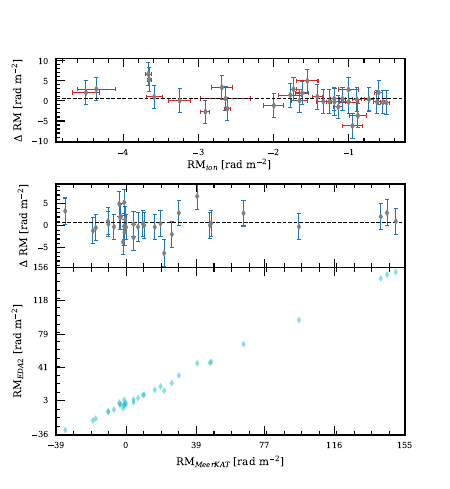}
\caption{The top panel of the figure shows the difference between RM$_{ISM}$ from EDA2 and MeerKAT telescopes ($\Delta$ RM) against the ionospheric RM (RM$_{ion}$) measured for EDA2 for each of those pulsars. The bottom panel shows the one-to-one RM$_{ISM}$ for pulsars in Table \ref{tab:rmtable} for EDA2 and MeerKAT. The central panel shows $\Delta$ RM vs RM$_{ISM}$ for MeerKAT. The dashed line in both the top and central panels indicates the mean value of $\Delta$ RM}
\label{fig:rmdiff}
\end{figure}

In Figure \ref{fig:rmdiff}, we show the difference between ionosphere-corrected RM values for each pulsar between EDA2 and MeerKAT telescopes. We see no particular trend between $\Delta$RM  which is the difference between RM$_{ISM}$ from EDA2 and MeerKAT telescopes and RM$_{ion}$ measured for each of the EDA2 observations. Moreover, the mean value of $\Delta$RM is close to zero. Since the contribution of ionospheric RM should be normally distributed around zero across multiple observations, this indicates no systematic trend and that variations in the $\Delta$RM is mainly due to fluctuation in ionospheric RM correction between these two telescopes.

\begin{figure*}[htbp!]
\includegraphics[width=\textwidth, height=8cm]{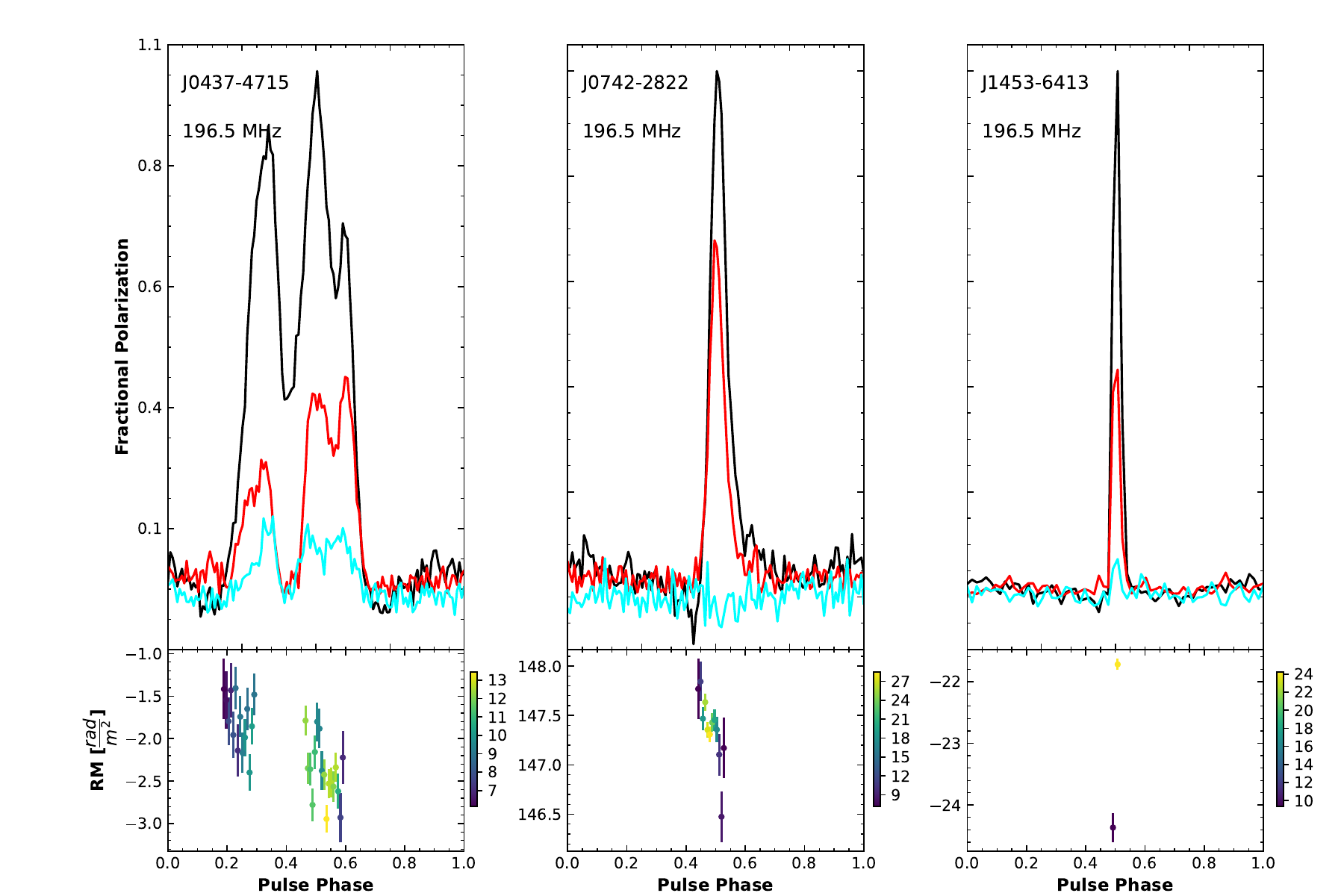}
\caption{Phase resolved RM variation for three pulsars. The top panel of each subplot shows the total intensity, fractional linear polarization, and fractional circular polarization in black, red, and cyan, respectively. The bottom panel shows RM values measured across the pulse phase, along with a colorbar which indicates the S/N of the FDF peak.}
\label{fig:rmphase}
\end{figure*}

Since we measure pulsar RM across the pulse phase bin for each pulsar, we are also able to measure a variation in RM over the pulse profile, indicating a magnetospheric origin as explained previously in Section \ref{sec:rmsynth}. We found phase-resolved RM variations in six pulsars, however in three of those there was only on RM measurement which deivated from the rest and the peak S/N of FDF peak was also low. Figure \ref{fig:rmphase} shows phase-resolved RM variations for the other pulsars. Among these, RM of J0437-4715 falls within the RMSF peak and hence could be biased. We also see a gradual variation in RM for J0742-2622, which has a scattered pulse profile. The most significant variations are seen in J1453-6413 which is far from RMSF peak and where the difference between RM measurements across the pulse phase is larger than the error bars on individual measurements. Moreover, both of these measurements for J1453-6413 FDF peak have a high S/N.

\subsection{Polarimetry}\label{sec:polarimetry}
Pulsar polarimetry can provide useful insights into the direction-dependent phase stability of the instrument and leakage between orthogonal polarizations of the telescope. This becomes particularly useful for prototype arrays like EDA2 using SKA-Low technologies, which require full validation and verification. Figure \ref{fig:stokesprof} shows the full Stokes integrated pulse profiles for 36 best pulsars with RM detections in Table \ref{tab:rmtable}.

We compared our polarimetric pulse profiles with those at similar frequencies in the EPN pulsar database\footnote{\url{http://www.jodrellbank.manchester.ac.uk/research/pulsar/Resources/epn/}}. We found the degree of linear polarization to be similar in most cases, against those in the EPN database, mainly comparing against the results at $\sim$ 150 MHz and $~\sim$400 MHz from \cite{Noutsos} and \cite{Gould_1998} respectively. For EPN pulsars with polarization profiles at 150 MHz, the degree of linear polarization was about 5-10$\%$ lower than the EDA2 measurements. This could be due to the large FWHM of the EDA2 beam observations leading to depolarization effects. Additionally, due to uncalibrated polarimetry and variation in dipole response with pointing direction, there could be leakage between the two orthogonal polarizations. To investigate this further, we looked at the variation in fractional linear polarization for J0630-2830 with elevation as shown in the bottom panel of Figure \ref{fig:elev}, where we get a variation of 2-3$\%$ for elevations above 40$^{\circ}$, and $\sim$4-7$\%$ below that. This indicates the presence of residual errors due to uncalibrated polarimetry, which needs to be investigated further. 

For profiles at higher frequencies, a direct comparison of linear polarization fraction with EDA2 measurements is not useful. This is due to the following, assuming radius-to-frequency mapping \citep[RFM,][]{Cordes}, where lower frequencies originate higher in the magnetosphere and curvature radiation framework with two propagating modes \citep{Wang}, higher frequencies are emitted from lower heights where the rotation is weaker and the region of the two modes overlap, leading to depolarization and lower fractional polarization at higher frequencies. This is evident from the higher linear polarization fraction for pulsars J1709-1640, J1935+1616, and J2155-3118 in EDA2 compared to 400 MHz profiles from \cite{Gould_1998} in EPN by a factor of 2. For 20 pulsars, we get multiple detections of pulsar RM across multiple EDA2 bands as shown in Figure \ref{fig:multipol}. However, we see no particular trend in fractional linear polarization, which may suggest that the prediction of \citep{Wang} do not hold for all pulsars or that we need to compare widely separated frequencies. Another possibility is that it occurs due to the uncalibrated instrumental XY phase.

\section{Discussions}\label{sec:discuss}

\subsection{Profile evolution}\label{sec:profevol}
We obtained pulse profiles for 120 pulsars in the frequency range of 50-250 for up to 5 narrow bands, which allows us to study various features of pulsars that evolve with frequency, including the pulse profile. Although many low-frequency pulsar censuses have been reported in the last decade with sensitive instruments like LOFAR and MWA \citep[see, e.g.,][]{bilous2016,Bhat_2023}, we have been able to obtain 32 new pulse profiles below 150 MHz, especially five below 100 MHz. These detections can help inform future studies of morphological assessment of pulse profiles, which can improve our understanding of pulsar emission\citep{Rankin}. Modeling of these pulse profiles, as stated in Section \ref{sec:calib}, also resulted in the detection of multiple components across pulse profiles. For 72, 36, and 11 pulsars, we detected the presence of 1,2, and 3 components, respectively, and 4 components in J1057-5226. In a few pulsars, the number of components required to model the pulse profiles changed across our observed band, which could be a result of a change in spectral index and hence S/N of different "core" and "conal" components with frequency \citep{basumitrageorge}. For single-component pulsars, the evolution of pulse width with frequency was generally consistent with the predictions of RFM, decreasing with frequency \citep{Kijak}, but remained almost constant in some cases. Availability of pulse profiles at multiple frequencies also enabled us to make a qualitative study of pulsar scattering. In our census, 11 pulsars show visible presence of scattering, among which the most notable are J1820-0427 and J1453-6413, which also show profile evolution. 

We observe an inverse correlation between pulse duty cycle (ratio of width and period) and pulsar period, except for the case of J2256-1024. Such broad correlations have been reported for pulsars in the past \citep{Lorimer1995}. For normal pulsars with period around 0.5\,s pulse duty is $\sim$4-6$\%$, however the scatter in duty cycle for normal pulsars is $\sim$ factor of 3 higher than those of MSPs. This impacts the fraction of visible pulsars, since pulsars with wider beams are more likely to intersect the LOS to Earth, increasing their chance of detection. Hence, these measurements have implications for pulsar population studies and pulsar beaming fraction \citep{Lyne1988}.

\begin{figure*}[htbp!]
\includegraphics[width=\textwidth]{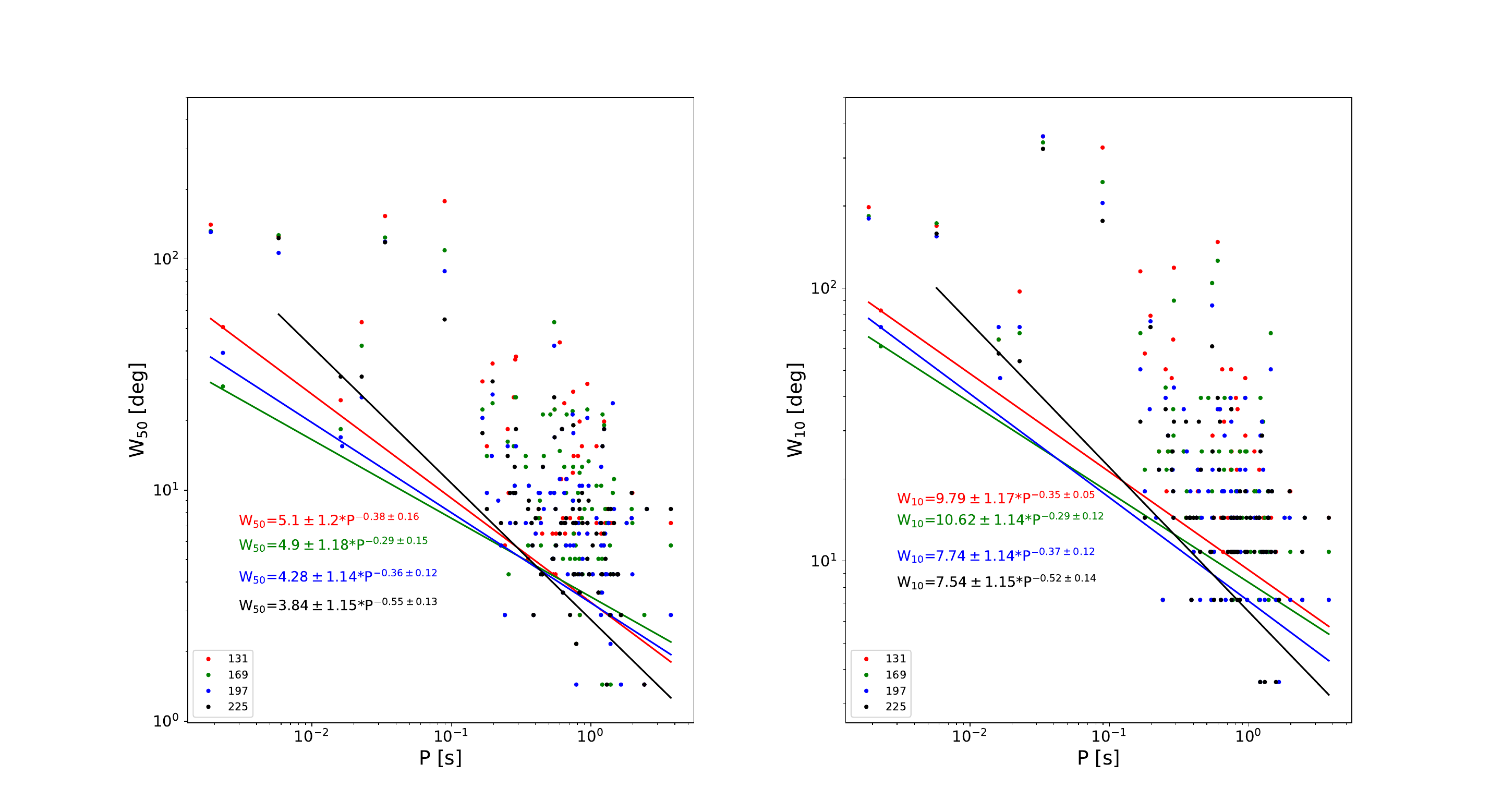}
\caption{Plot of pulse width versus pulsar period, along with the Lower boundary line in solid lines (Left) using W50 (Right) using W10. The colors represent respective frequencies in the legend.}
\label{fig:edabound}
\end{figure*}

 \cite{olszanskimitra} showed that there exists a relation between profile width and period and how that evolves with frequency. This is depicted using a Lower boundary line (LBL) of the form W$_{B}$$*$P$^{-d}$, where W$_{B}$ is the boundary width and P is the period of the pulsar, where they obtained d $\sim$ -0.5. This width is thought to correspond to narrowest emission beam and can help estimate the inclination angle. We perform this analysis for our census measurement using quantile regression. Figure \ref{fig:edabound} shows the result of this analysis along with parametric fits to the above relation. Although the value of d at our highest frequency is $\sim$-0.5$\pm$0.1, consistent with those above, at lower frequency it deviates to $\sim$-0.3$\pm$0.1, suggesting frequency evolution. Similarly, considering the W$_{50}$ case, the value of boundary width at the highest census frequency is 3.84$\pm$1.15$^{\circ}$, consistent with 327 MHz measurement of 2.37$^{\circ}$ from \cite{skrzypczak}; however, it increases at lower frequencies. Assuming a power law evolution of boundary width with frequency, we find the power law index of -0.35$\pm$0.03 comparing the value with the value at 327 MHz. 

\subsection{Flux densities}\label{sec:measured flux}
\begin{figure}[htbp!]
\includegraphics[width=0.45\textwidth,height=10cm]{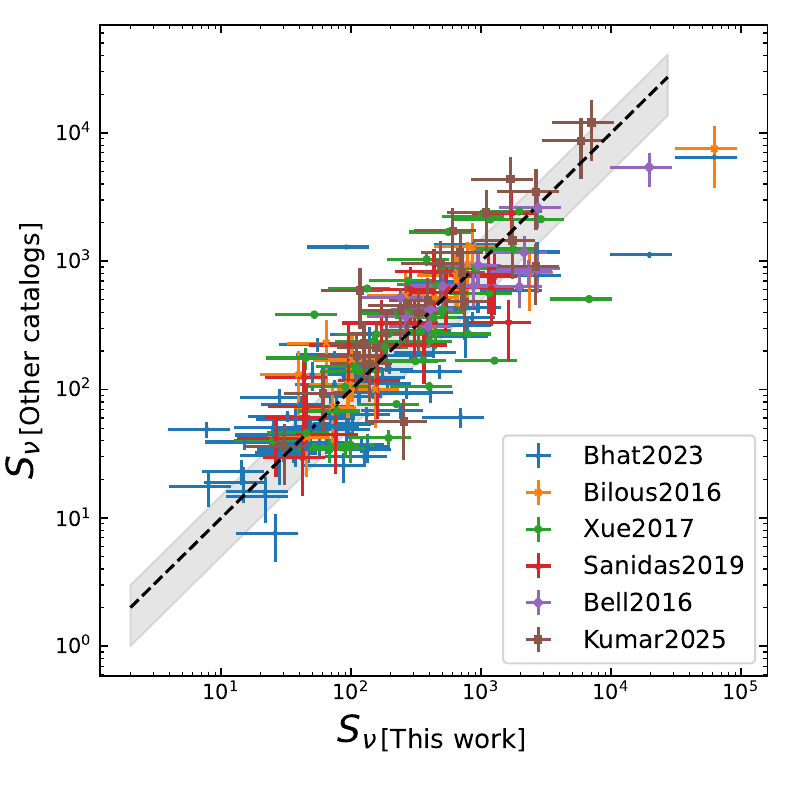}
\caption{Comparison of our census measurement of mean flux densities with those at similar low frequency observations. Colors in the plot represent the comparison of our census measurement at a frequency close to other catalogs, as indicated in the legend. The dashed line indicates equal flux values, and the shaded grey region represents the 50 and 150 $\%$ region of the equal flux value.}
\label{fig:fluxcomp}
\end{figure}

We measured mean flux density for 120 pulsars at multiple frequencies as described before in Section \ref{sec:calib}, providing several new measurements, including five below 100 MHz and 28 below 150 MHz, where no measurements had been reported before. Additionally, many of these measurements have an overlap with prior low-frequency census measurements at similar frequencies and comparable bandwidths, which can be used for independent verification of our method. Figure \ref{fig:fluxcomp} shows the comparison of our measurements with other low-frequency catalogs, where we only compared measurements at nearby frequencies. The plot shows detection down to a flux density level of $\sim$15-20 mJy, indicative of our sensitivity limit as discussed in Section \ref{sec:obs2}. We compared census measurements at $\sim$168 MHz with \cite{Bhat_2023}, \cite{bilous2016} and \cite{bell} at 154, 149 and 154 MHz respectively, census measurements at $\sim$ 131 MHz with 135 MHz in \cite{sanidaslotass}, census measurements at $\sim$198 MHz with 185 MHz in \cite{Xue_2017}, and census measurements at $\sim$71 MHz with 79.2 MHz in \cite{Kumar2025}. For \cite{bhat2004}, \cite{bell}, and \cite{Xue_2017} observations were made with similar bandwidths, whereas census bandwidth was higher compared to \cite{Kumar2025}, and \cite{sanidaslotass}, and significantly lower compared to \cite{bilous2016}. However, in Figure \ref{fig:fluxcomp} we see that almost all measurements are in agreement within a factor of 2, with large deviations for a few very bright pulsars irrespective of their DMs, which validates our measurements of mean flux densities. However, our measurements are slightly biased towards the higher values around the equality line, which could be due to some of these other census measurements being derived from imaging observations which measure period averaged flux density and are biased towards pulsars with large duty cycle.
Five pulsars show a factor of 5 increment in our census flux compared to other catalogs. Among which the variation in J0835-4510 (Vela) and J0534+2200 (Crab) is likely due to highly variables scattering from the associated nebula affecting profile modeling including J0742-2822 which also has a scattered profile. Whereas the variation in J0437-4715, and J2145-0750 both of which are nearby MSPs with a DM of 2.6348 and 9.0008 pc cm$^{-3}$ respectively, is likely due to scintillation. The rms of the relative difference between these flux density measurements from multiple census, except the outlier pulsars, is 118 $\%$. However, the measurements of flux density are almost equally distributed around the equality line of Figure \ref{fig:fluxcomp}. 

\subsection{Spectral Analysis}\label{sec:specanalysis}
\begin{figure*}[htbp!]
\includegraphics[width=\textwidth, height=6cm]{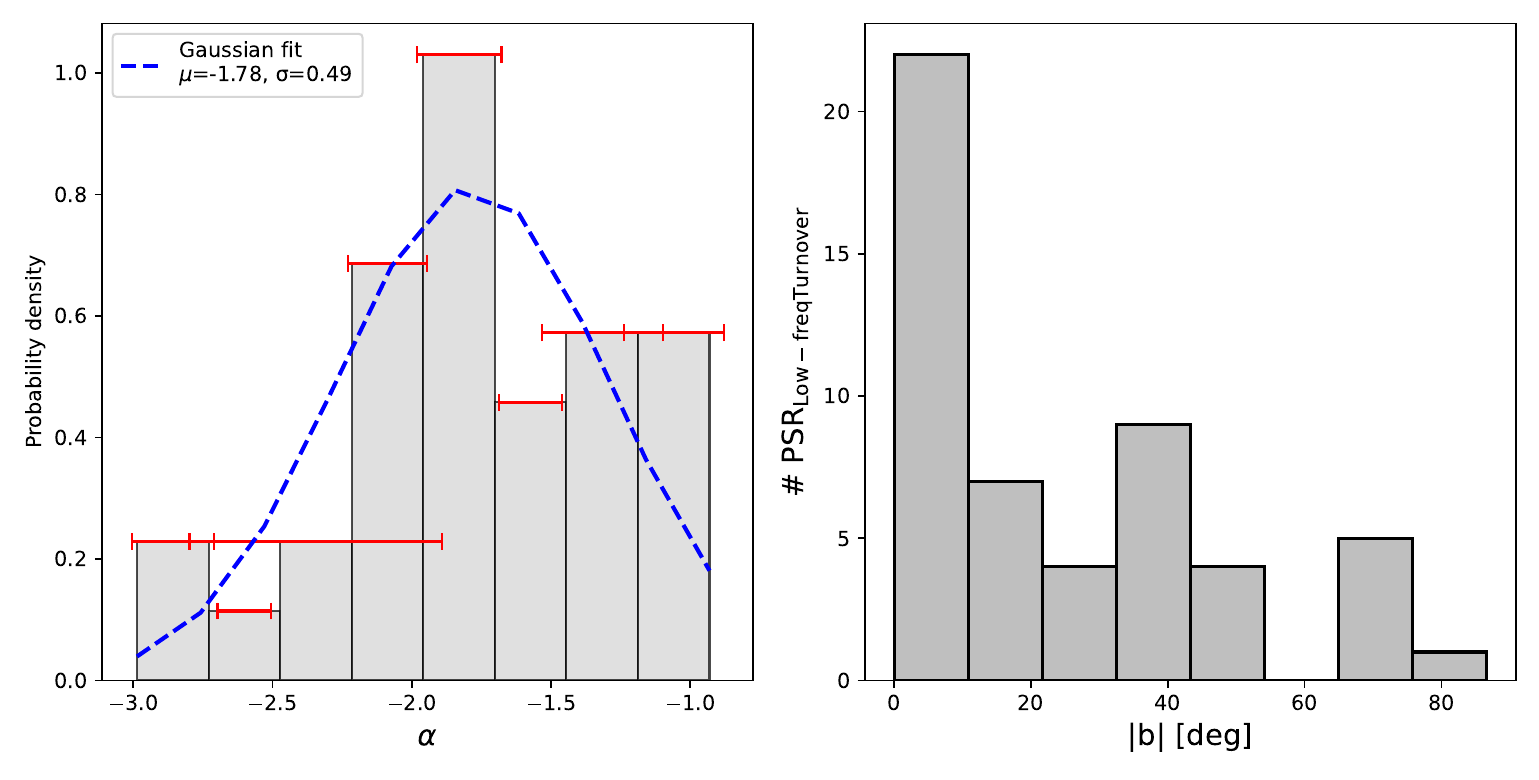}
\caption{Left: Distribution of spectral index for pulsars classified as PL, along with corresponding error of each bin indicated in red. We also show a Gaussian fit to the data in blue. Right: Distribution of distance from the Galactic plane, as absolute value of Galactic latitude ($|$b$|$) for pulsars classified as PLL and DT}
\label{fig:specsimple}
\end{figure*}

Among the census pulsars 34 follow a simple power law with a mean and median spectral index of -1.78$\pm$0.17 and -1.79$\pm$0.15, which is steeper than the empirical value of -1.41 reported by \cite{bates} near 1 GHz and but closer to the weighted spectral index of -1.6$\pm$0.54 of \cite{jankowski441psr}, however better constrained. Figure \ref{fig:specsimple} shows the spectral index distribution for these pulsars, along with a Gaussian fit to the data with mean and sigma values of -1.78 and 0.49, respectively. This value is in better agreement with measurements of \cite{jankowski441psr}. To validate the normality of our Gaussian fit, we perform a Pearson $\chi^2$ test. We find that the null hypothesis is accepted with a p-value of 0.53 with a $\chi^2$ of 1.28, suggesting that the spectral indices are normally distributed. Additionally, we do not find any correlation between spectral index and Galactic latitude or longitude to suggest that the measured spectral indices are affected by the Galactic plane region.

In our sample, 52 pulsars showed a turnover at low frequencies, with 14 and 38 classified as DT and PLL, respectively, with a mean value for $\nu_{peak}=$115$\pm$10 MHz in agreement with \cite{Izvekova} but higher than the measurements in \cite{Kumar2025}. Given the small sample of MSPs in our census, it is difficult to deduce if they show turnover at low frequencies. Since some of these MSPs turnover in agreement with the results of \cite{rahulsharan}; however, others do not follow and are consistent with the suggestion of \cite{Kuzmin}. Prior studies of turnover in spectra have also suggested the origin of these turnovers to absorption in the ISM \citep[see, e.g.][]{Sieber, Malov}. To investigate this further, we looked for any correlation between the turnover frequency and the Galactic location of the pulsars but found none. Figure \ref{fig:specsimple} shows the Galactic latitude distribution of these 52 census pulsars, indicating an excess in the Galactic plane region, which could be indicative of absorption as stated above. However, this could also be due to selection bias since most of our pulsars are detected at small Galactic latitudes as shown in Figure \ref{fig:distribution}. 

\begin{figure*}[htbp!]
\includegraphics[width=\textwidth, height=6cm]{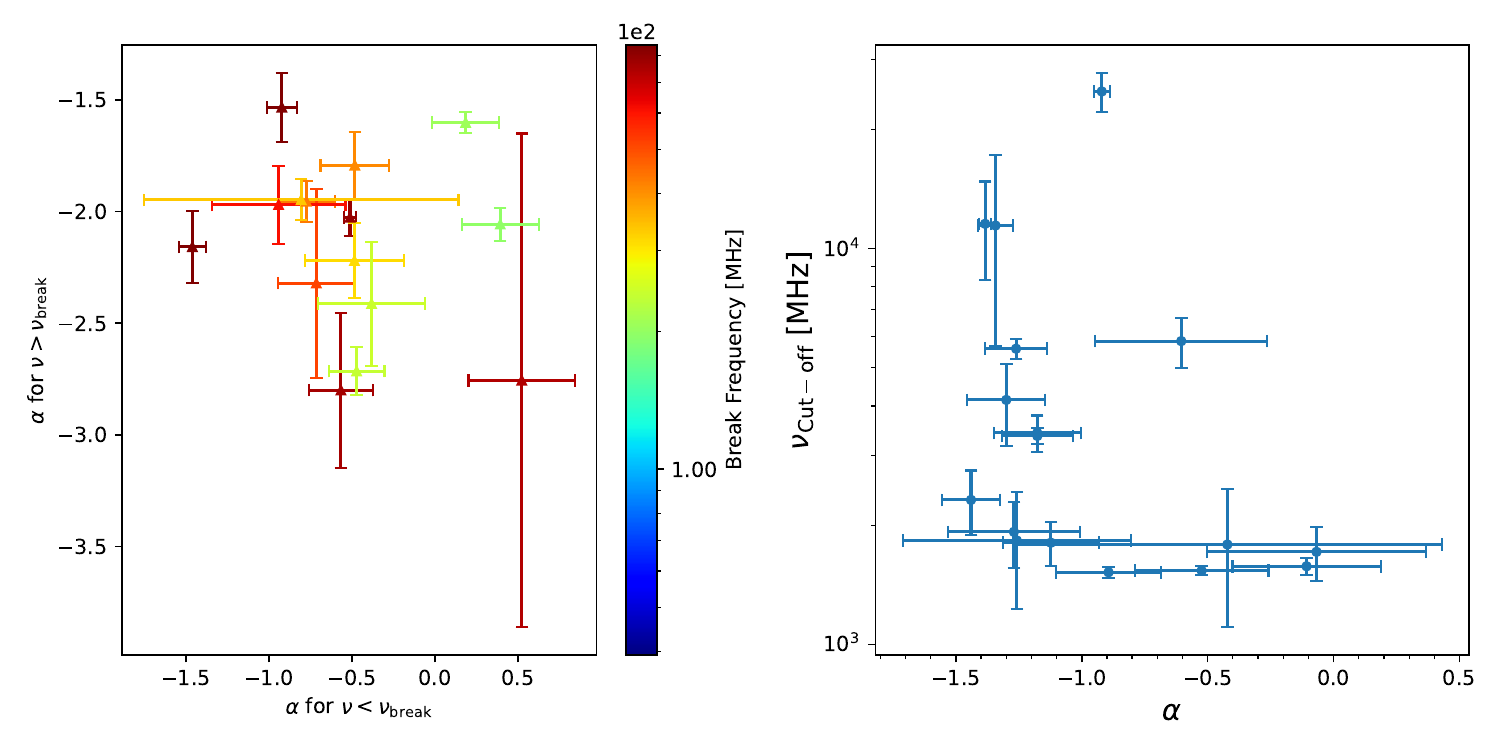}
\caption{Left: Spectral index below and above the spectral break for 16 pulsars classified as BPL, along with the break frequency indicated by the colorbar, except for outlier pulsar J0459-0210. Right: Plot of spectral index versus high frequency cut-off for 17 PLH pulsars in our census.}
\label{fig:specbreak}
\end{figure*}

For 16 BPL pulsars of our census left panel of Figure \ref{fig:specbreak} shows the plot of spectral index below and above the break, along with the break frequency ($\nu_{break}$), except for J0459-0210 which is an outlier due to the large errorbar on its spectral index since all except one of its flux density measurements are below 300 MHz. From the figure we can see the general trend where the spectral index flattens below the break frequency as the break frequency moves towards lower frequencies, which could be indicative of low frequency turn over in pulsars with lower $\nu_{break}$ and high frequency cut-off in pulsars with higher $\nu_{break}$ frequency. 

Additionally, 17 census pulsars show a high frequency cut-off ($\nu_{cut-off}$)and are classified as PLH. For these, the right panel of Figure \ref{fig:specbreak} shows the plot of 
$\nu_{cut-off}$ versus the spectral index, where we see two trends and a couple of outliers. First, pulsars show a cut-off at $\nu \sim1.5$ GHz due to thermal free-free absorption in the pulsar surrounding \citep{Kijak}, but with a wide range of spectral index values. Second, we see pulsars with an order of magnitude spread in $\nu_{cut-off}$ above 1.5 GHz but with a spectral index around $\sim$ -1.35, close to the spectral index value of -1.41 reported by \cite{bates} at higher frequencies, suggesting that this value may be the correct representative value at these frequencies. However, it is difficult to conclude with any certainty given the small sample of our census pulsars. 

\subsection{DM variations}\label{sec:dmvar}
Pulsars can have large proper motions and are among the fastest-moving objects in the Galaxy\citep{Hobbsvel}. Due to the relative motion of the pulsar in ISM, the pulsar DM along any LOS can change with time depending on ISM variability, proper motion of the pulsar, and span of observations. Hence, long-term pulsar monitoring has resulted in the measurement of DM variability, often linear and with additional periodic or stochastic variations \citep[see, e.g.][]{Coles2015,Kumar2025}. Sometimes these variations arise due to the presence of discrete structures like bubbles and nebulae\citep{Ocker}. However, it is possible to detect the stochastic signals due to DM variations if the deterministic signals along the LOS red due to discrete structures can be corrected, although it is often difficult \citep{Lammodel}. 
Analysis of these variations can cast light on the nature of turbulent ISM, as shown by \cite{Armstrong}, where they showed that DM variation in ISM on long time scales has a steep red spectrum. Due to the inverse square dependence of DM on observation frequency, low-frequency pulsar observations provide high precision DM measurements, useful for investigating these variations. Even though DM monitoring would be most suitable to investigate such effects, comparison of our census measurements to those in pulsar catalogs like ATNF can provide some crude estimates.
We define the rate of DM variation and associated error by comparing our measurements as stated in Section \ref{sec:dm}, with those of ATNF catalog values as,
\begin{equation}\label{eq:12}
    \left|\frac{dDM}{dt}\right| = \left|\frac{\Delta DM}{\Delta MJD/365.25}\right|
\end{equation}

where $\Delta DM$ is the change in DM between census and ATNF and $\Delta MJD$ is the difference in epochs at which those measurements were taken. Similarly, the error is given by,

\begin{equation}\label{eq:13}
    \epsilon_{\left|\frac{dDM}{dt}\right|} = \left|\frac{\epsilon_{\Delta DM}}{\Delta MJD/365.25}\right|
\end{equation}

where $\epsilon_{\Delta DM}$ is the quadrature sum of DM errors in the census and ATNF catalogs. Only those pulsars were chosen where both the census and ATNF had a reported uncertainty. Additionally, for final analysis, we rejected the pulsars where the value of $\epsilon_{\Delta DM}$ was bigger than the value of $\Delta DM$. This gave us a final sample of 43 pulsars with DMs in the range 5-180 pc cm$^{-3}$, with a time span of 8 to 38 years between the two measurements. 

\begin{figure}[htbp!]
\includegraphics[width=0.45\textwidth]{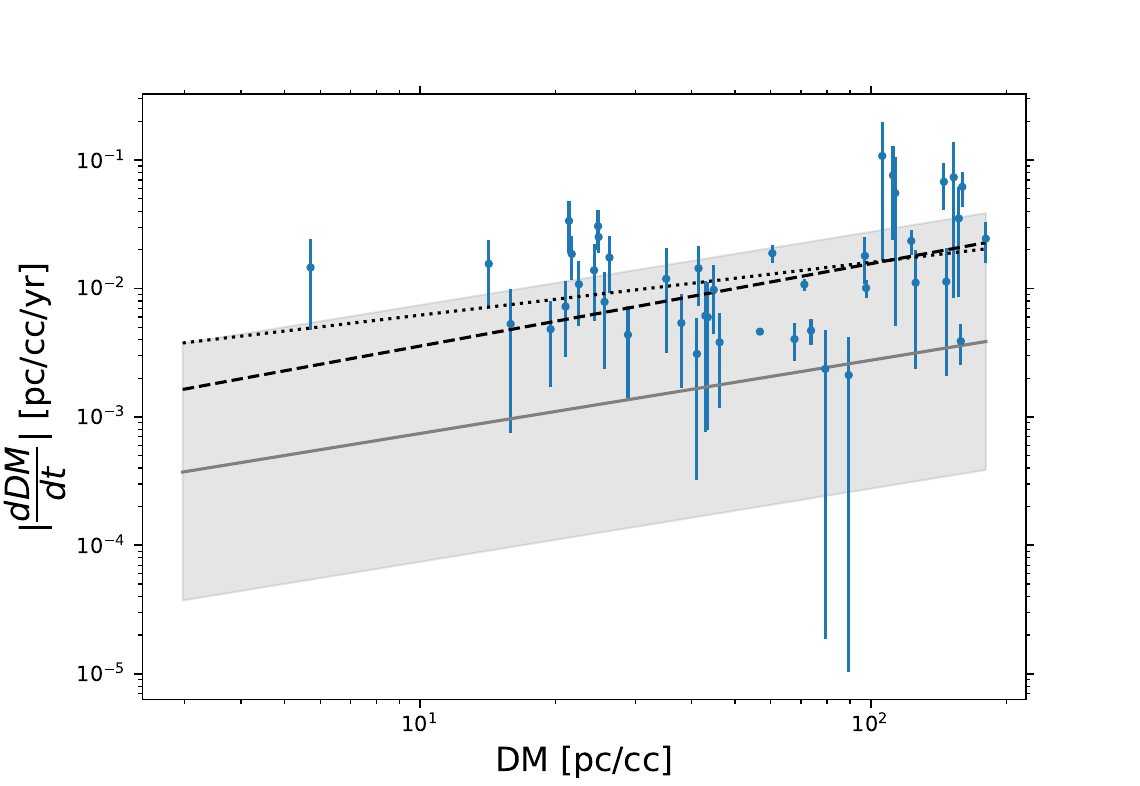}
\caption{Plot of DM variation versus mean pulsar DMs. Blue points indicate the measured DM rate with error bars, and the dotted and dashed line show the best-fit results, unweighted and weighted by $\left|\frac{dDM}{dt}\right|$/$\epsilon_{\left|\frac{dDM}{dt}\right|}$ respectively. The solid black line is the relation from \citet{Hobbs} with an order of magnitude scatter in the shaded grey region.}
\label{fig:dmrate}
\end{figure}

Prior studies on a sample of 13 pulsars \citep{Backer} with DM in the range 2-200 pc cm$^{-3}$ proposed a wedge model with the relation $\left|\frac{dDM}{dt}\right| \sim \sqrt{DM}$. Similarly, a study of 100 pulsars in the DM range 3-600 pc cm$^{-3}$ spanning 6 to 34 years using pulsar timing method \citep{Hobbs} in frequency range of 0.4-1.4 GHz resulted in a relation $\left|\frac{dDM}{dt}\right| = 0.0002\times DM^{0.57\pm0.09}_{\mathrm{pc~cm}^{-3}}$ pc cm$^{-3}$ yr$^{-1}$. In both of these cases, the spread in $\left|\frac{dDM}{dt}\right|$ was an order of magnitude around the best fit measurements. We compare our measurements against \cite{Hobbs} indicated by the solid black line with an order of magnitude spread as shown by the shaded region in grey in Figure \ref{fig:dmrate}. We fit the function $\left|\frac{dDM}{dt}\right| = A\times DM^{B}$ to our weighted and unweighted data, weighted by $\left|\frac{dDM}{dt}\right|$/$\epsilon_{\left|\frac{dDM}{dt}\right|}$. For the weighted case, A$\approx$0.0008 and B=0.64$\pm$0.18 , and for the unweighted A$\approx$0.0002 and B=0.41$\pm$0.20, which is closer to the \cite{Hobbs} relation, albeit slightly steeper. It is interesting to note that for pulsars with similar DMs, our DM rate is about an order of magnitude higher than that of \cite{Hobbs} as can also be seen in Figure \ref{fig:dmrate}. There can be two explanations, either the turbulence is higher due to the large sampled volume of ISM at low frequencies or due to the possible bias in measurement caused by profile evolution at low frequencies. Parametric values of our unweighted fits are also similar to those reported at low frequencies by \cite{bilous2016}, but steeper than \cite{Kumar2025}. 

\subsection{RM and Polarization}\label{sec:rmandpol}
Considering 28 pulsars whose RM measurements are unaffected by the instrumental RMSF peak, pulsars with the most significant change in their RM are Vela and J2155-2118. While the former shows variations of ${\Delta}$RM=6.7, likely due to the supernova remnant and the Gum Nebula \citep{velaXue} with an associated ionospheric correction of -3.66, J2155-3118 has no such associated structure. Since the ${\Delta}$RM value for this is -6.24 with an ionospheric correction of -0.95, it is unlikely to be due to ionospheric variation; hence, this is a new, improved RM measurement. Whereas for pulsars J1543-0620, J1825-0935, and J0922+0638, the magnitude of ${\Delta}$RM is greater than RM$_{ion}$ by one. Since the ionospheric corrections at high frequencies are generally smaller than one, RM measurements for these three pulsars are also improved due to the high sensitivity of low-frequency observations to pulsar RMs. For all other pulsars, the magnitude of ${\Delta}$RM is comparable to RM$_{ion}$, hence any change is only due to the random fluctuations in ionospheric RMs. The phase-resolved RM variation in J1453-6413 shown in Figure \ref{fig:rmphase} is consistent with the measurements reported in Figure 1 of \cite{Noutsos2009}. This suggests the presence of magnetospheric RM component in this pulsar, caused the propagation of pulsar radiation through its own plasma. Detection of such RM variations has implications for pulsar emission modeling.

\subsection{Census completeness}\label{sec:yeild}
Out of the 240 observed pulsars, we have detected 120 sources, among which 33 detections were for the first time below 150 MHz, including some below 100 MHz. Additionally, compared to \cite{Bhat_2023} and \cite{Xue_2017}, we detected five pulsars undetected by these two MWA censuses at frequencies between 150 and 200 MHz. Even though the detection rate is only 50$\%$, we have to consider the bias due to limited sensitivity and high scattering at low frequencies.  

\begin{figure*}[htbp!]
\includegraphics[scale=0.8]{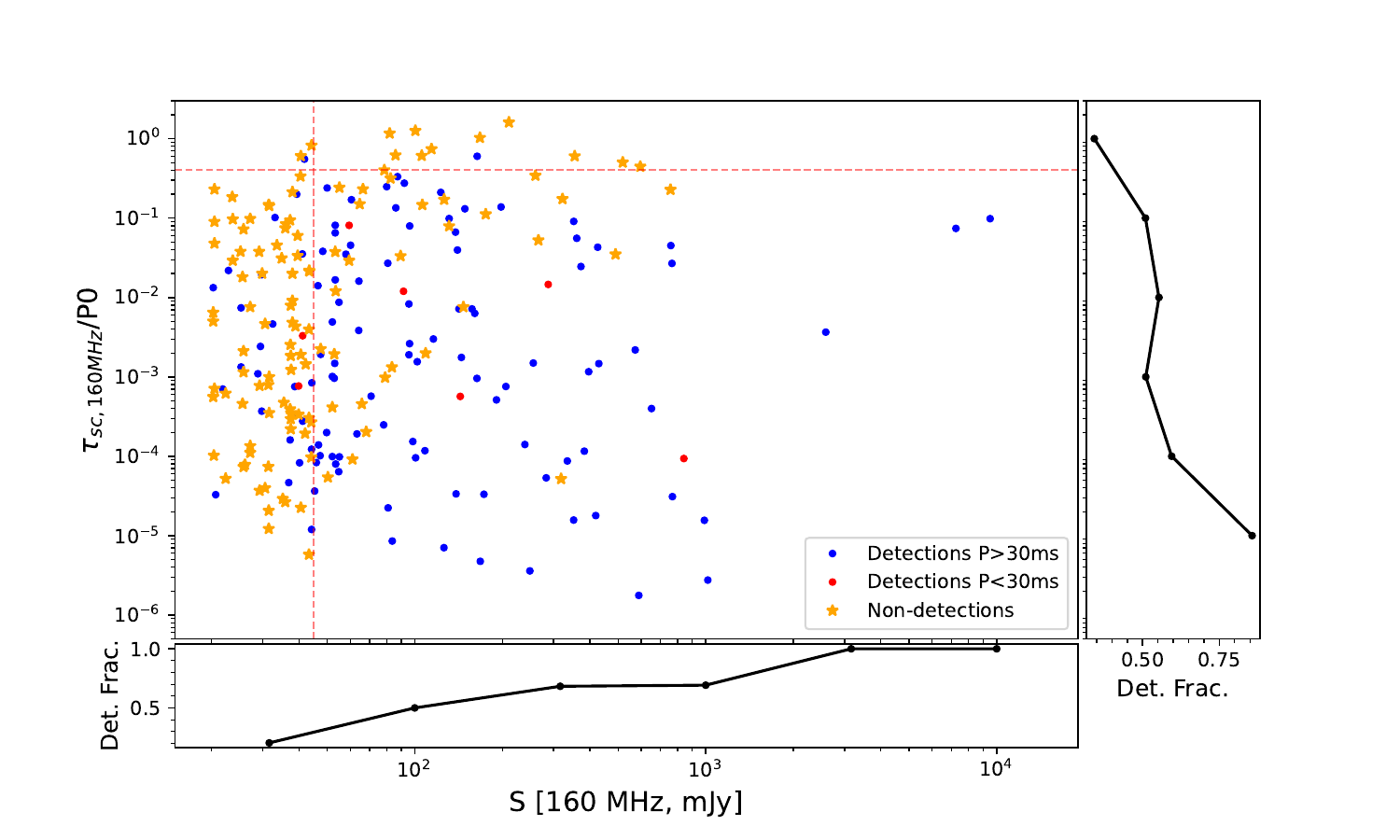}
\caption{Center: Plot of scattering time at 160 MHz from the YMW16 model \citep{Yaoscat} normalized by the pulsar period versus the expected pulsar flux density at 160 MHz. Blue and red dots represent the detected pulsars with period above and below 30 ms, and orange stars represent the non-detected pulsars. The dashed vertical  and horizontal line represent expected flux density value of 45 mJy and ratio of scattering time to pulsar period of 0.4 both at 160 MHz. The right and bottom panels represent the fraction of detected pulsars as a function of ratio of scattering time to period and expected flux density at 160 MHz respectively.}
\label{fig:yeild}
\end{figure*}

\begin{figure*}[htbp!]
\includegraphics[scale=0.8]{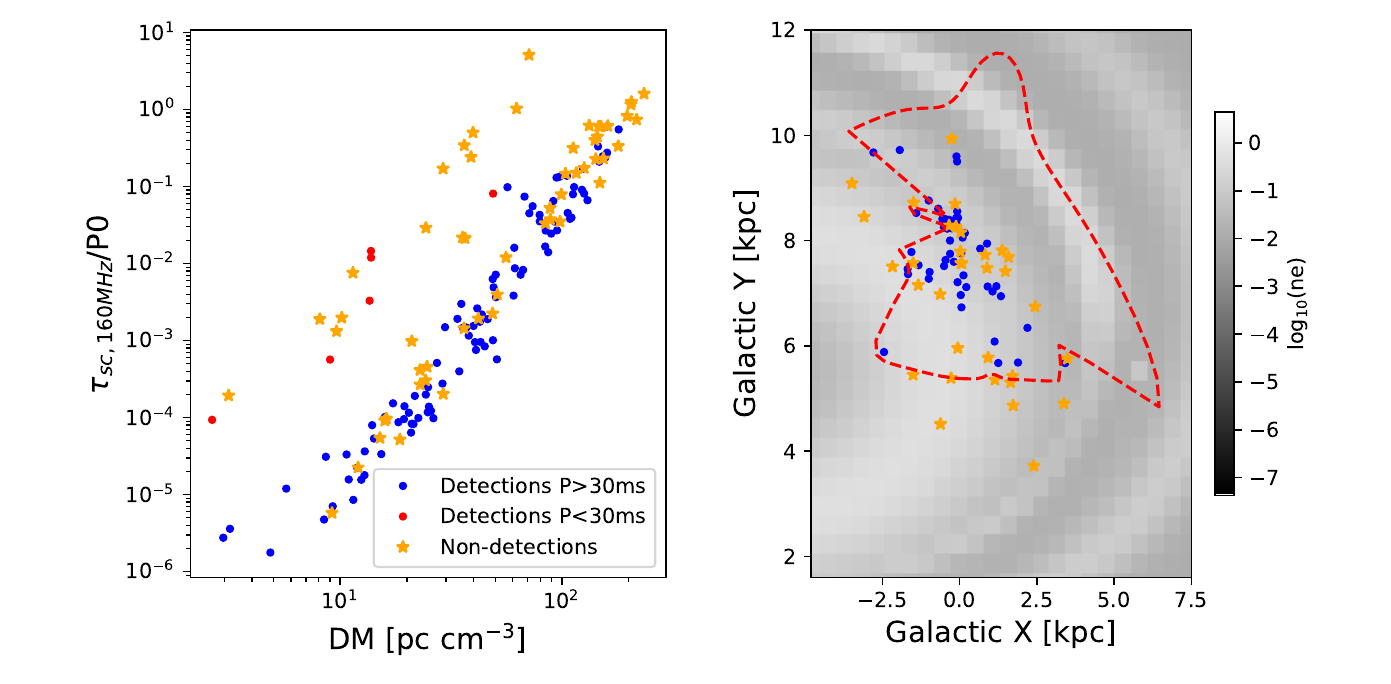}
\caption{Left: Normalized scattering time in units of pulse period versus the pulsar DM. Right: The grey-scale represent the electron density model of YMW16 in the Galactic plane. Blue points and orange stars represent pulsars distance derived from YMW16 model for the detected and undetected pulsars of census respectively, for pulsars with distance to Galactic plane less than 300 pc. The dotted red line is a contour of DM 150 pc cm$^{-3}$. Only pulsars with expected flux density $>40$ mJy at 160 MHz are plotted.}
\label{fig:complete}
\end{figure*}

Figure \ref{fig:yeild} shows the detected and non-detected pulsars in our census against the expected flux density at 160 MHz obtained by scaling literature measurements with a spectral index of -1.4 and using the \textit{PULSAR SPECTRA}, and the expected scattering delay at 160 MHz obtained from the YMW16 model \citep{Yaoscat} using \textit{PyGEDM} \citep{Pygedm} normalized by the pulsar period. The two smaller plots in the figure represent the fraction of detected pulsars as a function of flux density and fractional scattering delay. As we can see clearly from the plot, most of our non-detections are from the low flux density regime, where we are sensitivity limited, or from the high fractional scattering. The left panel of Figure \ref{fig:complete} shows the correlation between DM and scattering and allows us to estimate the maximum DM at which we can detect pulsars, which is 180 pc cm$^{-3}$ for our census. We can also see two distinct trend lines for normal and millisecond pulsars, where the latter is only detectable up to a slightly lower scattering regime. The right panel of Figure \ref{fig:complete} shows the location of our census targets in the Galactic plane, with Earth at the origin. Since DM is a proxy for distance, this allows us to estimate the maximum distance up to which pulsars are detectable, as indicated by the axes of the right panel of Figure \ref{fig:complete}. We can see that most of our detections come within the DM range of 150 pc cm$^{-3}$, indicating that low-frequency observations are mostly limited to the nearby pulsar population.

\section{Summary and Future Outlook}\label{sec:summary}
We present the most comprehensive census of pulsars in the southern hemisphere at low frequencies with a SKA-Low prototype station. We have performed an all-sky census of pulsars, mainly southern sky, down to a sensitivity limit of $\sim$ 20 mJy in modest observing times of 0.5-2 hrs. We measured flux densities of pulsars in five sub-bands in the frequency range of 50-250 MHz, allowing us not only to robustly validate the system sensitivity by comparing these measurements against other prior low-frequency pulsar census but also to improve spectral characterization of pulsars at low frequencies, and provide improved DM and RM measurements. Below is a summary of the most important results with concluding remarks.
\begin{enumerate} 
    \item We detected 120 pulsars in the frequency range of 50-250 MHz, many of which were detected in multiple sub bands centered at $\sim$ 71, 131, 169, 197, and 225 MHz.
    \item We report the lowest frequency detection of 29 pulsars, which includes 23 pulsars below 150 MHz, and five pulsars, namely J1057-5226, J1116-4112, J1430-6623, J1456-6843, and J1534-5350, detected below 100 MHz.
    \item Multi-band pulse profiles are provided for all census pulsars, where 72 and 36 pulsars show profiles with 1 and 2 components, respectively. Additionally, 11 pulsars have a visible presence of scattering at low frequencies. 
    \item Analysis of pulsars with PL spectra using our census measurement shows that the mean spectral index is -1.78$\pm$0.17, consistent with the results of \cite{jankowski441psr}. Moreover, 52 pulsars have a turnover at low frequencies with a mean turnover frequency of 115$\pm$10, consistent with results from \cite{Izvekova}. For pulsars with PLH, the high-frequency cutoff occurs around 1 GHz, consistent with the predictions of \cite{Kijak}.
    \item Due to the high precision and sensitivity of low-frequency observations to pulsar DM and variation due to the change in LOS, we provide improved DM measurement for 110 pulsars with a median absolute offset with respect to ATNF values of 0.1 pc cm $^{-3}$ with the most notable DM change of -0.85, -0.83, 0.76, and 0.59 pc cm$^{-3}$ in PSR J1711-5350, J1240-4124, J0856-6137, and J1418-3921 respectively. 
    \item Using these DM variation we model the fluctuation of electron density in the ISM, where we obtain the relation $\left|\frac{dDM}{dt}\right| = 0.0008\times DM^{0.64\pm0.18}_{\mathrm{pc~cm}^{-3}}$ pc cm$^{-3}$ yr$^{-1}$ which is slightly steeper than the result of \citep{Hobbs} but consistent with \cite{Kumar2025} obtained below 100 MHz. This may suggest that the turbulence in ISM is higher at low frequencies.
    \item We also performed RM-synthesis on our census pulsars, where we obtained RM measurements for 40 pulsars, mostly consistent with the prior values reported at higher frequencies. However, for PSR J2155-3118, J1543-0630, J1825-0935, and J0922+0638, we obtained improved RM measurements after rejecting the possibility of a change in RM due to variations in ionospheric RMs. For PSR J1453-6413, we also obtained a phase-resolved variation in RM.
\end{enumerate}

This study offers a compelling demonstration of the pulsar and ISM science that will be enabled by the future SKA‑Low telescope. Since many of the detections reported here are the first at frequencies below 150 MHz, these data are particularly valuable for exploring spectral turnover behavior in pulsar emission. With the enhanced sensitivity of SKA-Low and observations of a larger sample of pulsars, spectral modeling can be extended across a broader population. This dataset can be used to improve existing pulsar population models, which will be valuable for future pulsar surveys aimed at searching for new pulsars with SKA-Low. Moreover, the variation in pulsar DM due to the ISM and RM due to the ionosphere presented in this work demonstrates how low‑frequency monitoring of pulsars can be gainfully employed to study and characterize the ISM and ionosphere. This can easily be achieved by the multi-beaming and sub-array capabilities of SKA-Low, by taking simultaneous observations of many pulsars.

\begin{acknowledgments}
 We thank the referee for constructive suggestions. EDA2 is hosted by the MWA under an agreement via the MWA External Instruments Policy. This scientific work makes use of the Murchison Radio astronomy Observatory, operated by CSIRO. We acknowledge the Wajarri Yamatji people as the traditional owners of the Observatory site. This work was further supported by funding from the Australian Government and the Government of Western Australia. The acquisition system was designed and purchased by INAF/Oxford University and the RX chain was design by INAF, as part of the SKA design and prototyping program. 
\end{acknowledgments}

\software{astropy \citep{2013A&A...558A..33A,2018AJ....156..123A,2022ApJ...935..167A} }, numpy (\citep{numpy}, matplotlib \citep{Hunter:2007}, scipy \citep{2020SciPy-NMeth}




\bibliography{census}{}
\bibliographystyle{aasjournalv7}



\appendix
\setcounter{figure}{0} 
\renewcommand{\thefigure}{A\arabic{figure}}
\renewcommand{\theHfigure}{A\arabic{figure}}
\renewcommand\theHtable{Appendix.\thetable}
\restartappendixnumbering

\section{Period and dispersion measure of pulsars detected by EDA2}
\input{tables/dmtable}
\section{Flux density and pulse widths}
\input{tables/fluxtable}
\section{Spectral index and other parameters of pulsars detected by EDA2}
\input{tables/spectratable}
\section{RM values for pulsars detected by EDA2}
\input{tables/rmtable}

\clearpage

\section{Multi-frequency pulse profiles of EDA2 detected pulsars}
\begin{figure*}[htbp!]
\includegraphics[width=\textwidth,height=20 cm]{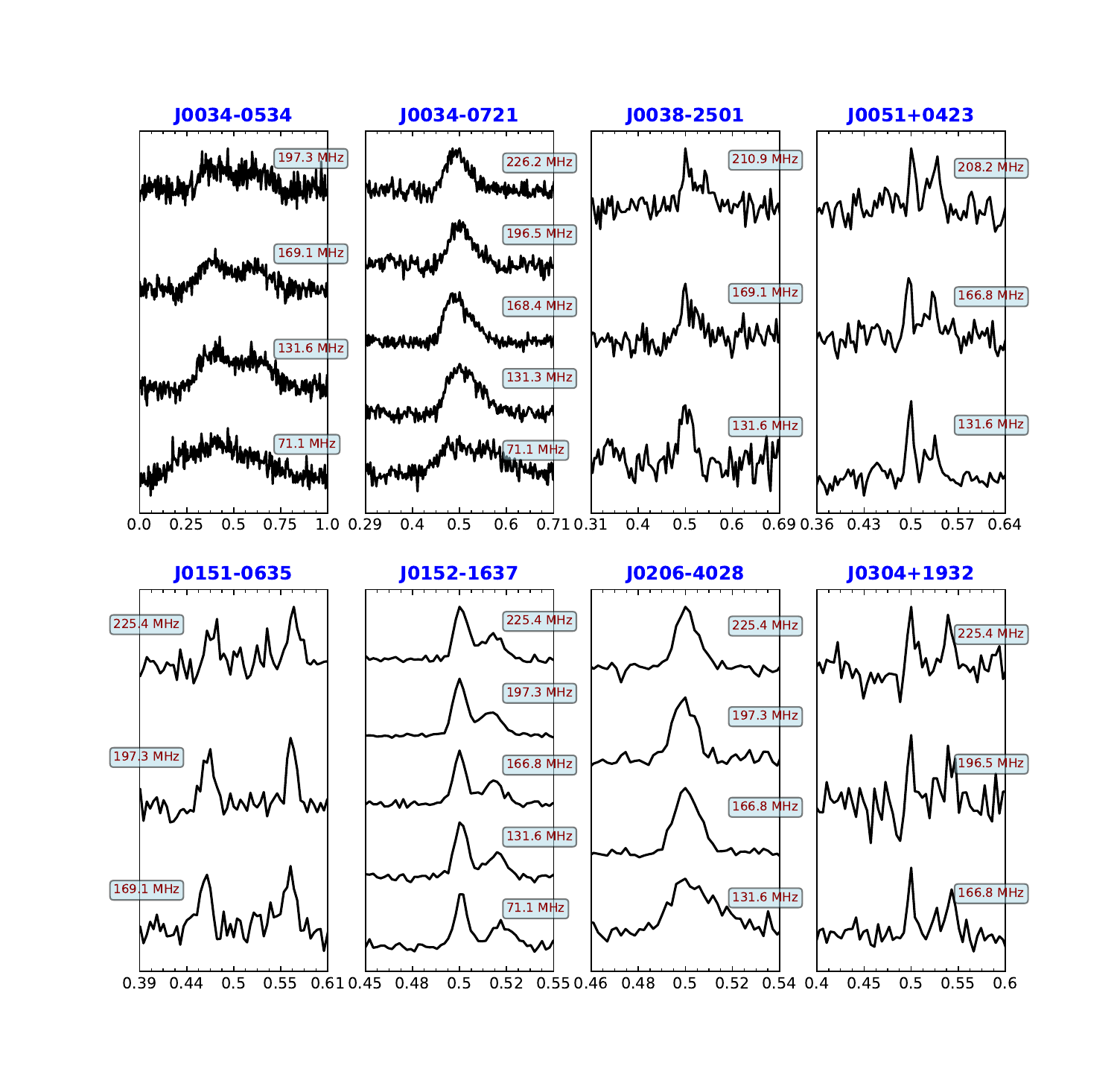}
\caption{Average integrated pulse profiles, with x-axis zoomed on the phase region with emission. The center frequency of each profile is shown alongside, and the corresponding bandwidth is shown in Table \ref{tab:fluxdensity}. Profiles are aligned using their peaks.}
\label{fig:psrprof}
\end{figure*}
\begin{figure*}[htbp!]
\centering
\includegraphics[width=\textwidth,height=22cm]{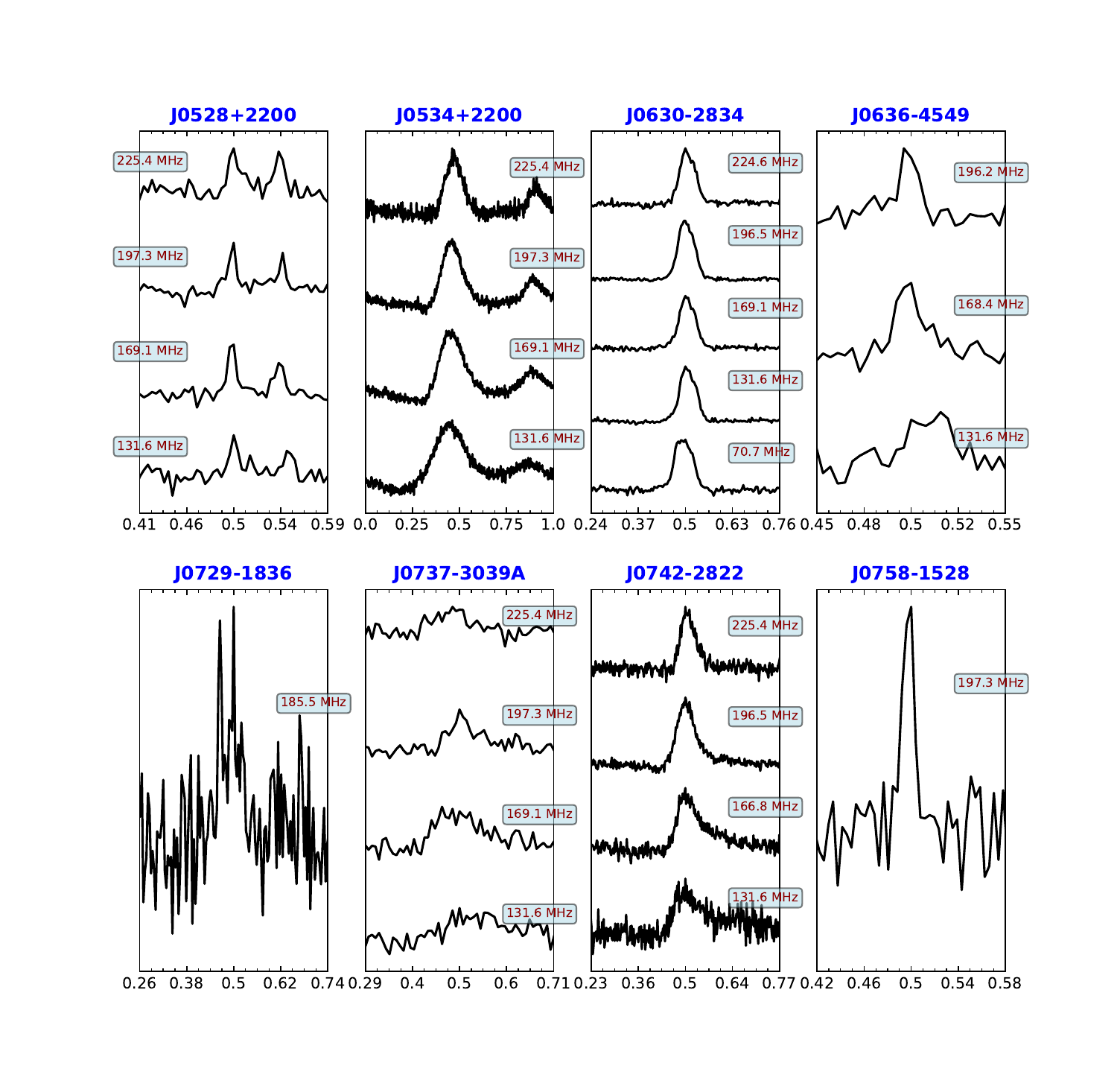}
\caption{Fig \ref{fig:psrprof} Continued}
\end{figure*}
\begin{figure*}[htbp!]
\centering
\includegraphics[width=\textwidth,height=22cm]{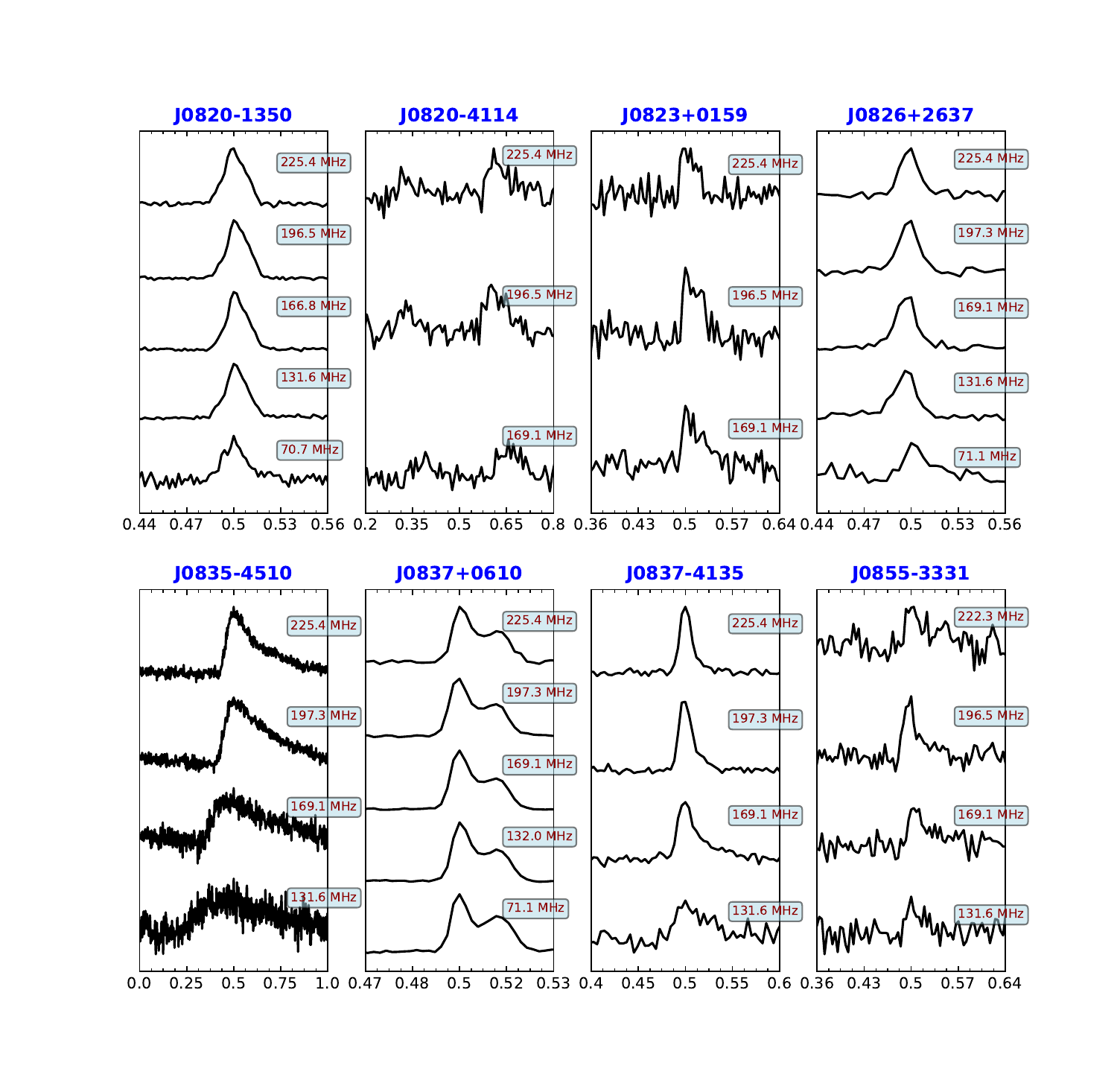}
\caption{Fig \ref{fig:psrprof} Continued}
\end{figure*}
\begin{figure*}[htbp!]
\centering
\includegraphics[width=\textwidth,height=22cm]{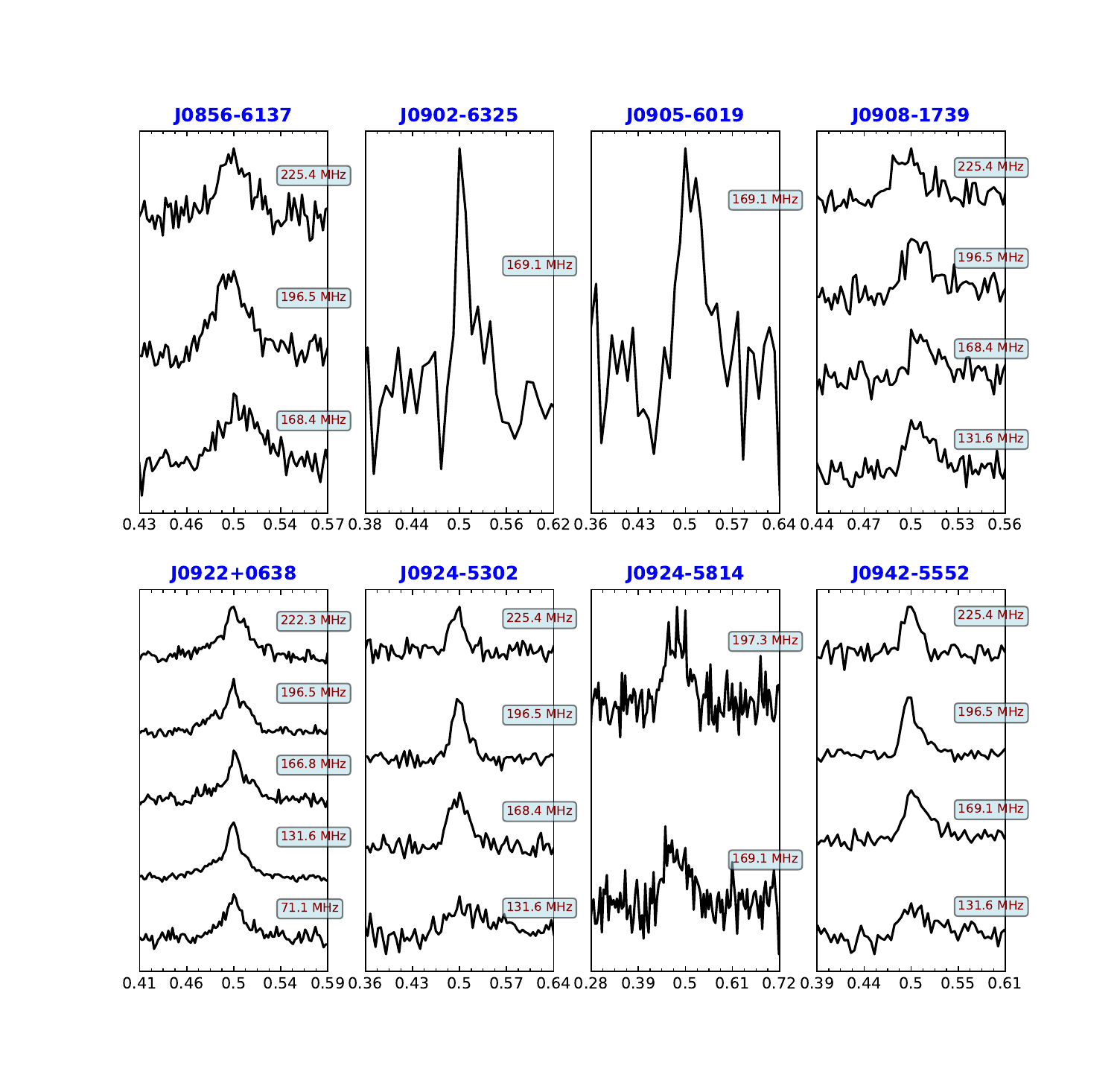}
\caption{Fig \ref{fig:psrprof} Continued}
\end{figure*}
\begin{figure*}[htbp!]
\centering
\includegraphics[width=\textwidth,height=22cm]{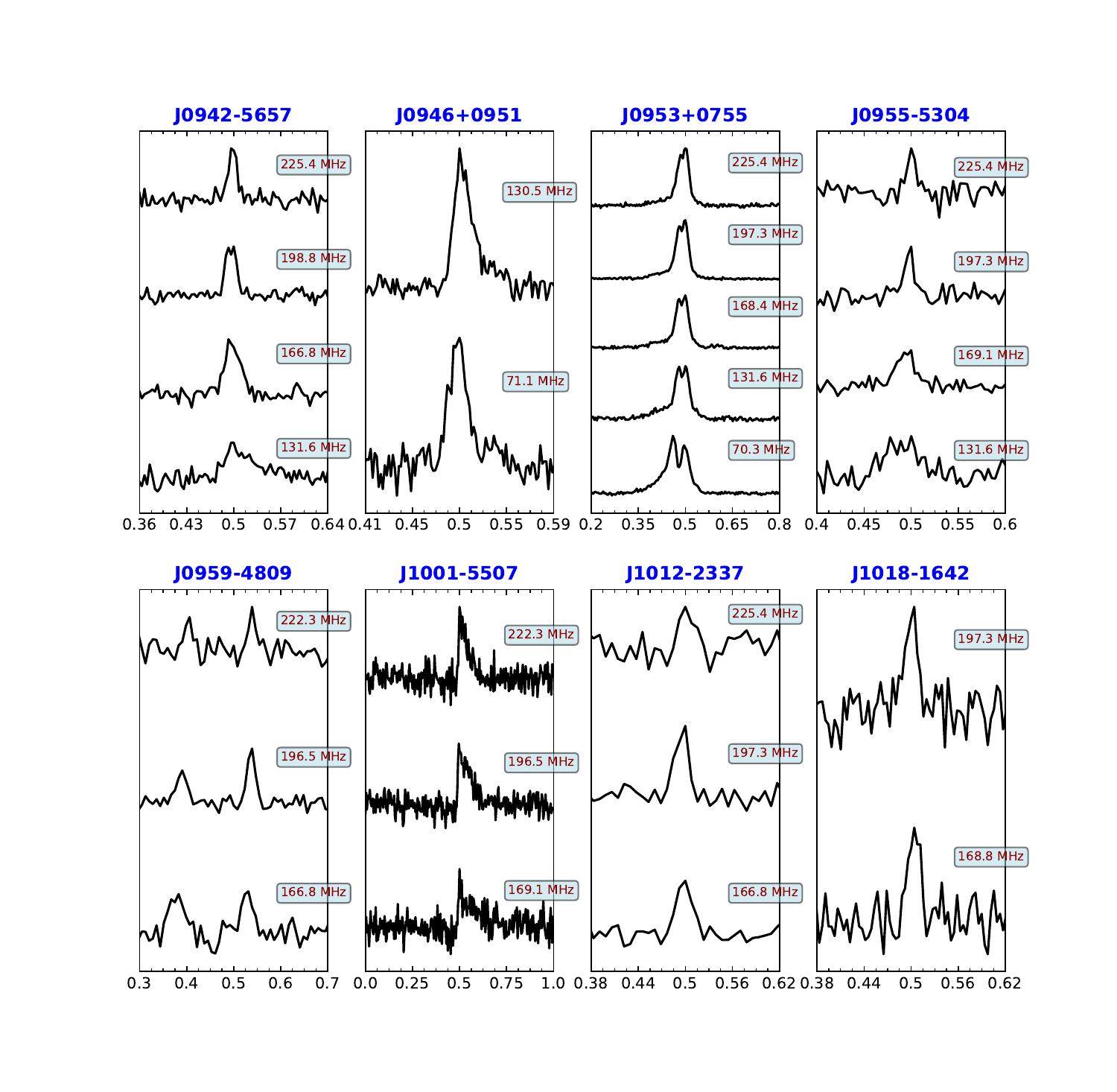}
\caption{Fig \ref{fig:psrprof} Continued}
\end{figure*}
\begin{figure*}[htbp!]
\centering
\includegraphics[width=\textwidth,height=22cm]{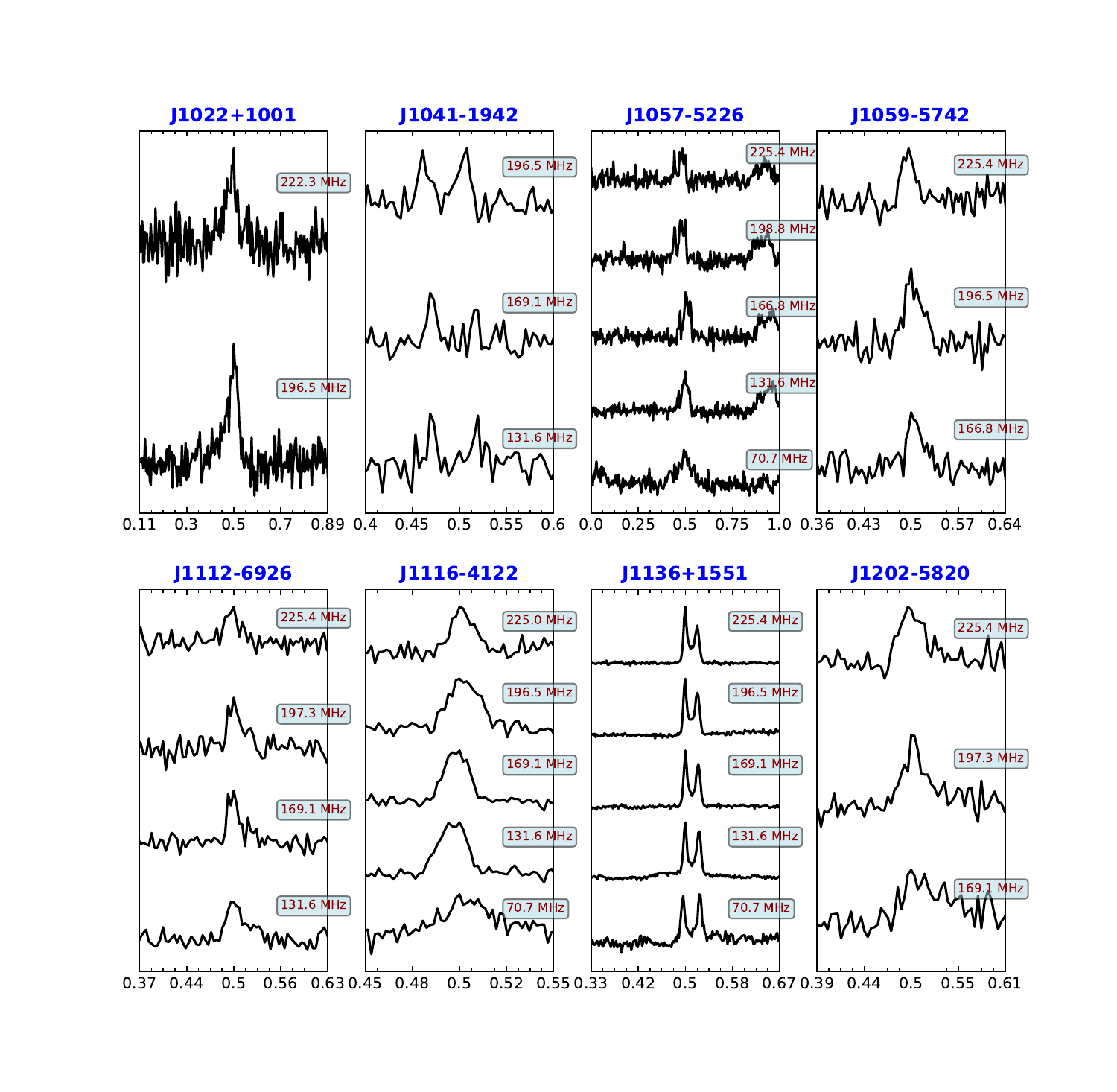}
\caption{Fig \ref{fig:psrprof} Continued}
\end{figure*}
\begin{figure*}[htbp!]
\centering
\includegraphics[width=\textwidth,height=22cm]{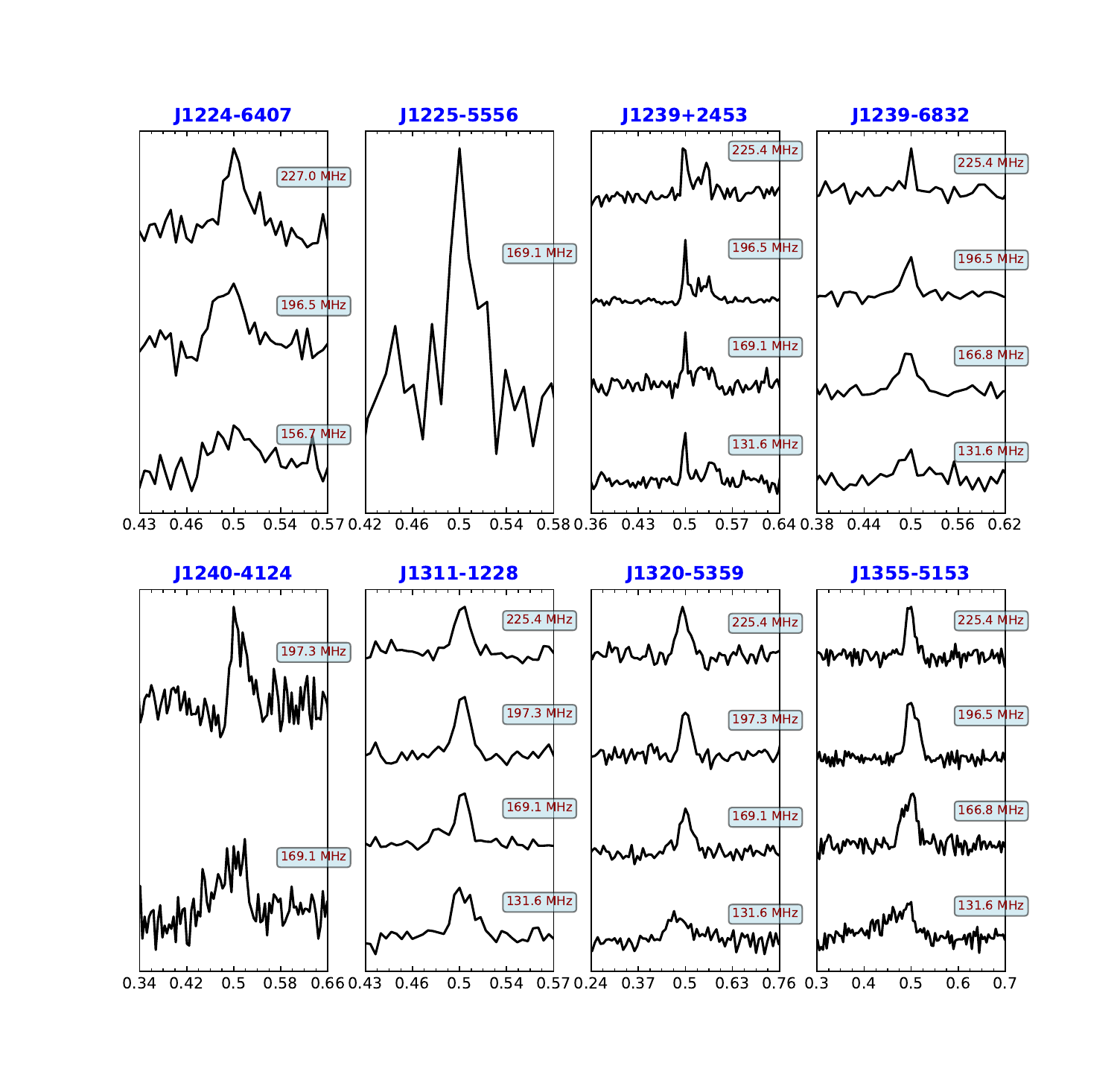}
\caption{Fig \ref{fig:psrprof} Continued}
\end{figure*}
\begin{figure*}[htbp!]
\centering
\includegraphics[width=\textwidth,height=22cm]{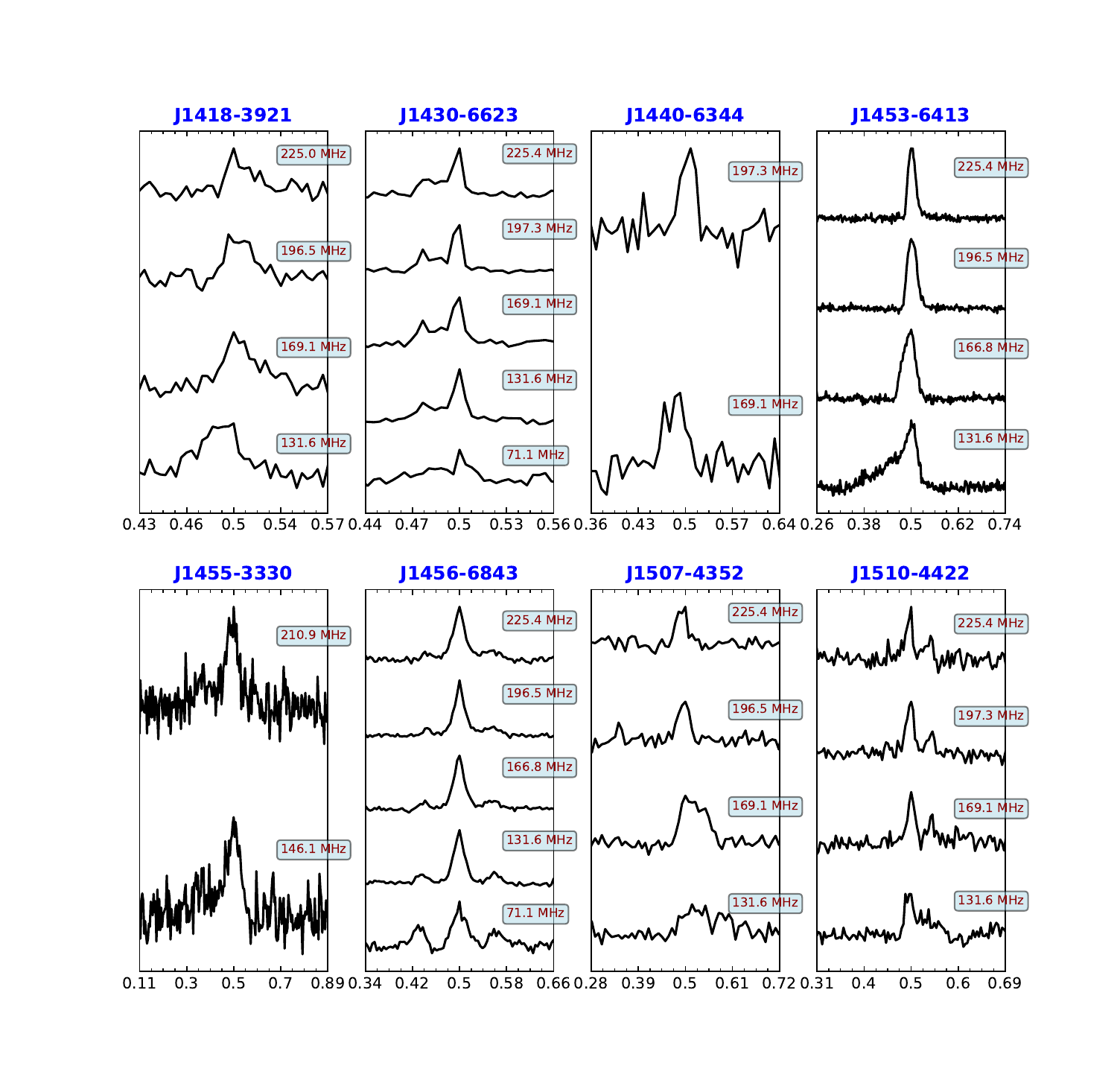}
\caption{Fig \ref{fig:psrprof} Continued}
\end{figure*}
\begin{figure*}[htbp!]
\centering
\includegraphics[width=\textwidth,height=22cm]{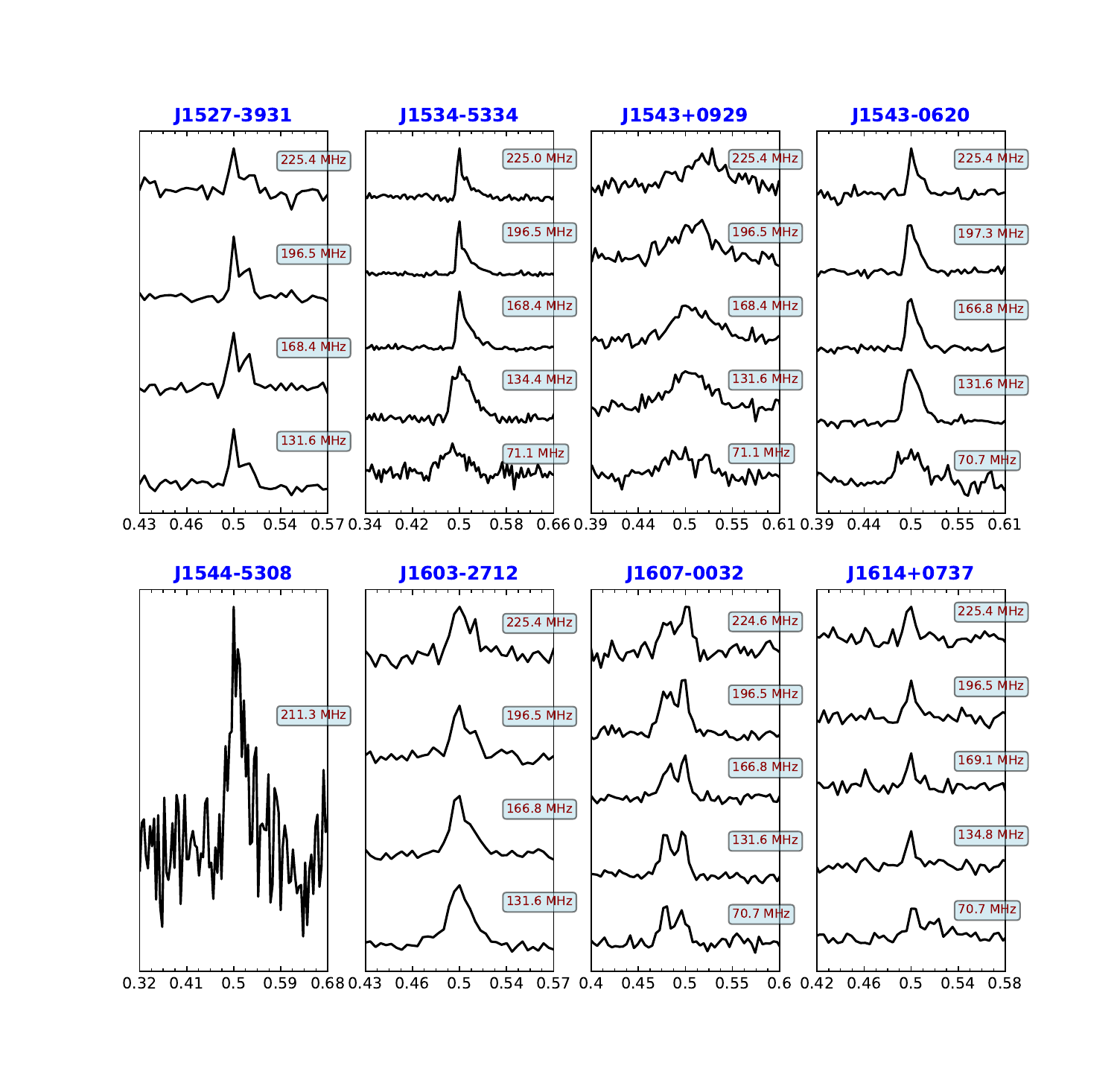}
\caption{Fig \ref{fig:psrprof} Continued}
\end{figure*}
\begin{figure*}[htbp!]
\centering
\includegraphics[width=\textwidth,height=22cm]{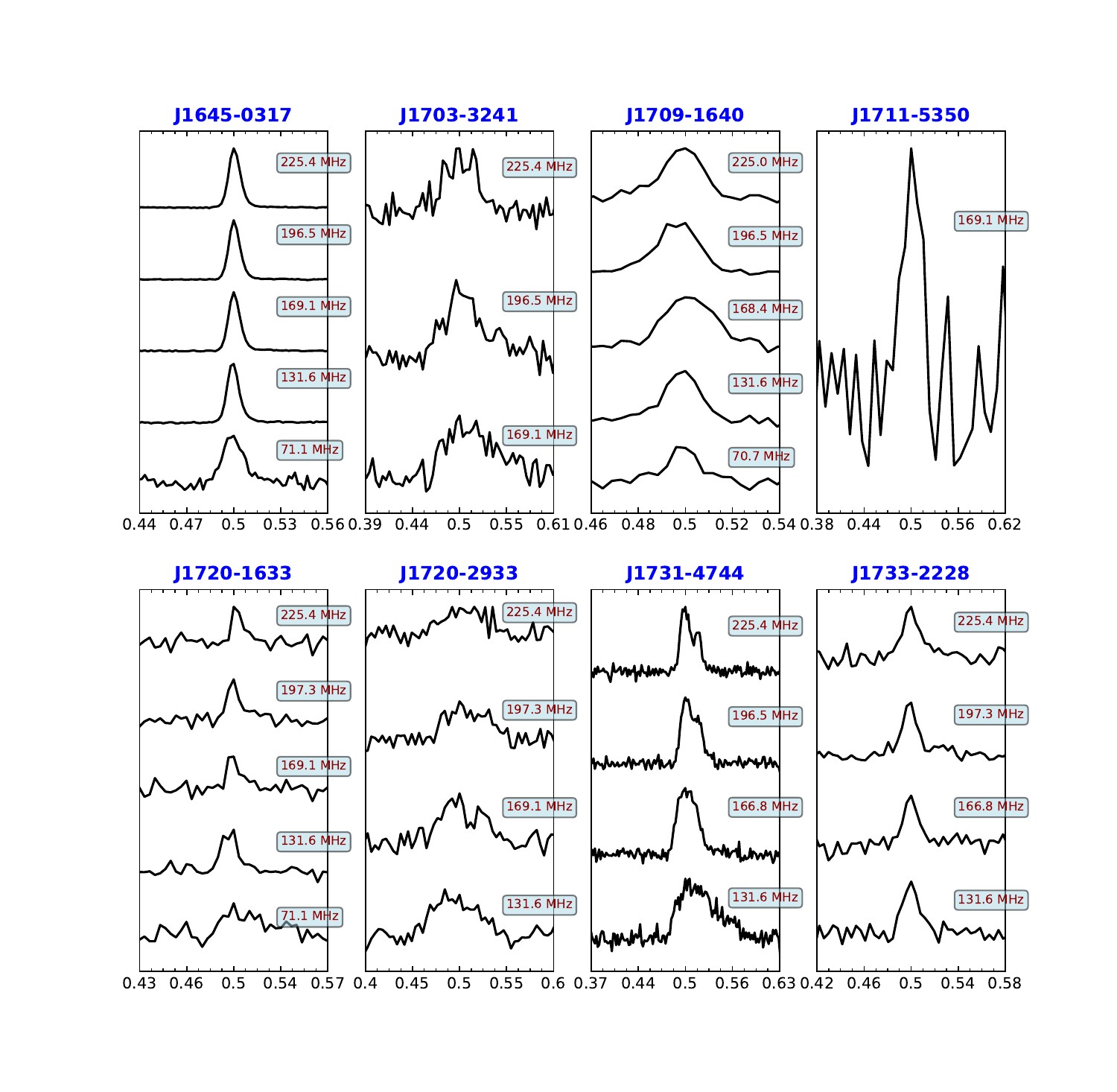}
\caption{Fig \ref{fig:psrprof} Continued}
\end{figure*}
\begin{figure*}[htbp!]
\centering
\includegraphics[width=\textwidth,height=22cm]{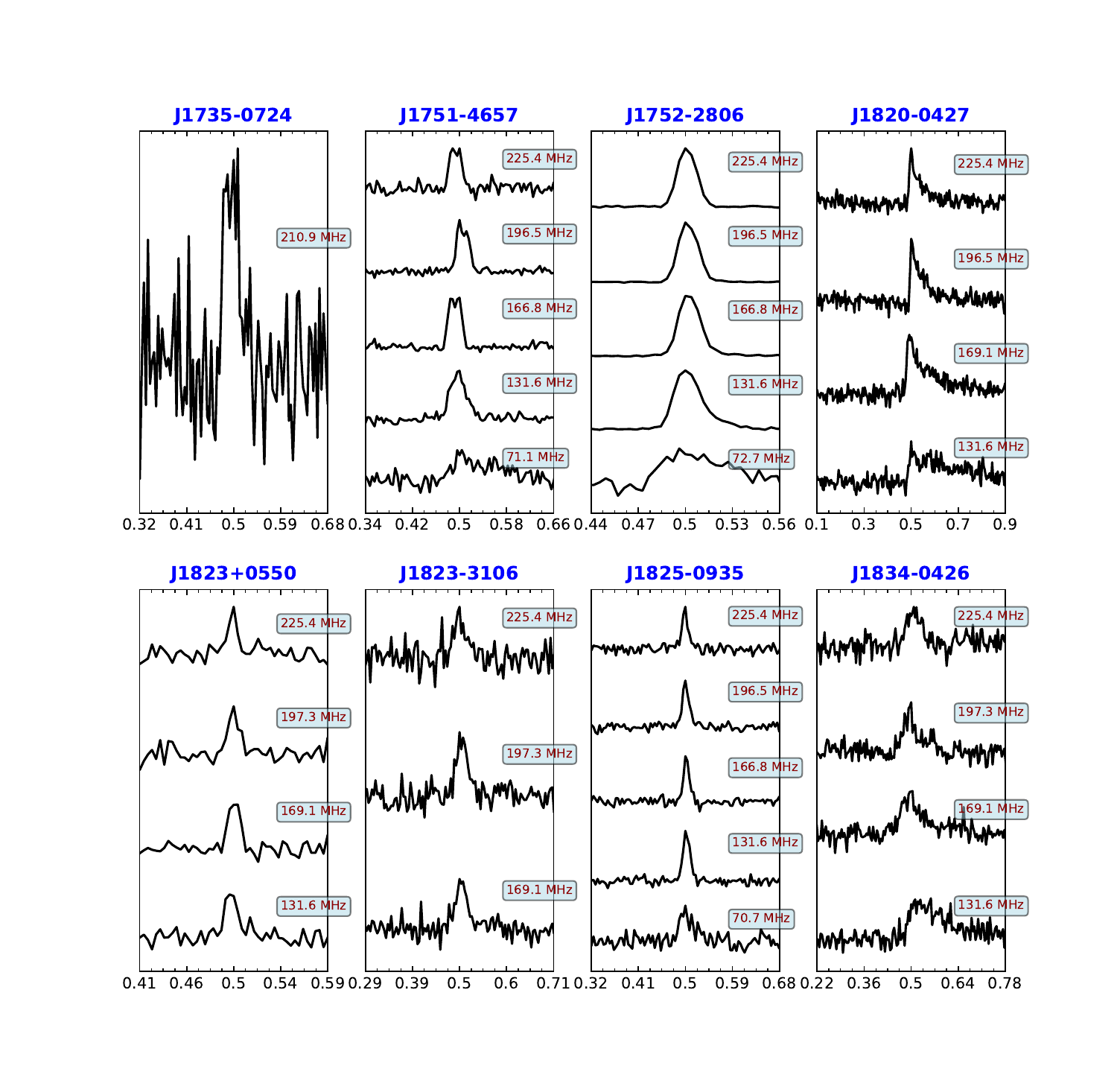}
\caption{Fig \ref{fig:psrprof} Continued}
\end{figure*}
\begin{figure*}[htbp!]
\centering
\includegraphics[width=\textwidth,height=22cm]{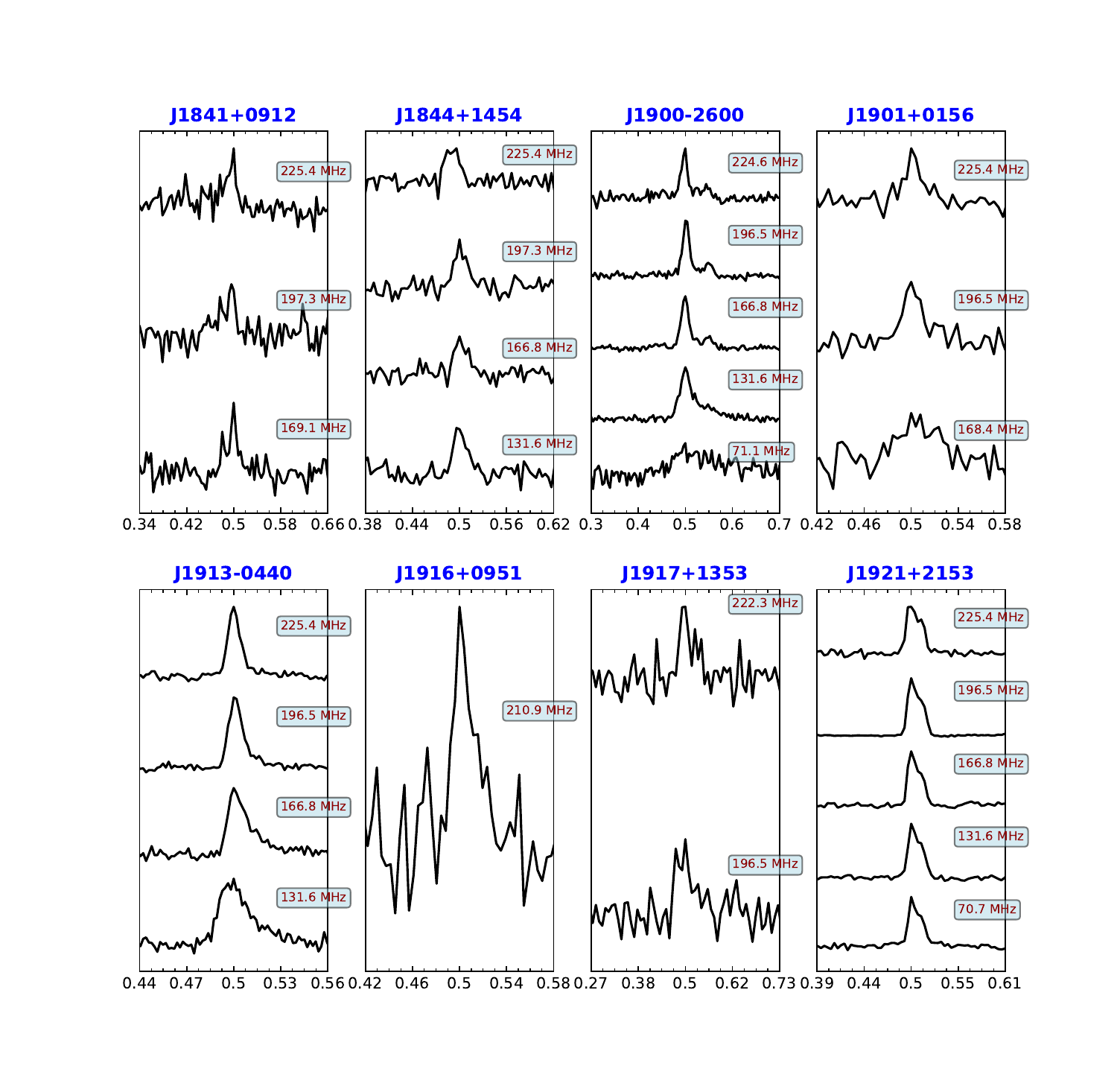}
\caption{Fig \ref{fig:psrprof} Continued}
\end{figure*}
\begin{figure*}[htbp!]
\centering
\includegraphics[width=\textwidth,height=22cm]{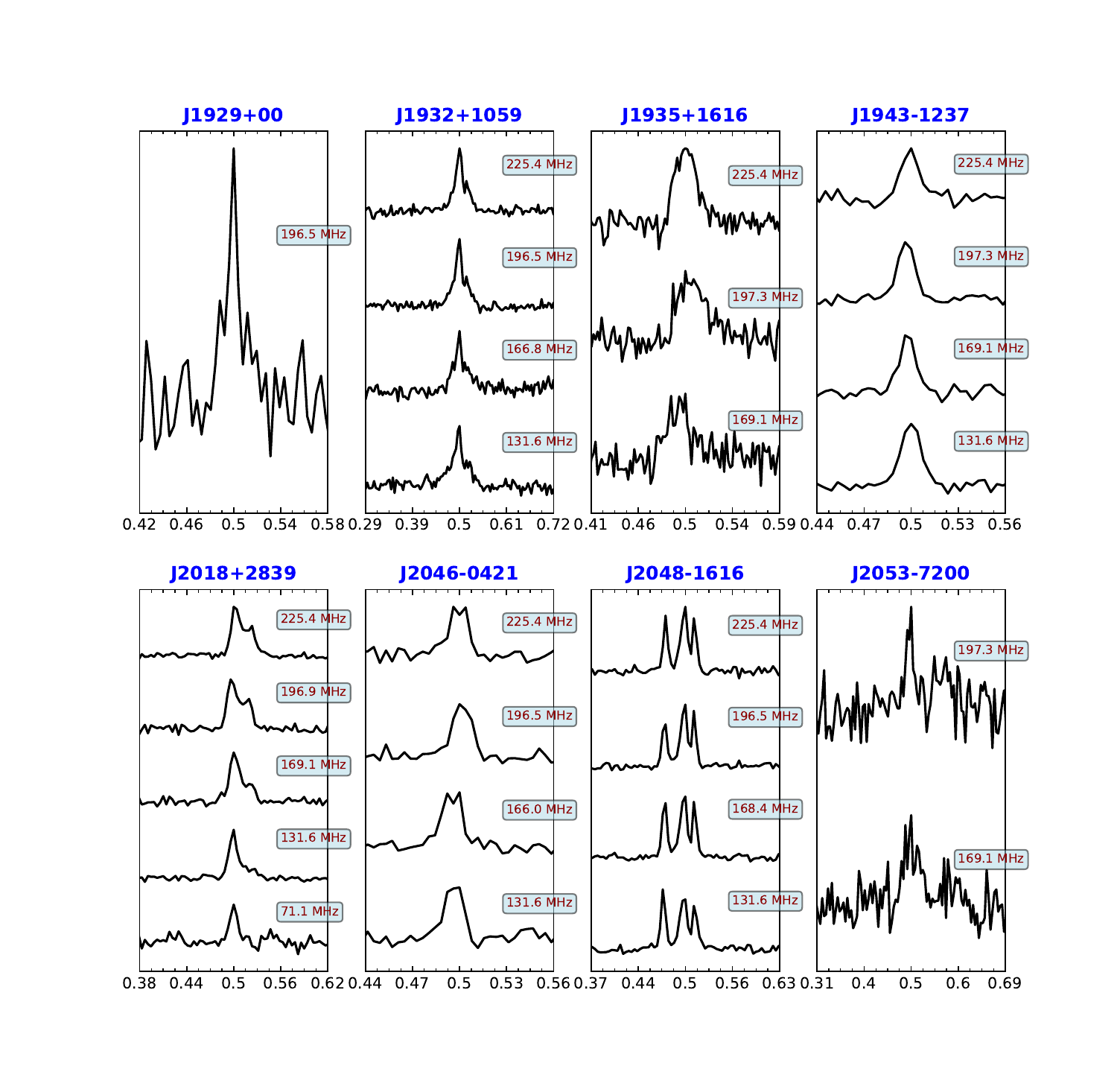}
\caption{Fig \ref{fig:psrprof} Continued}
\end{figure*}
\begin{figure*}[htbp!]
\centering
\includegraphics[width=\textwidth,height=22cm]{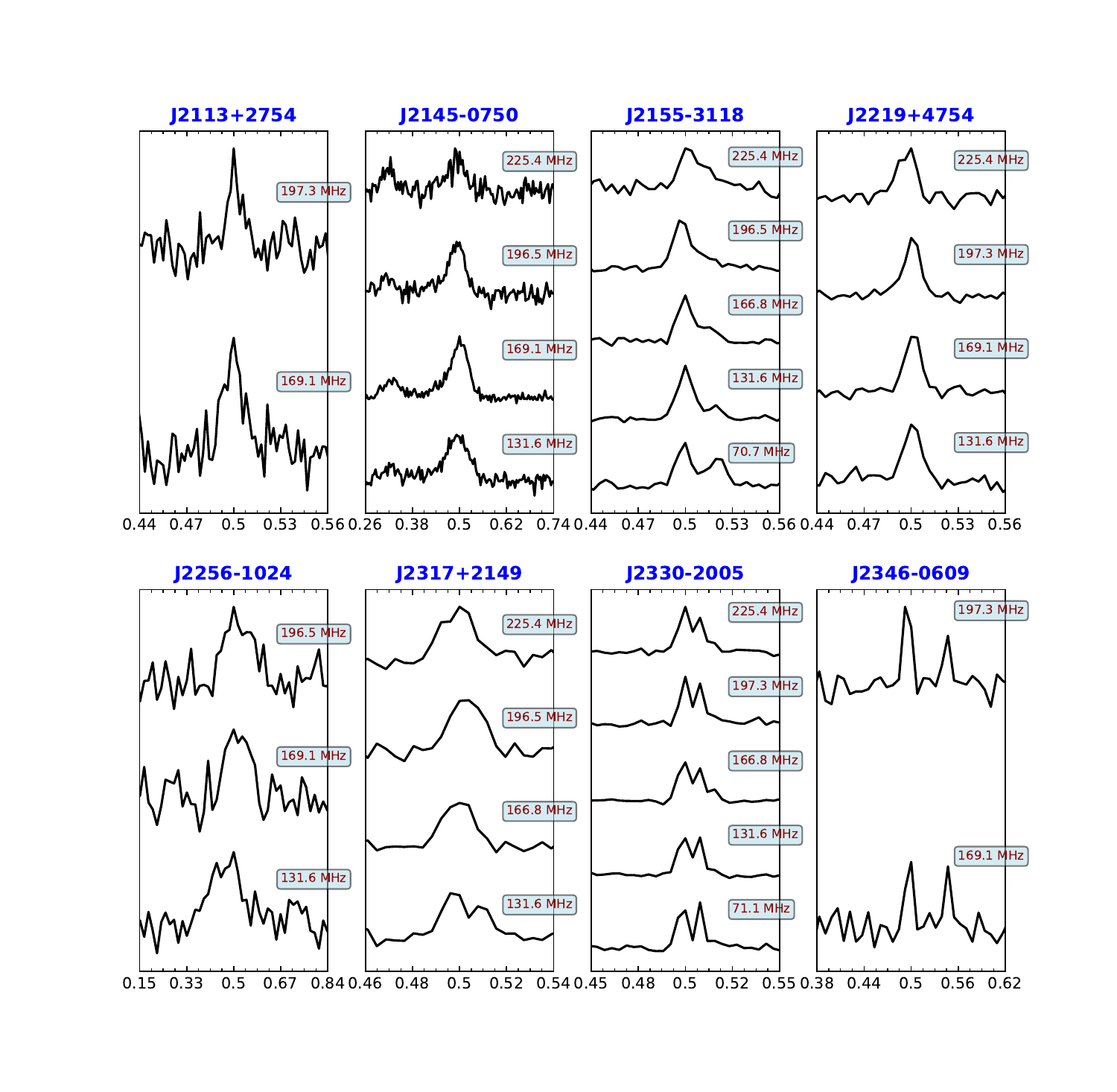}
\caption{Fig \ref{fig:psrprof} Continued}
\end{figure*}
\begin{figure*}[htbp!]
\centering
\includegraphics[width=\textwidth,height=22cm]{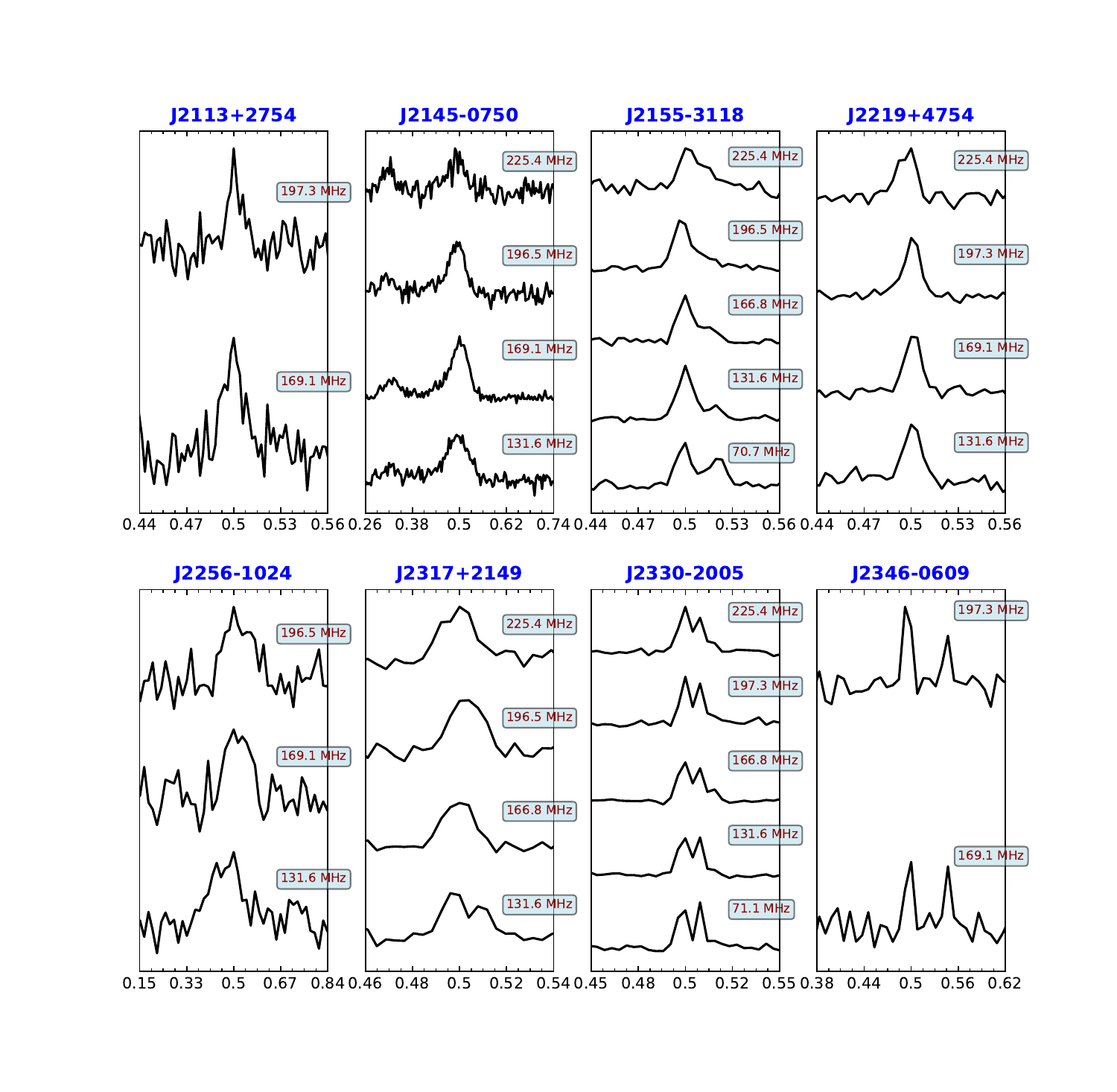}
\caption{Fig \ref{fig:psrprof} Continued}
\end{figure*}
\clearpage
\section{Spectra of EDA2 detected pulsars}
\begin{figure*}[htbp!]
\includegraphics[width=\textwidth,height=20 cm]{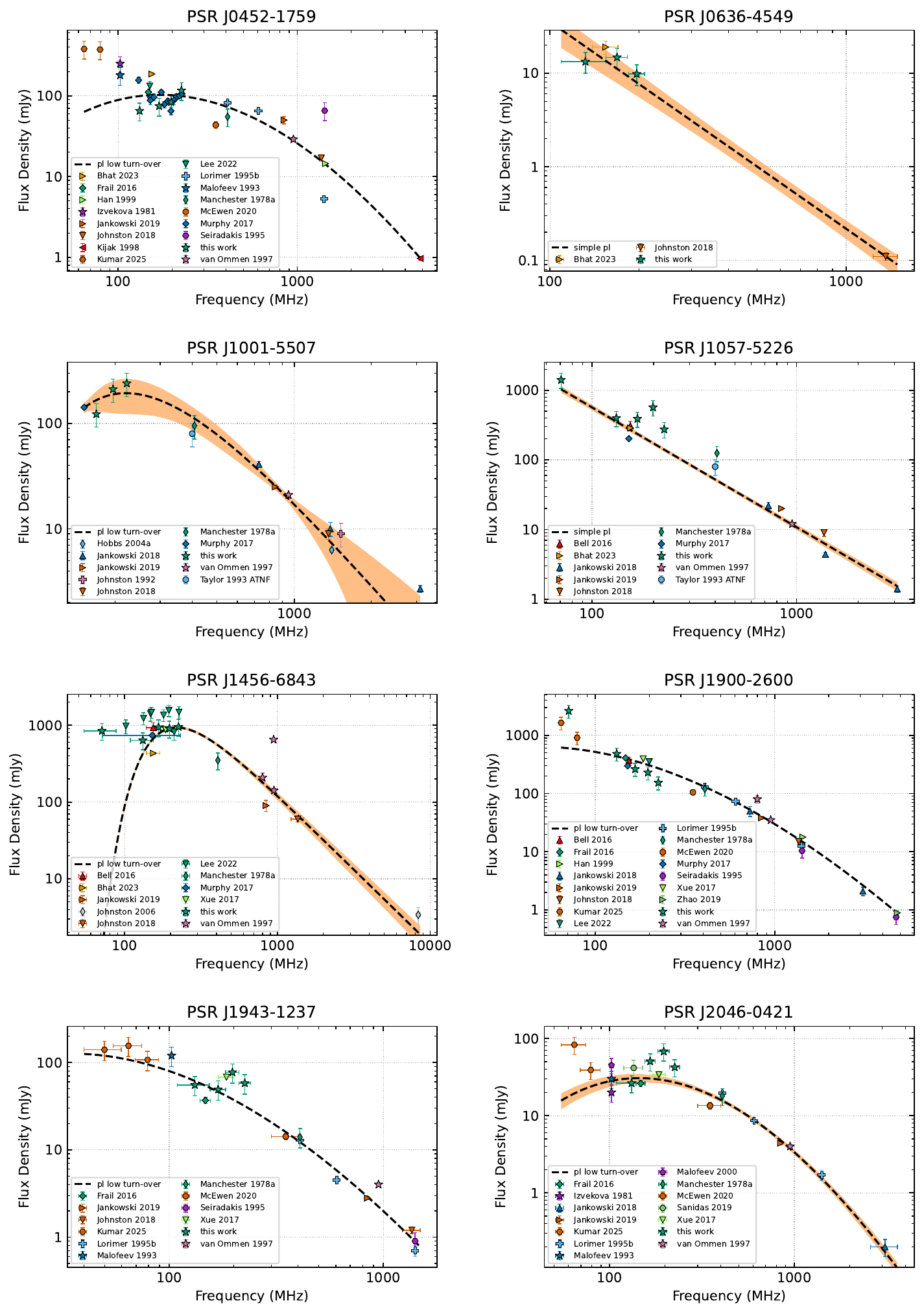}
\caption{Flux density spectra for detected census pulsars, where measurements from this work are shown in green stars. Black dashed lines show the best fit model along with the 1$\sigma$ uncertainty in the orange envelope.}
\label{fig:spectra}
\end{figure*}
\begin{figure*}[htbp!]
\centering
\includegraphics[width=\textwidth,height=22cm]{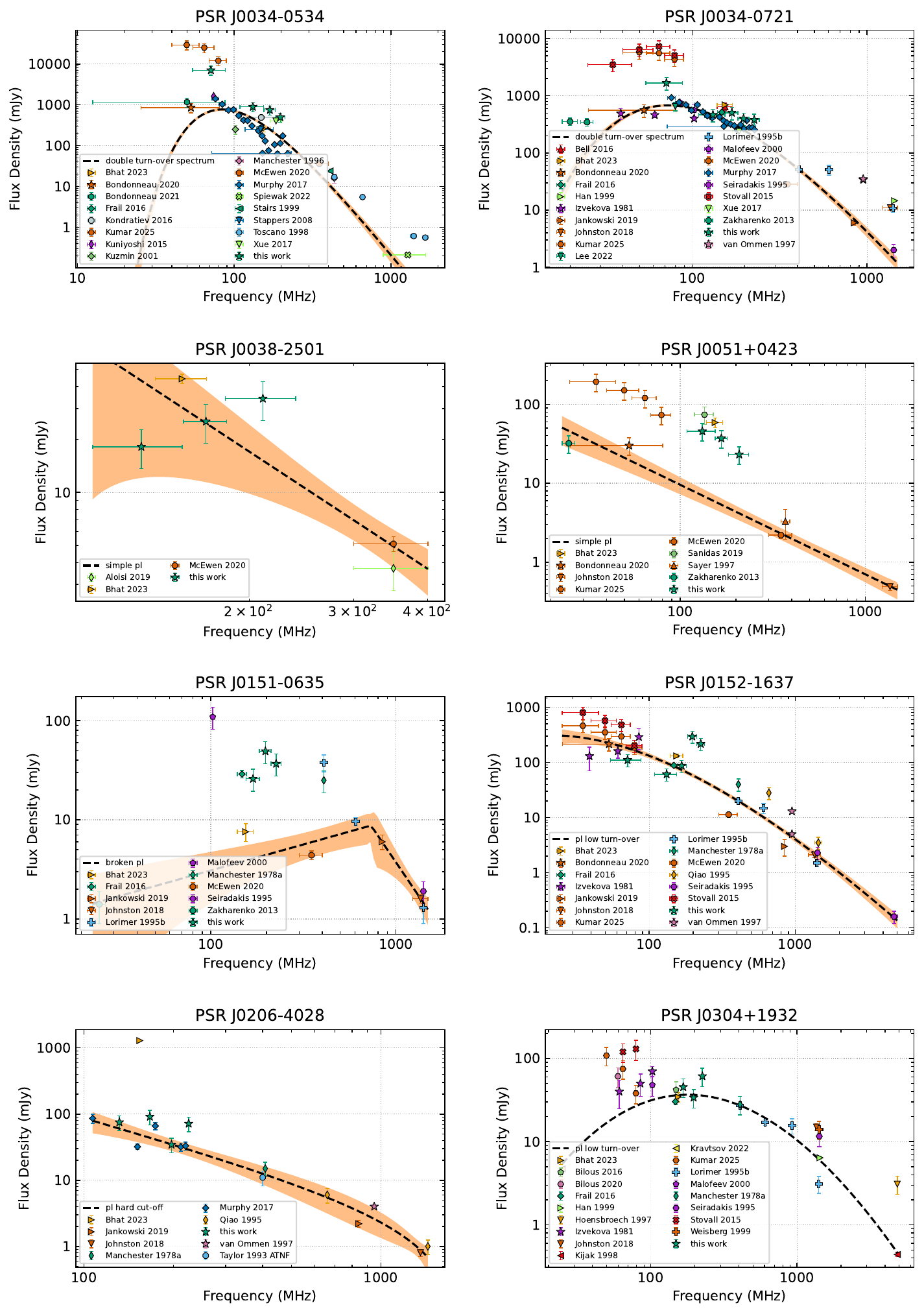}
\caption{Fig \ref{fig:spectra} Continued}
\end{figure*}
\begin{figure*}[htbp!]
\centering
\includegraphics[width=\textwidth,height=22cm]{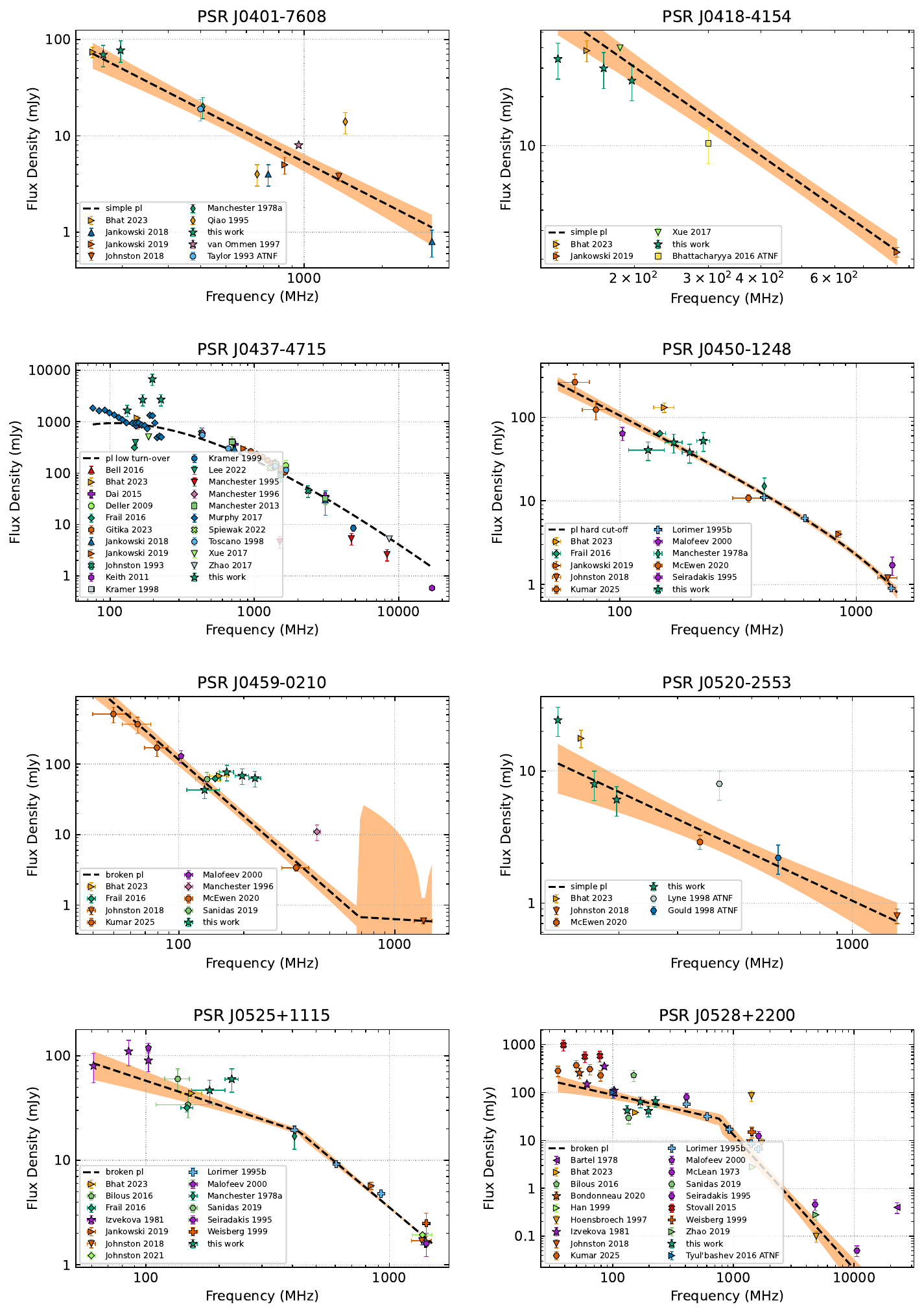}
\caption{Fig \ref{fig:spectra} Continued}
\end{figure*}
\begin{figure*}[htbp!]
\centering
\includegraphics[width=\textwidth,height=22cm]{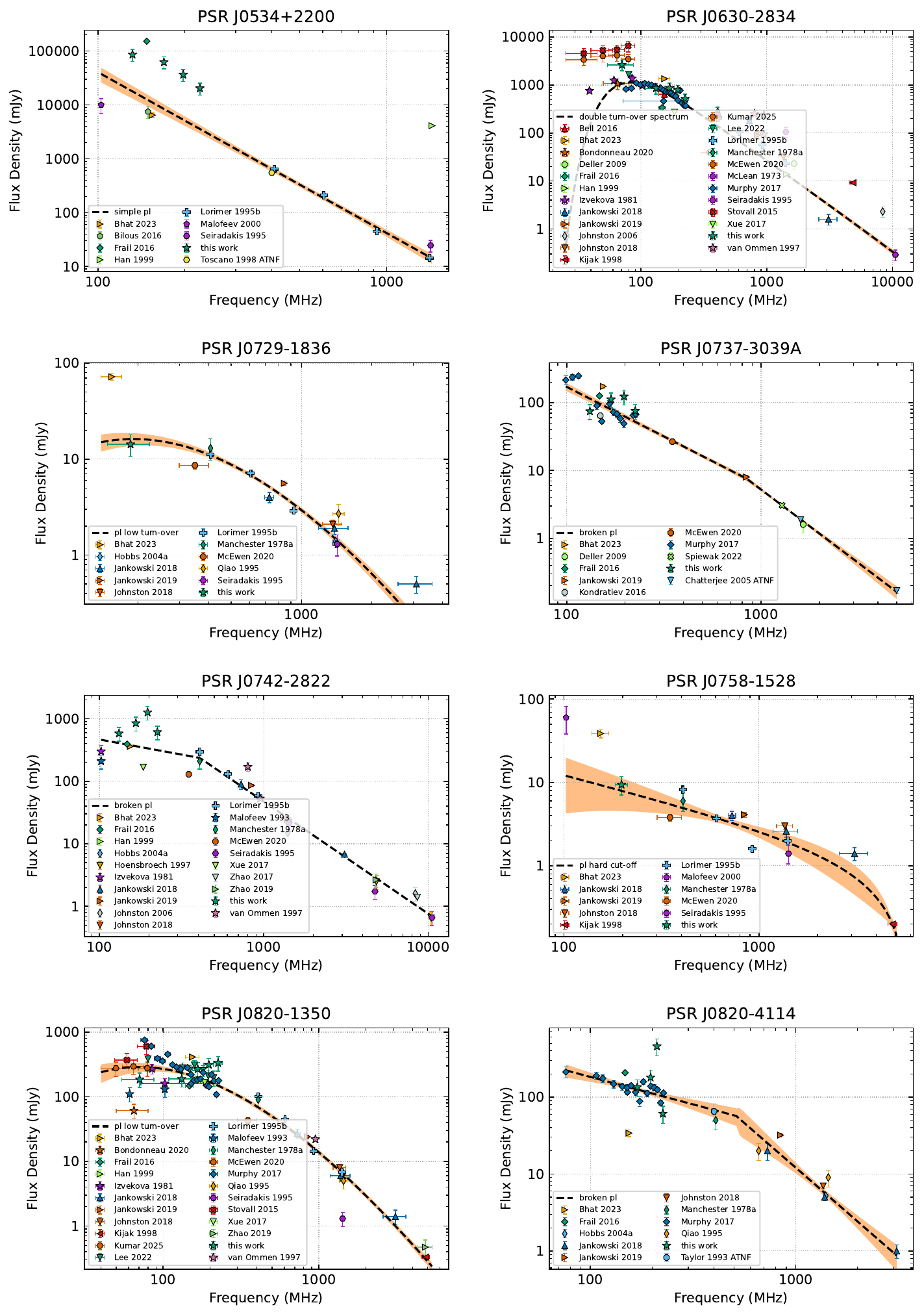}
\caption{Fig \ref{fig:spectra} Continued}
\end{figure*}
\begin{figure*}[htbp!]
\centering
\includegraphics[width=\textwidth,height=22cm]{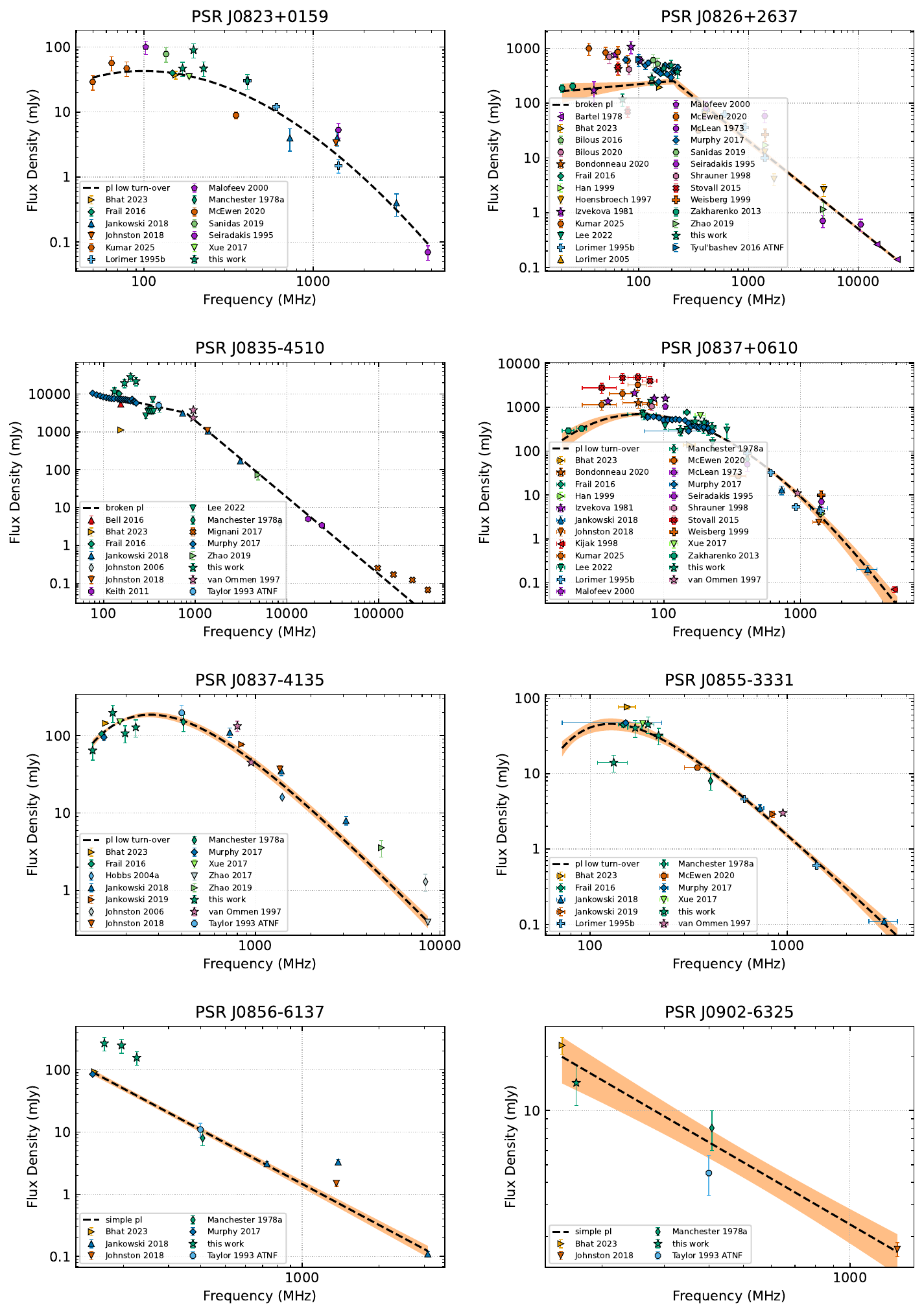}
\caption{Fig \ref{fig:spectra} Continued}
\end{figure*}
\begin{figure*}[htbp!]
\centering
\includegraphics[width=\textwidth,height=22cm]{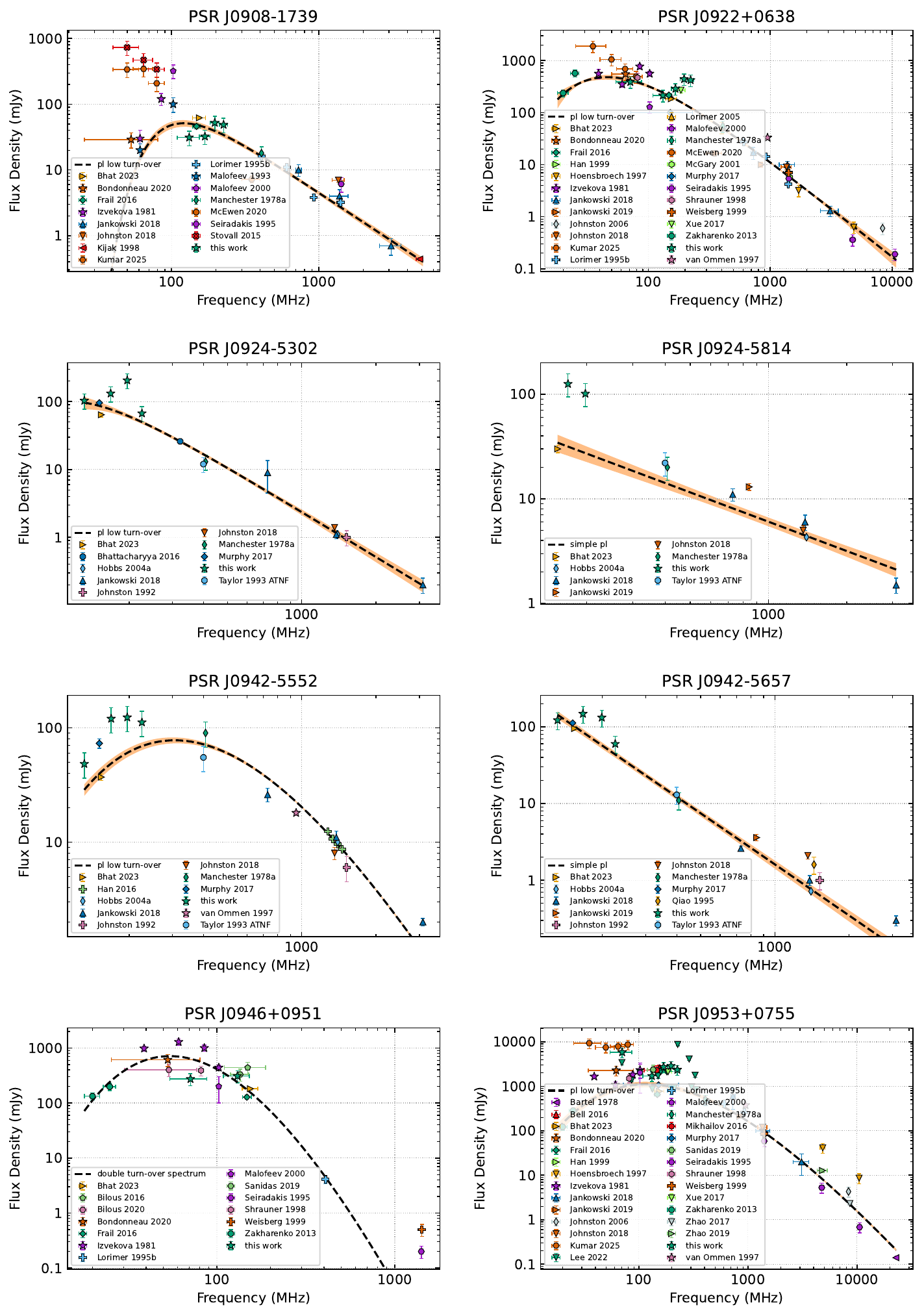}
\caption{Fig \ref{fig:spectra} Continued}
\end{figure*}
\begin{figure*}[htbp!]
\centering
\includegraphics[width=\textwidth,height=22cm]{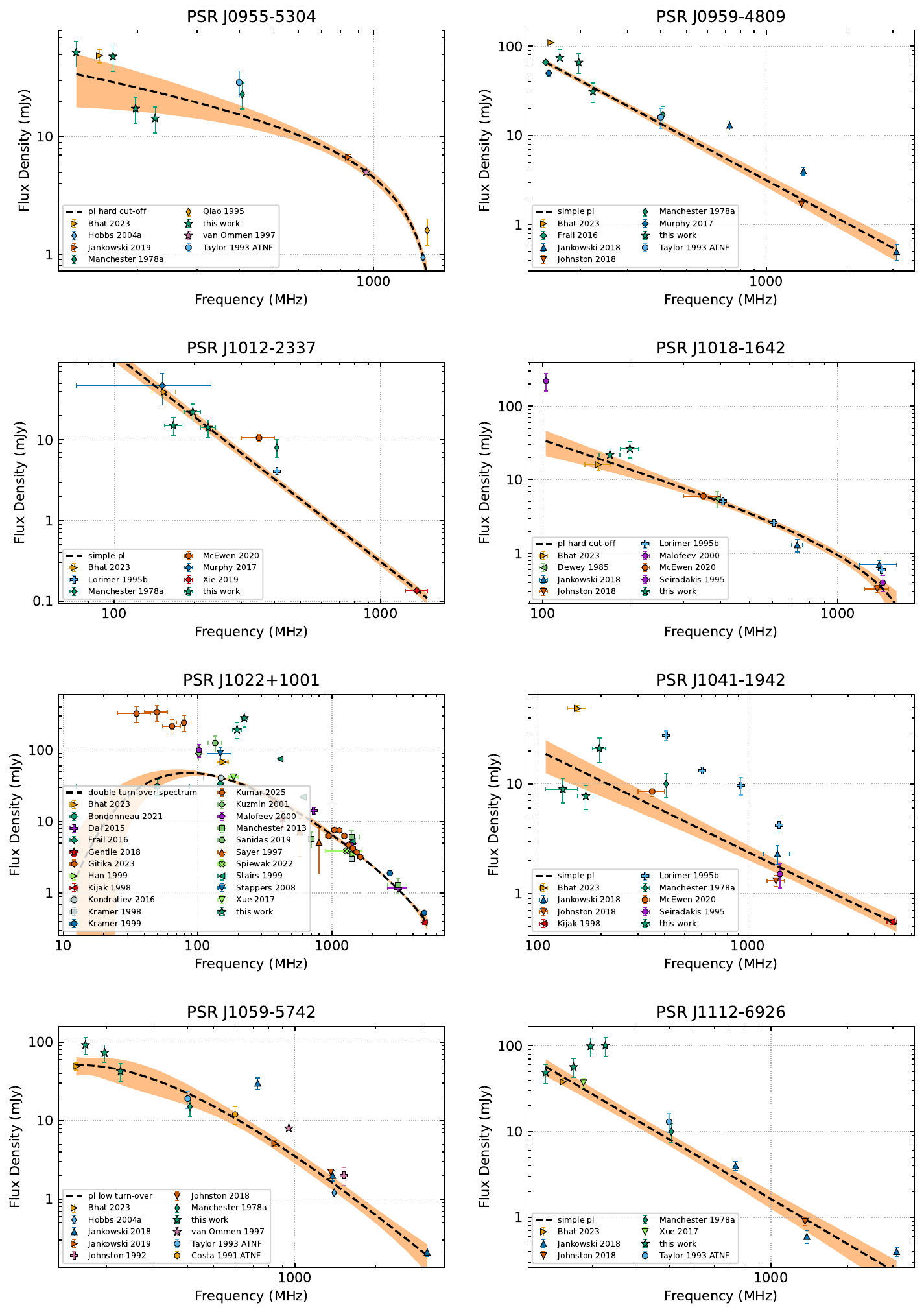}
\caption{Fig \ref{fig:spectra} Continued}
\end{figure*}
\begin{figure*}[htbp!]
\centering
\includegraphics[width=\textwidth,height=22cm]{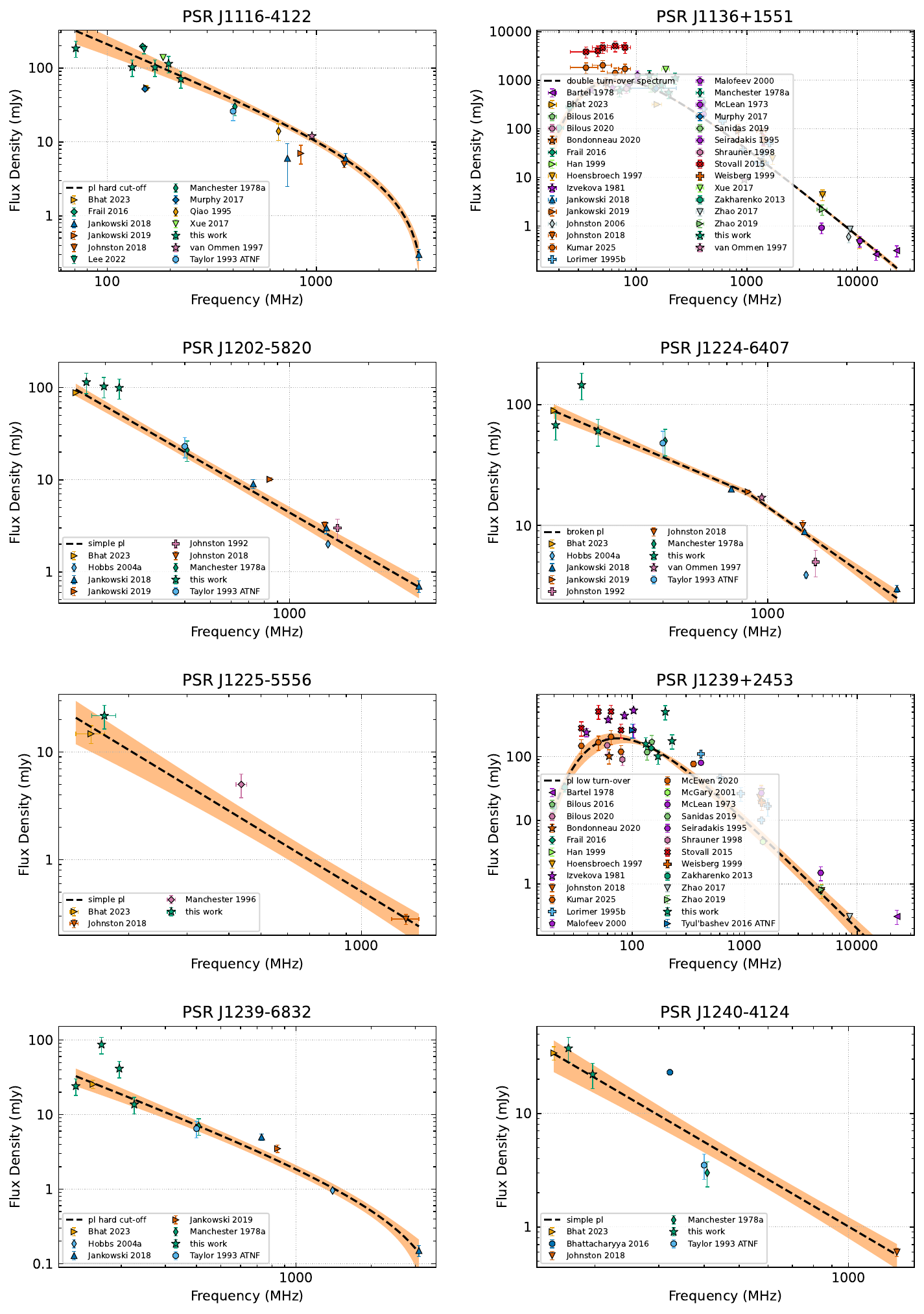}
\caption{Fig \ref{fig:spectra} Continued}
\end{figure*}
\begin{figure*}[htbp!]
\centering
\includegraphics[width=\textwidth,height=22cm]{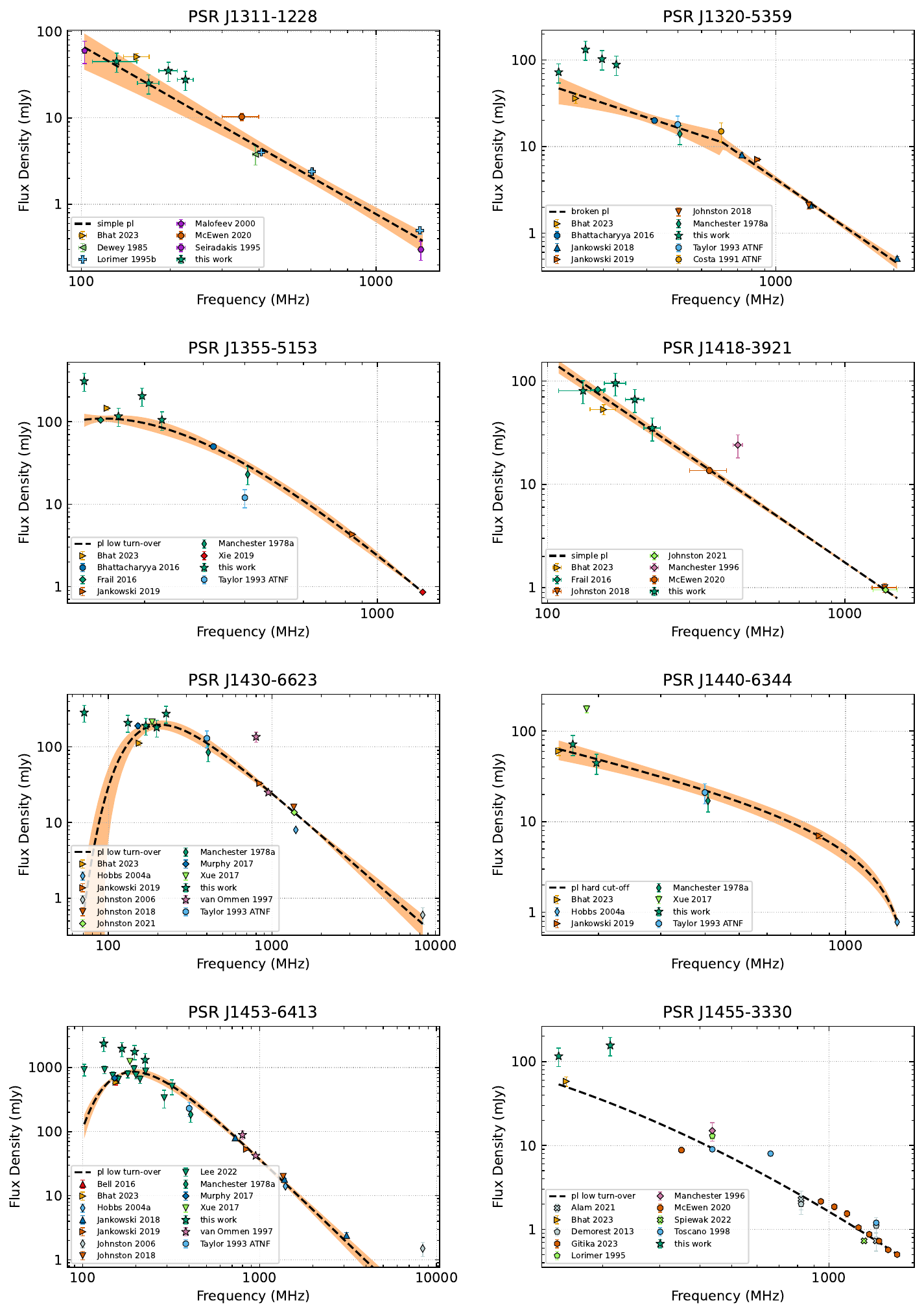}
\caption{Fig \ref{fig:spectra} Continued}
\end{figure*}
\begin{figure*}[htbp!]
\centering
\includegraphics[width=\textwidth,height=22cm]{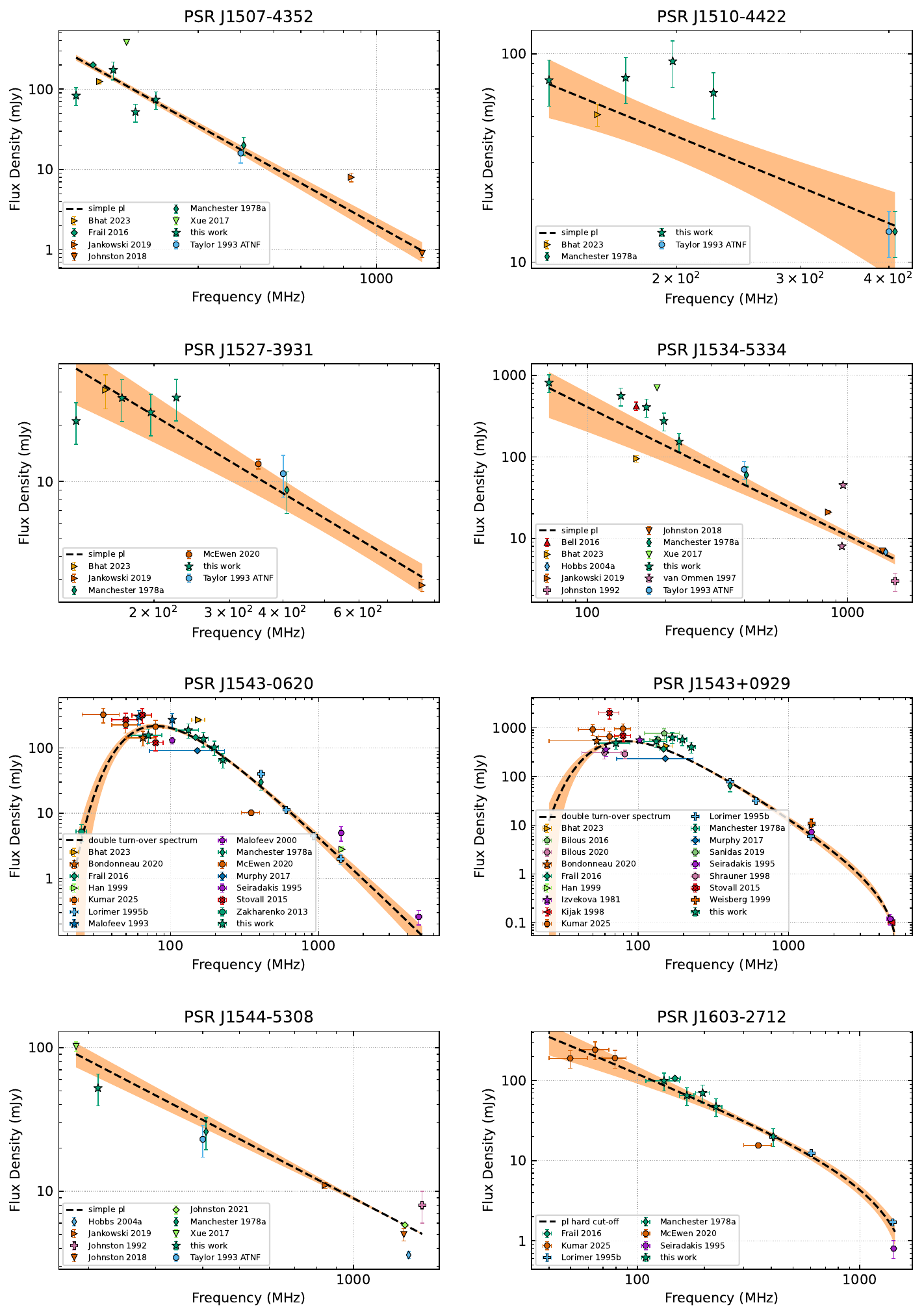}
\caption{Fig \ref{fig:spectra} Continued}
\end{figure*}
\begin{figure*}[htbp!]
\centering
\includegraphics[width=\textwidth,height=22cm]{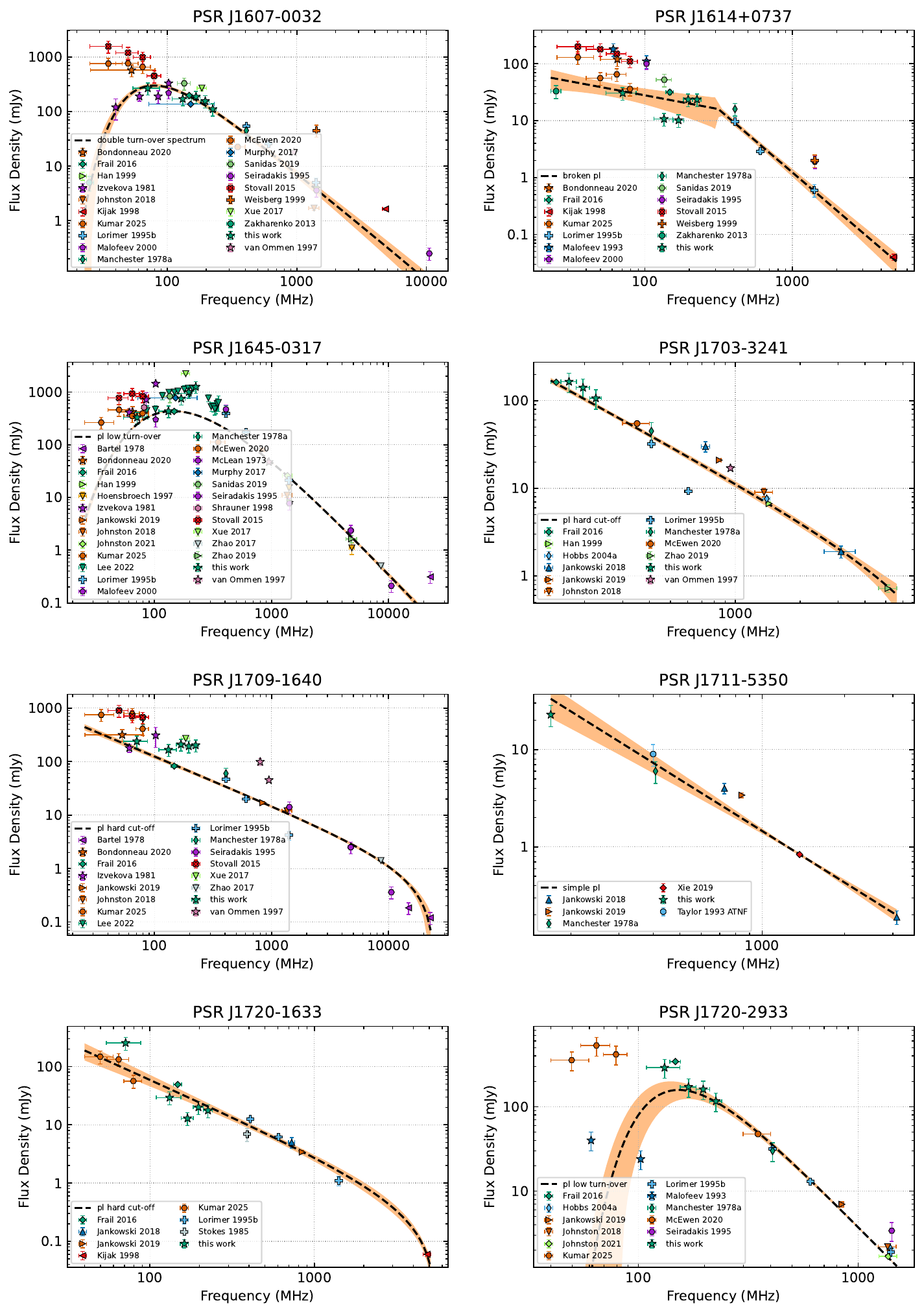}
\caption{Fig \ref{fig:spectra} Continued}
\end{figure*}
\begin{figure*}[htbp!]
\centering
\includegraphics[width=\textwidth,height=22cm]{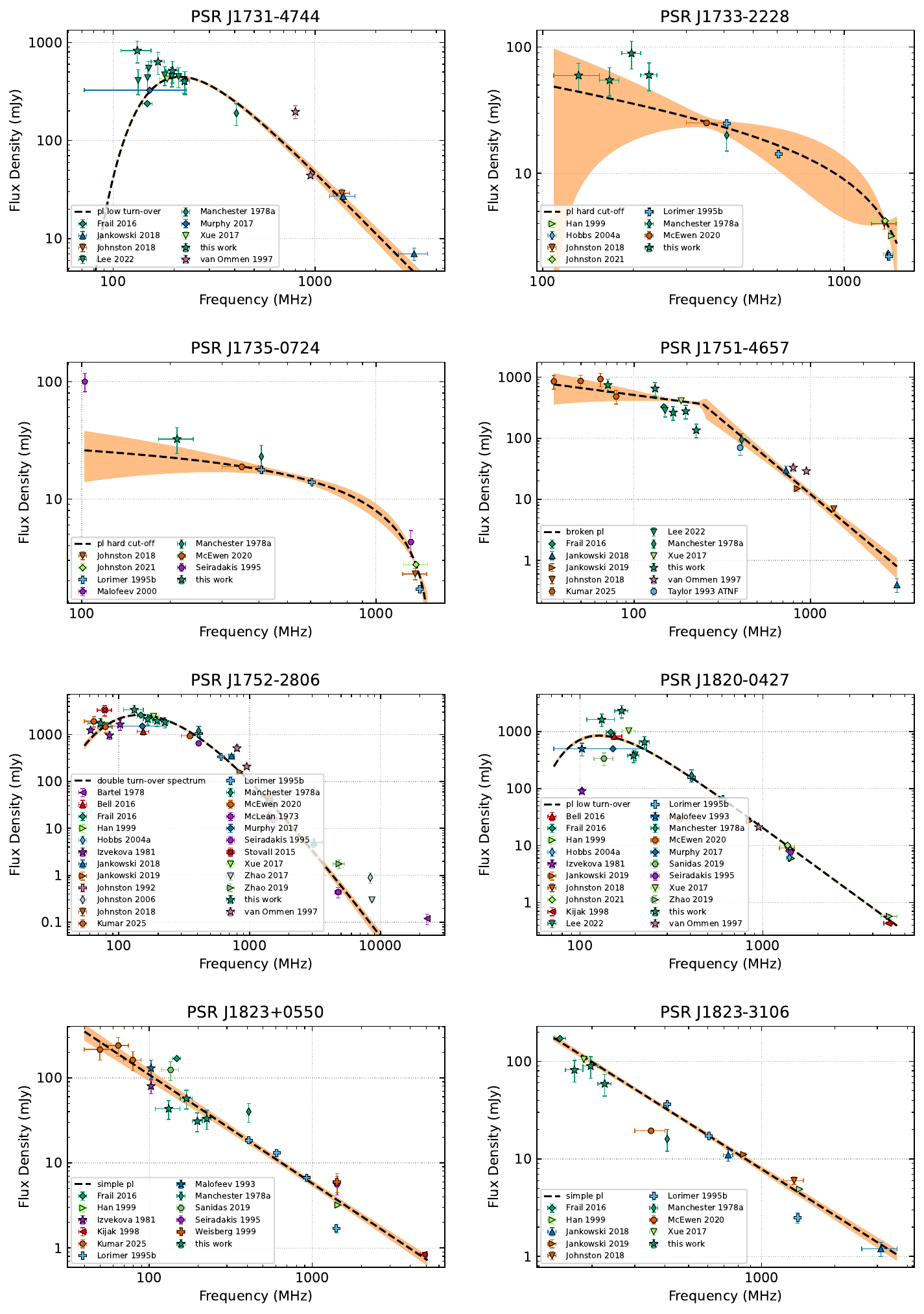}
\caption{Fig \ref{fig:spectra} Continued}
\end{figure*}
\begin{figure*}[htbp!]
\centering
\includegraphics[width=\textwidth,height=22cm]{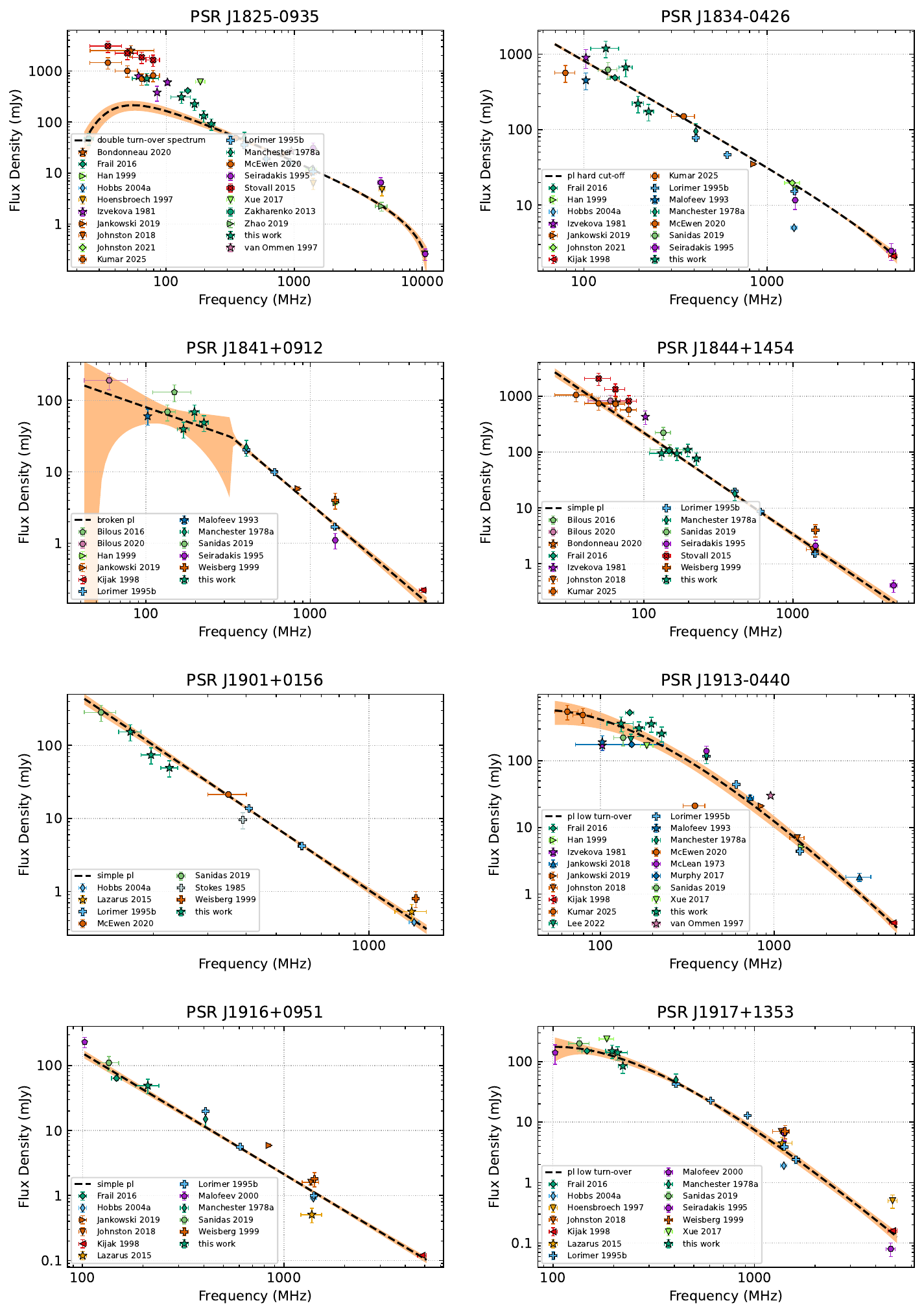}
\caption{Fig \ref{fig:spectra} Continued}
\end{figure*}
\begin{figure*}[htbp!]
\centering
\includegraphics[width=\textwidth,height=22cm]{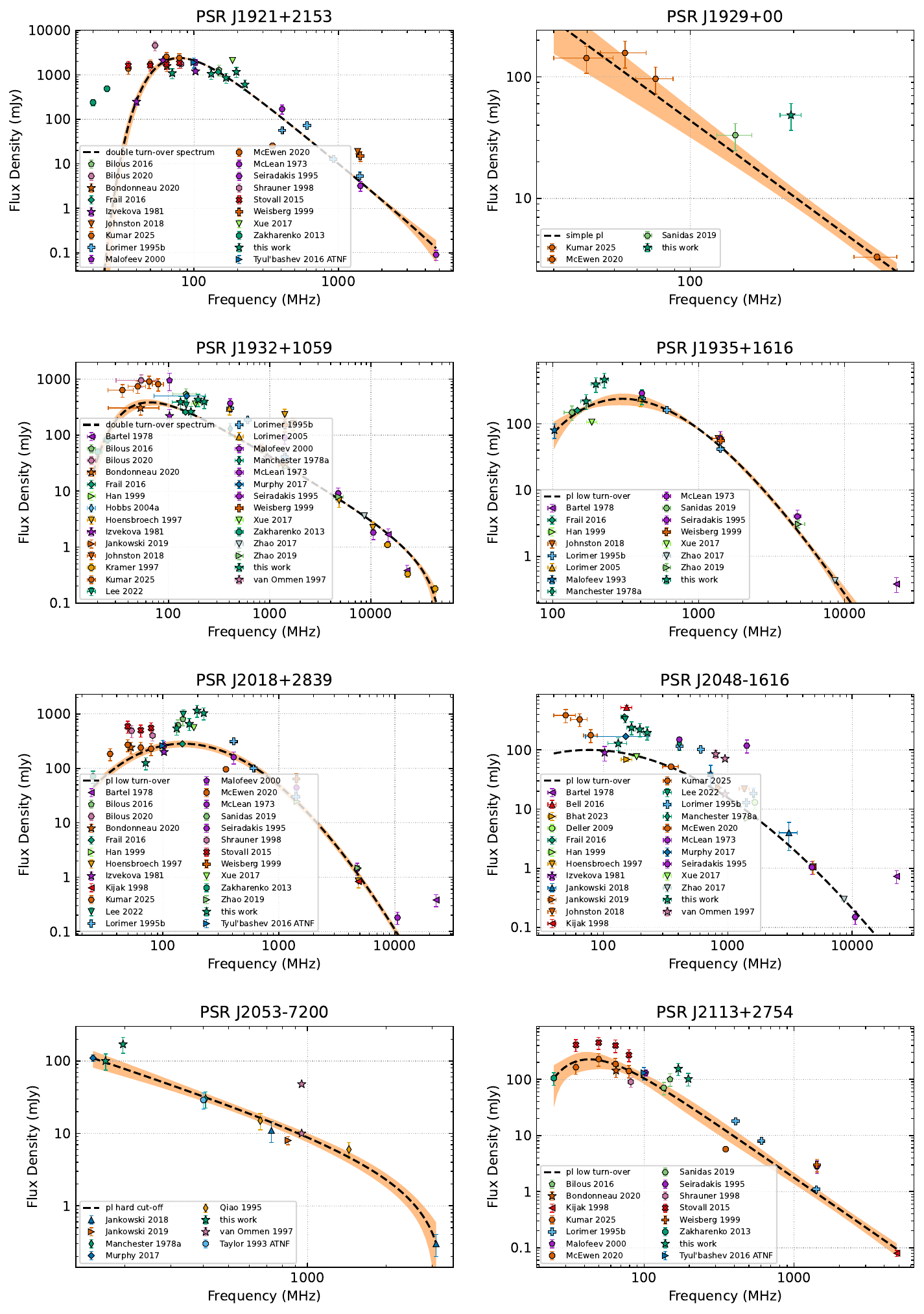}
\caption{Fig \ref{fig:spectra} Continued}
\end{figure*}
\begin{figure*}[htbp!]
\centering
\includegraphics[width=\textwidth,height=22cm]{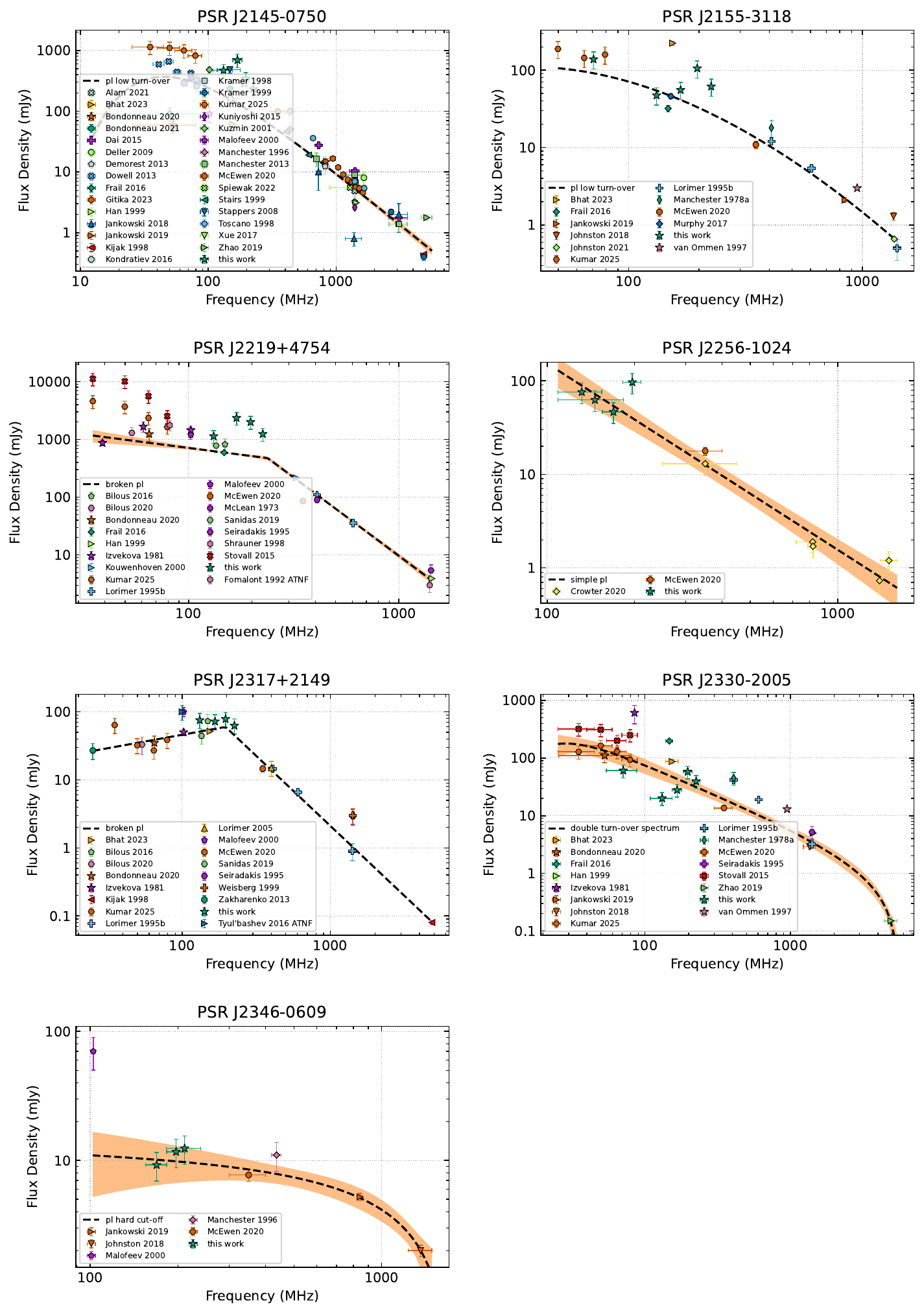}
\caption{Fig \ref{fig:spectra} Continued}
\end{figure*}

\section{Best Polarimetric profile of 36 pulsars}
\begin{figure*}[htbp!]
\includegraphics[width=1.1\textwidth,height=20 cm]{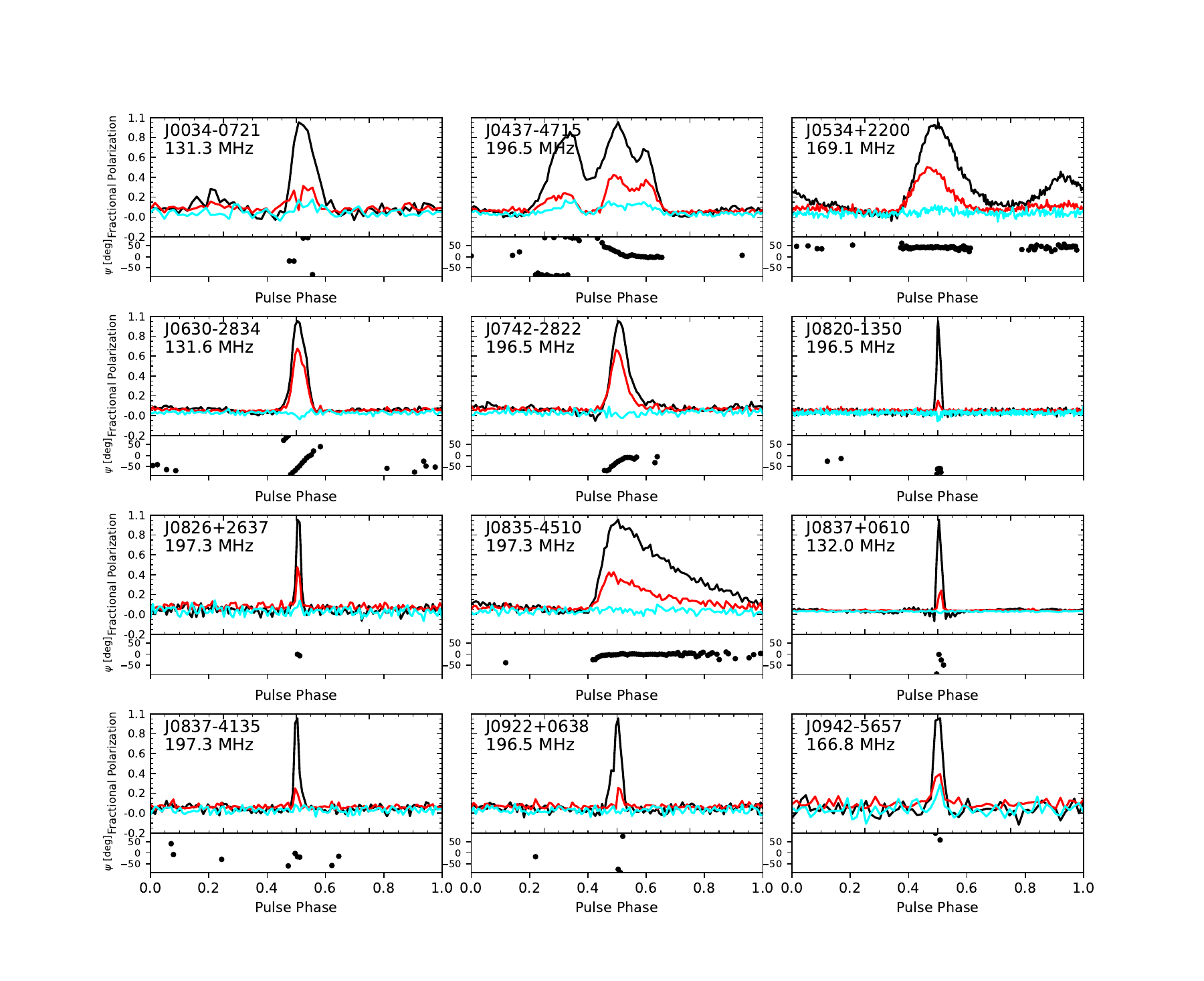}
\caption{Full Stokes best polarimetric pulse profiles for pulsars with RM detection in Table \ref{tab:rmtable}. For each subplot, the top panel shows the fractional linear and circular polarization in red and cyan, respectively, and the total intensity in black. The bottom panel shows the position angle.}
\label{fig:stokesprof}
\end{figure*}

\begin{figure*}[htbp!]
\centering
\includegraphics[width=\textwidth,height=22cm]{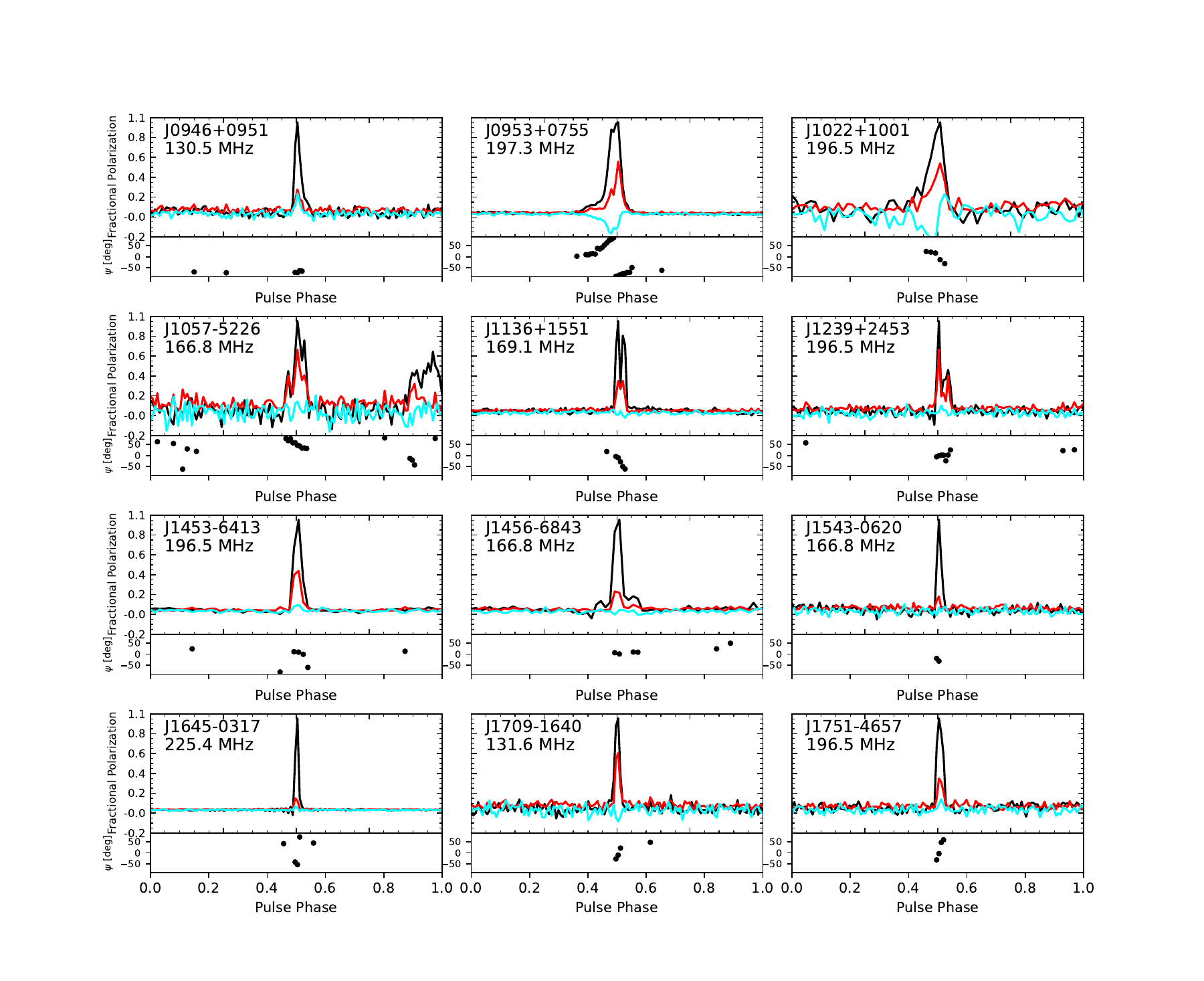}
\caption{Fig \ref{fig:stokesprof} Continued}
\end{figure*}

\begin{figure*}[htbp!]
\centering
\includegraphics[width=\textwidth,height=22cm]{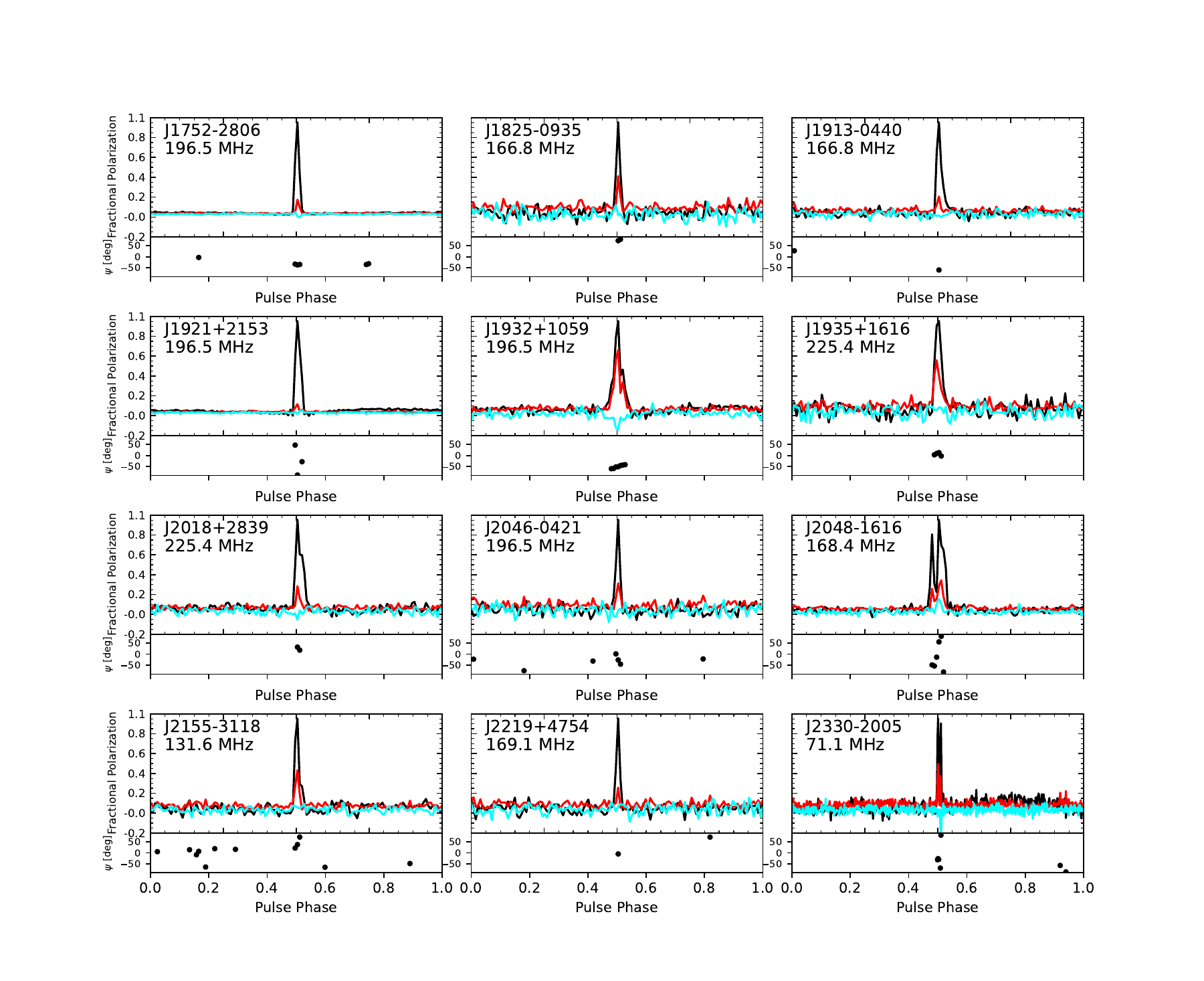}
\caption{Fig \ref{fig:stokesprof} Continued}
\end{figure*}

\section{Multi-frequency polarimetric profiles for 20 pulsars}
\begin{figure}[htbp!]
\includegraphics[width=\textwidth,height=20 cm]{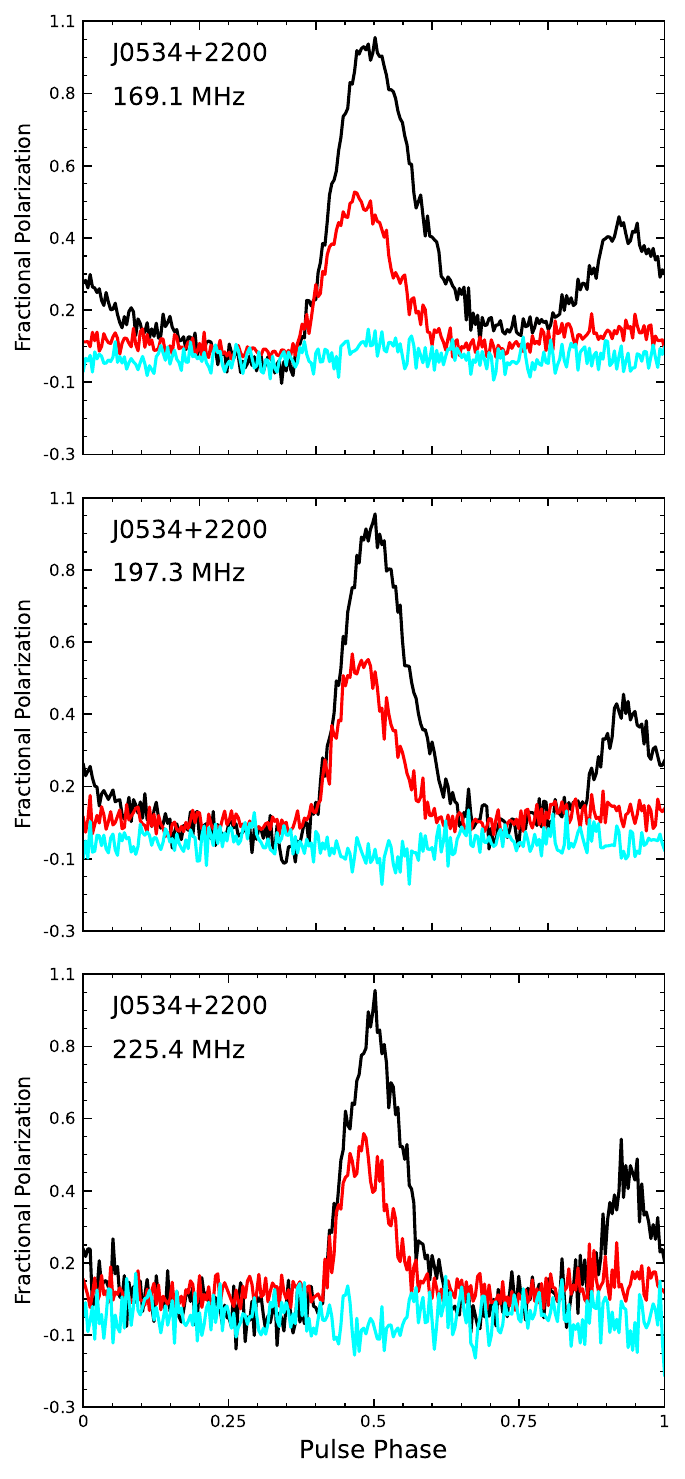}
\caption{Multi-frequency polarization profiles for EDA2 pulsar, where black is the total intensity, and red and cyan show the fraction of linear and circular polarization.}
\label{fig:multipol}
\end{figure}
\begin{figure*}[htbp!]
\centering
\includegraphics[width=\textwidth,height=22cm]{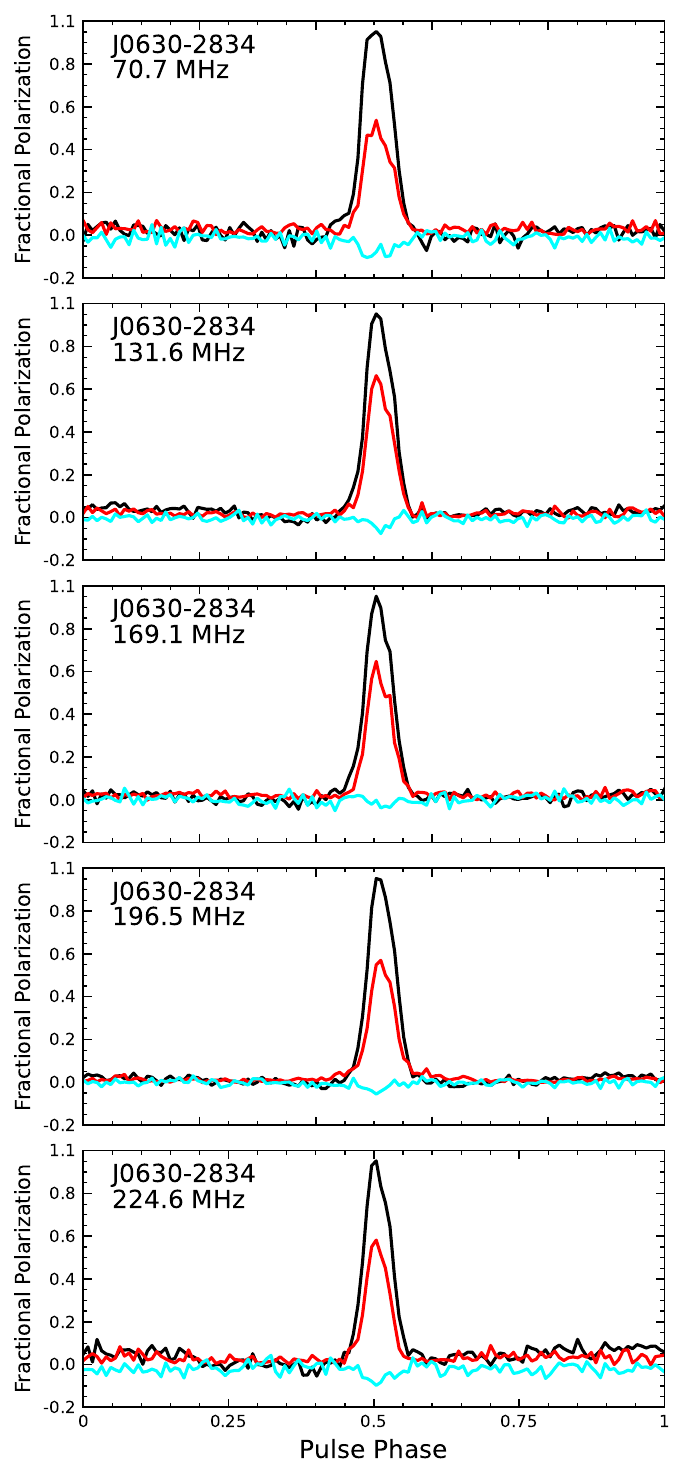}
\caption{Fig \ref{fig:multipol} Continued}
\end{figure*}
\begin{figure*}[htbp!]
\centering
\includegraphics[width=\textwidth,height=22cm]{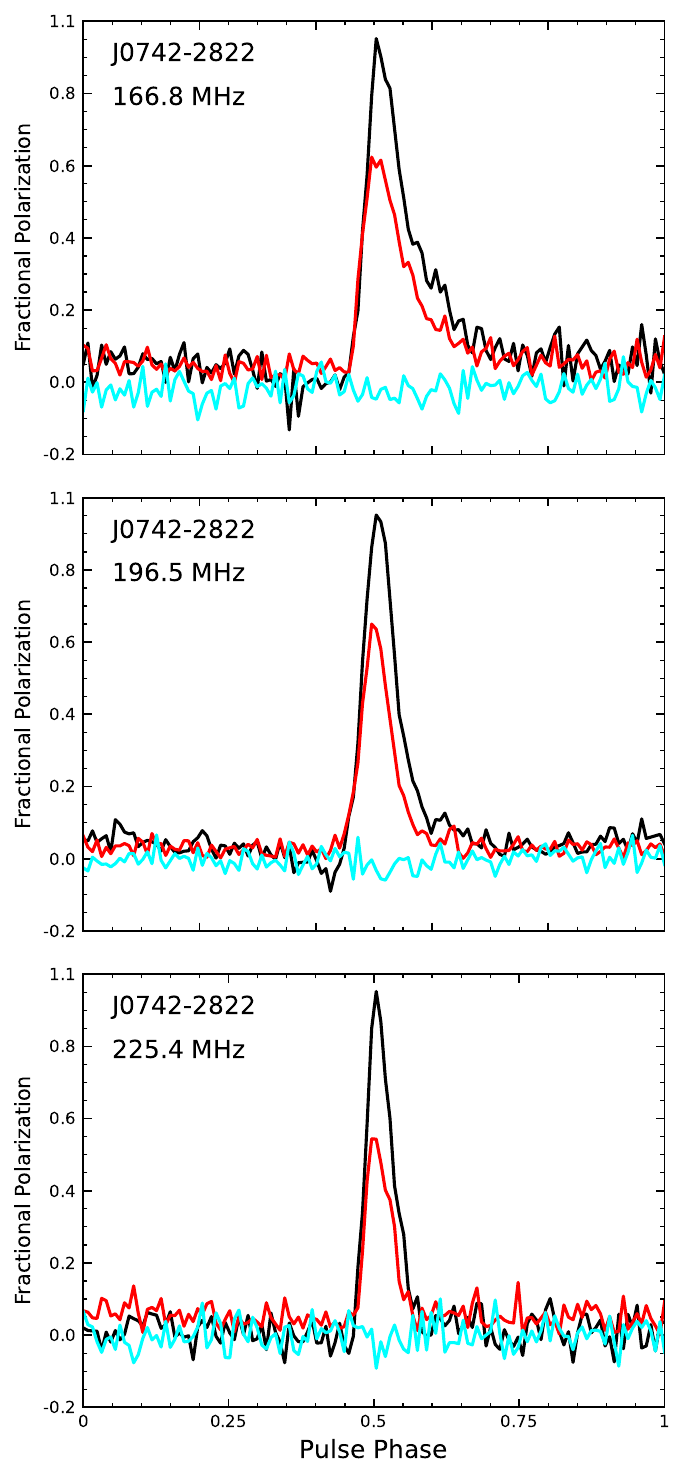}
\caption{Fig \ref{fig:multipol} Continued}
\end{figure*}
\begin{figure*}[htbp!]
\centering
\includegraphics[width=\textwidth,height=22cm]{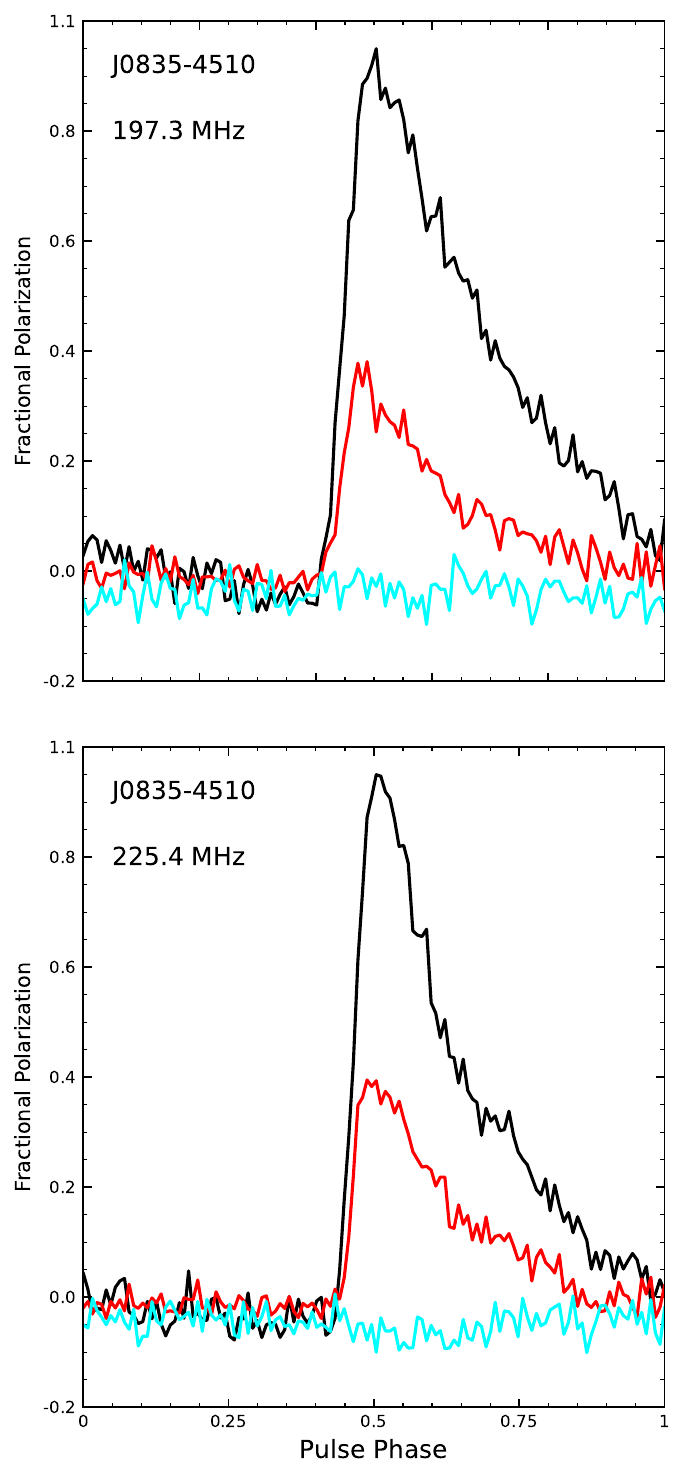}
\caption{Fig \ref{fig:multipol} Continued}
\end{figure*}
\begin{figure*}[htbp!]
\centering
\includegraphics[width=\textwidth,height=22cm]{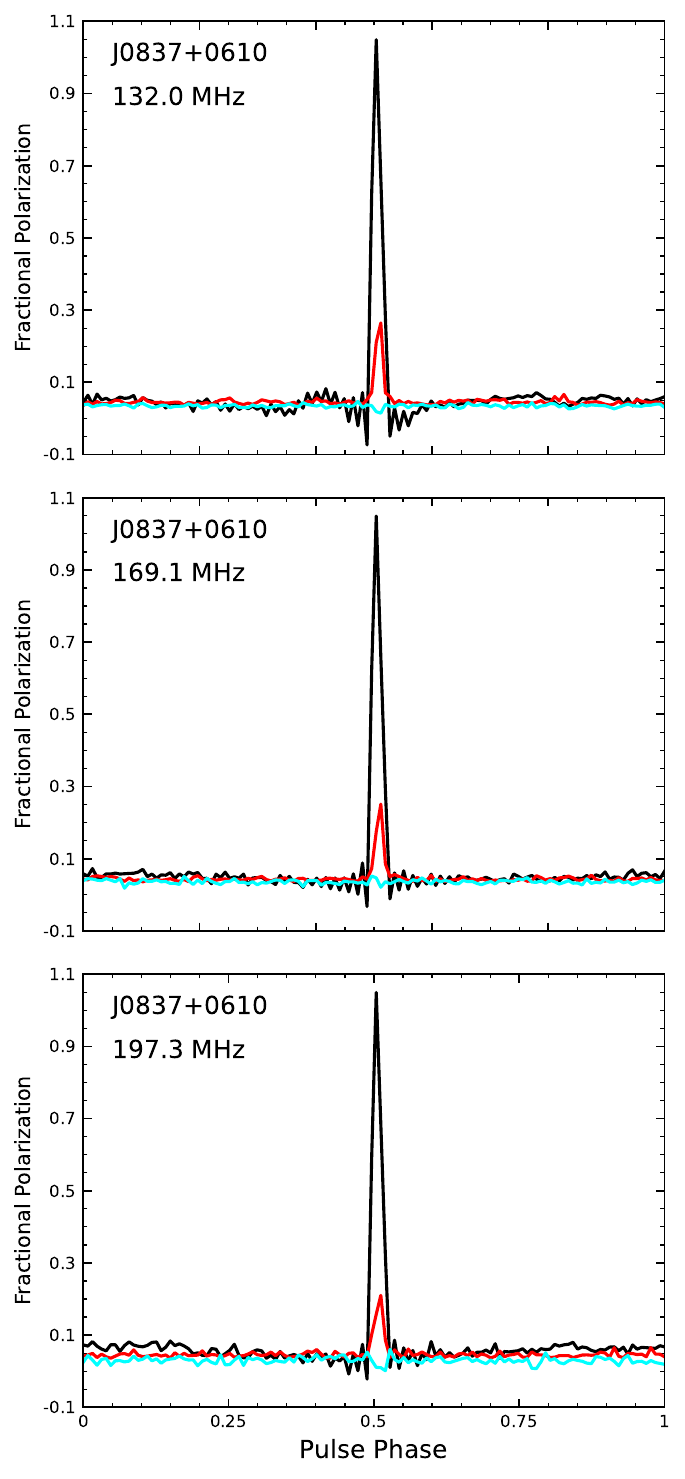}
\caption{Fig \ref{fig:multipol} Continued}
\end{figure*}
\begin{figure*}[htbp!]
\centering
\includegraphics[width=\textwidth,height=22cm]{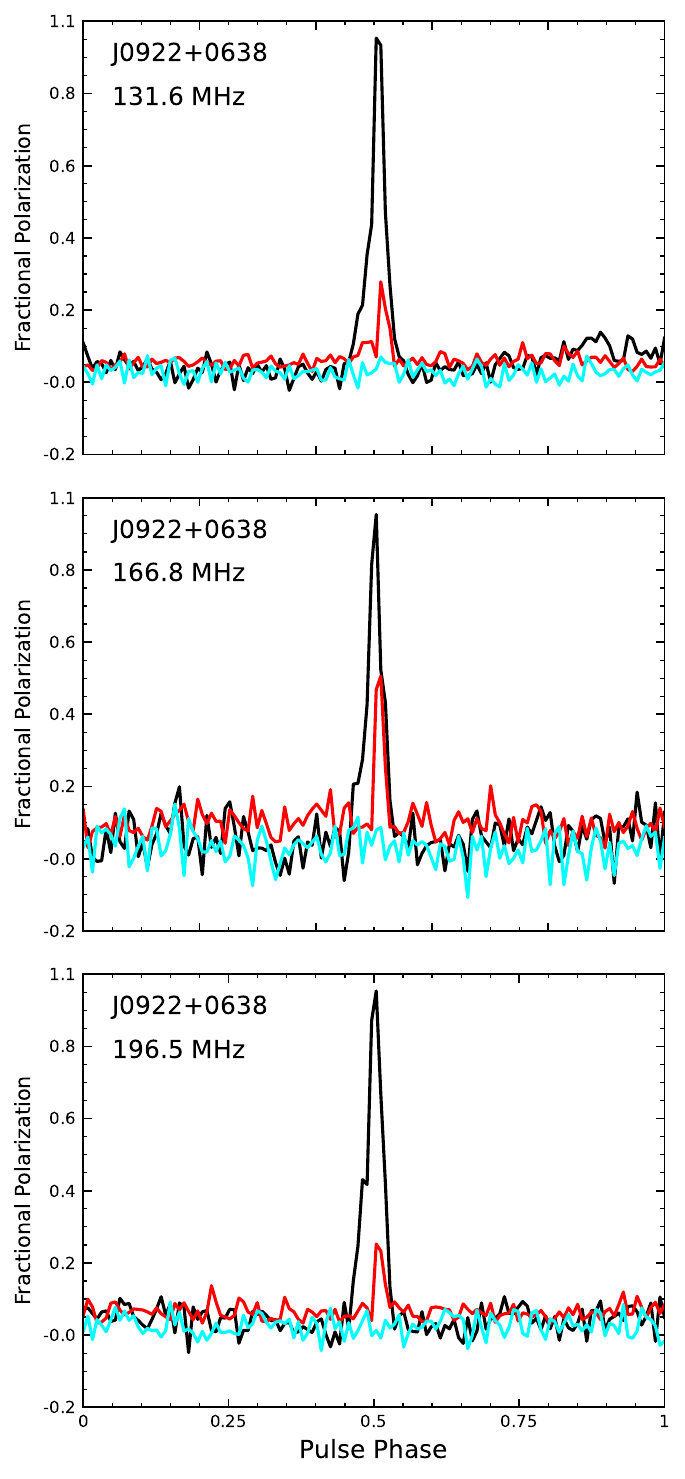}
\caption{Fig \ref{fig:multipol} Continued}
\end{figure*}
\begin{figure*}[htbp!]
\centering
\includegraphics[width=\textwidth,height=22cm]{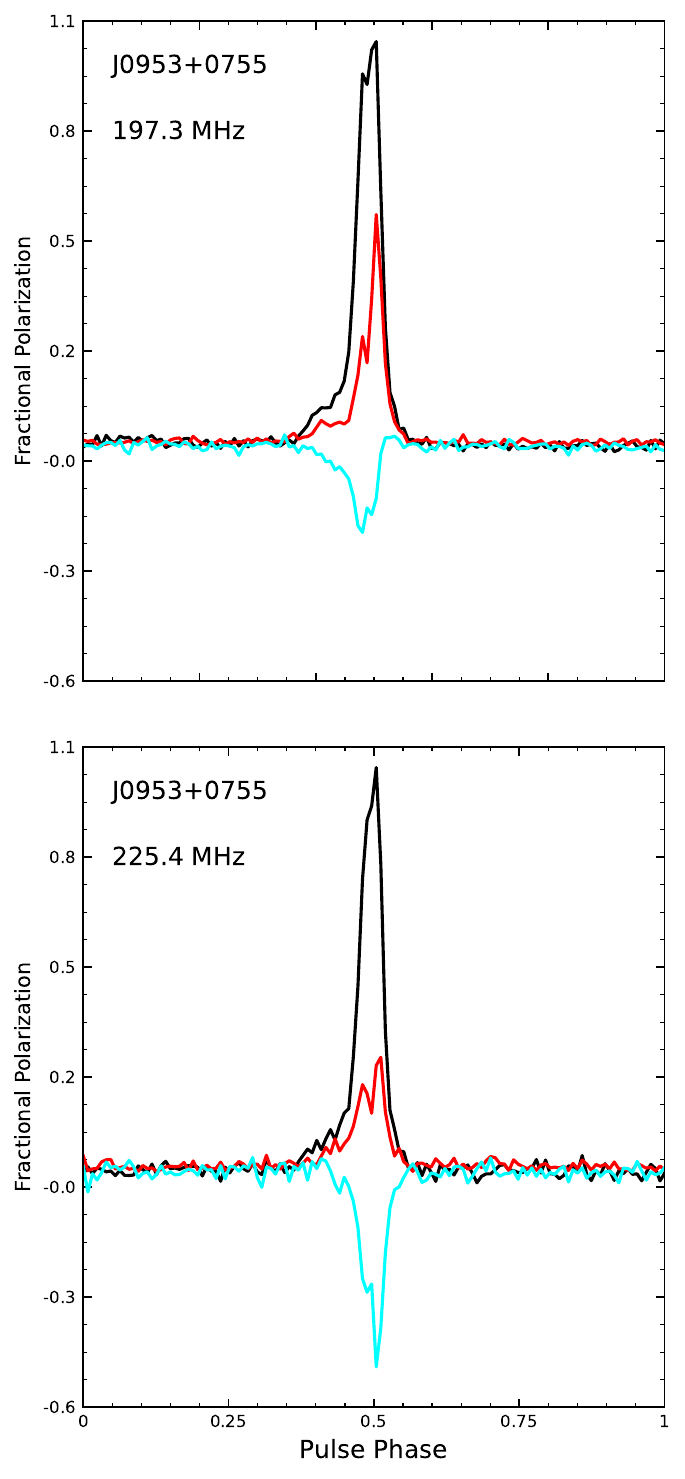}
\caption{Fig \ref{fig:multipol} Continued}
\end{figure*}
\begin{figure*}[htbp!]
\centering
\includegraphics[width=\textwidth,height=22cm]{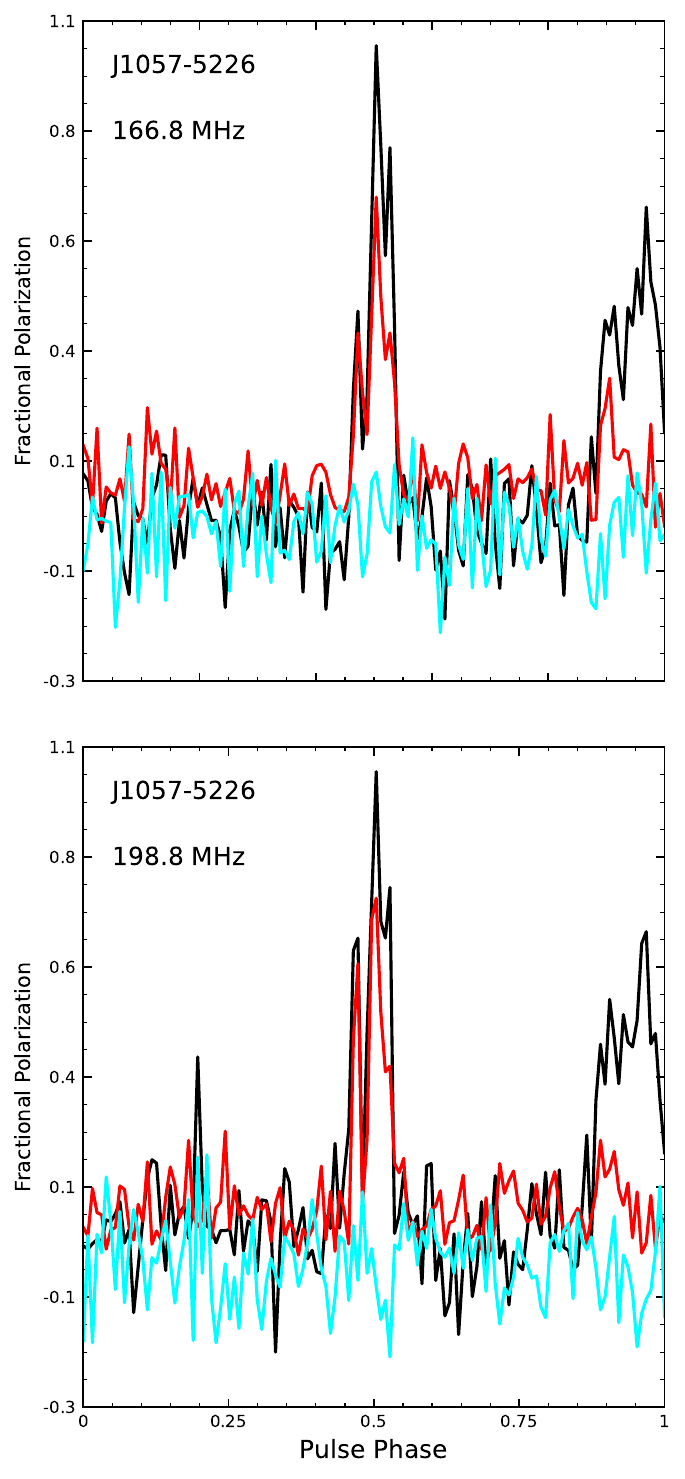}
\caption{Fig \ref{fig:multipol} Continued}
\end{figure*}
\begin{figure*}[htbp!]
\centering
\includegraphics[width=\textwidth,height=22cm]{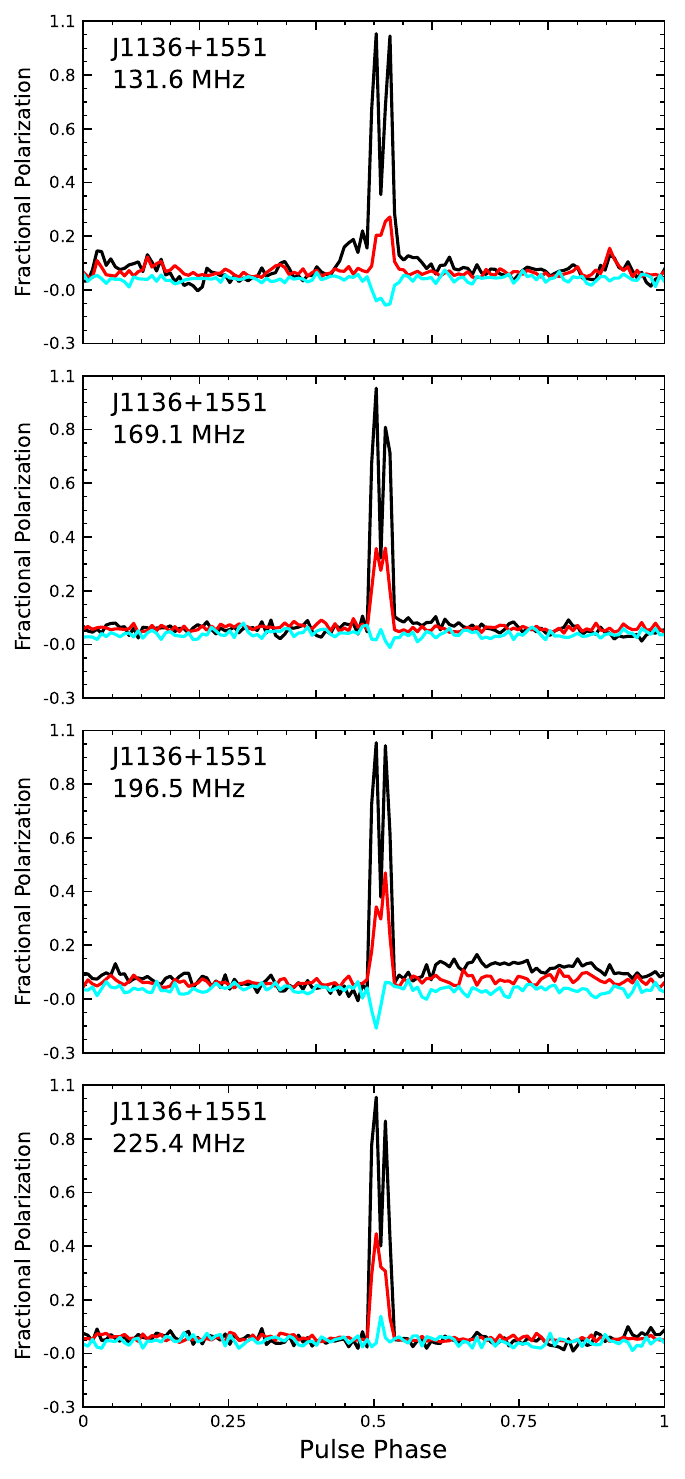}
\caption{Fig \ref{fig:multipol} Continued}
\end{figure*}
\begin{figure*}[htbp!]
\centering
\includegraphics[width=\textwidth,height=22cm]{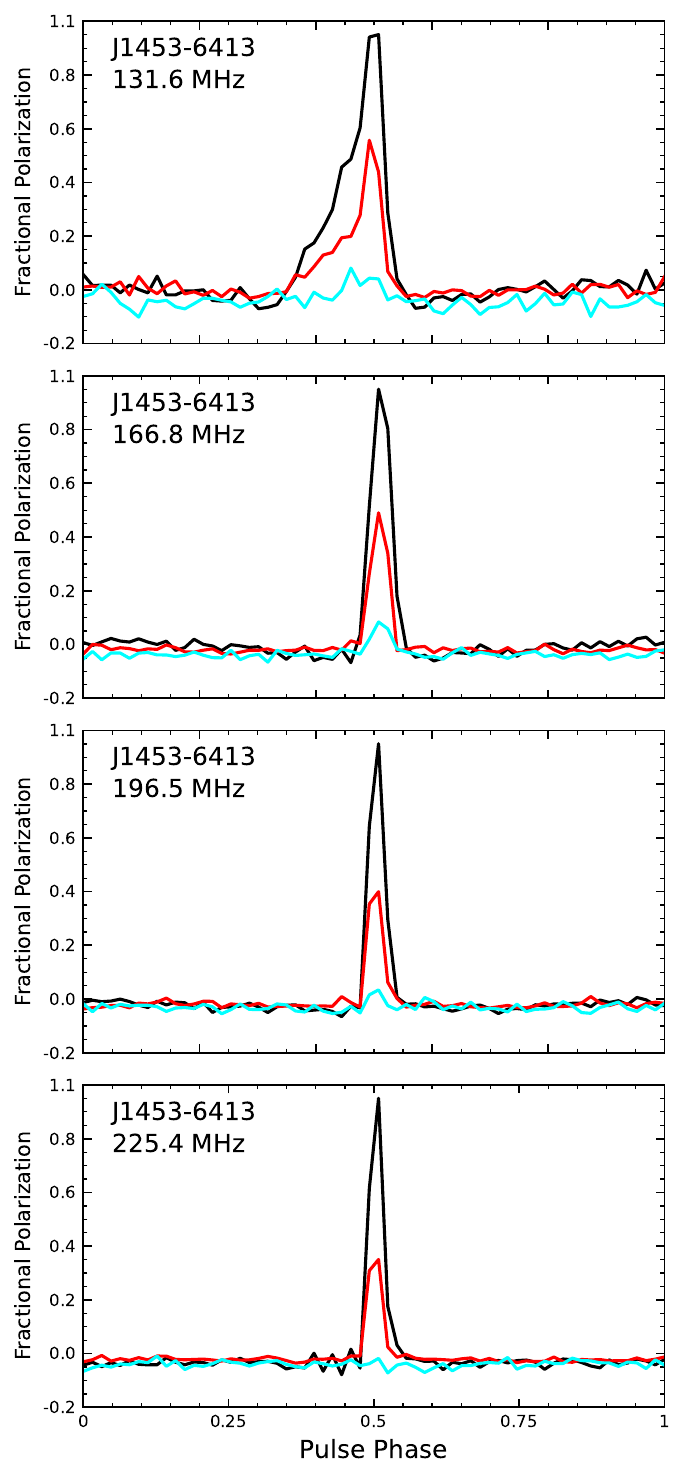}
\caption{Fig \ref{fig:multipol} Continued}
\end{figure*}
\begin{figure*}[htbp!]
\centering
\includegraphics[width=\textwidth,height=22cm]{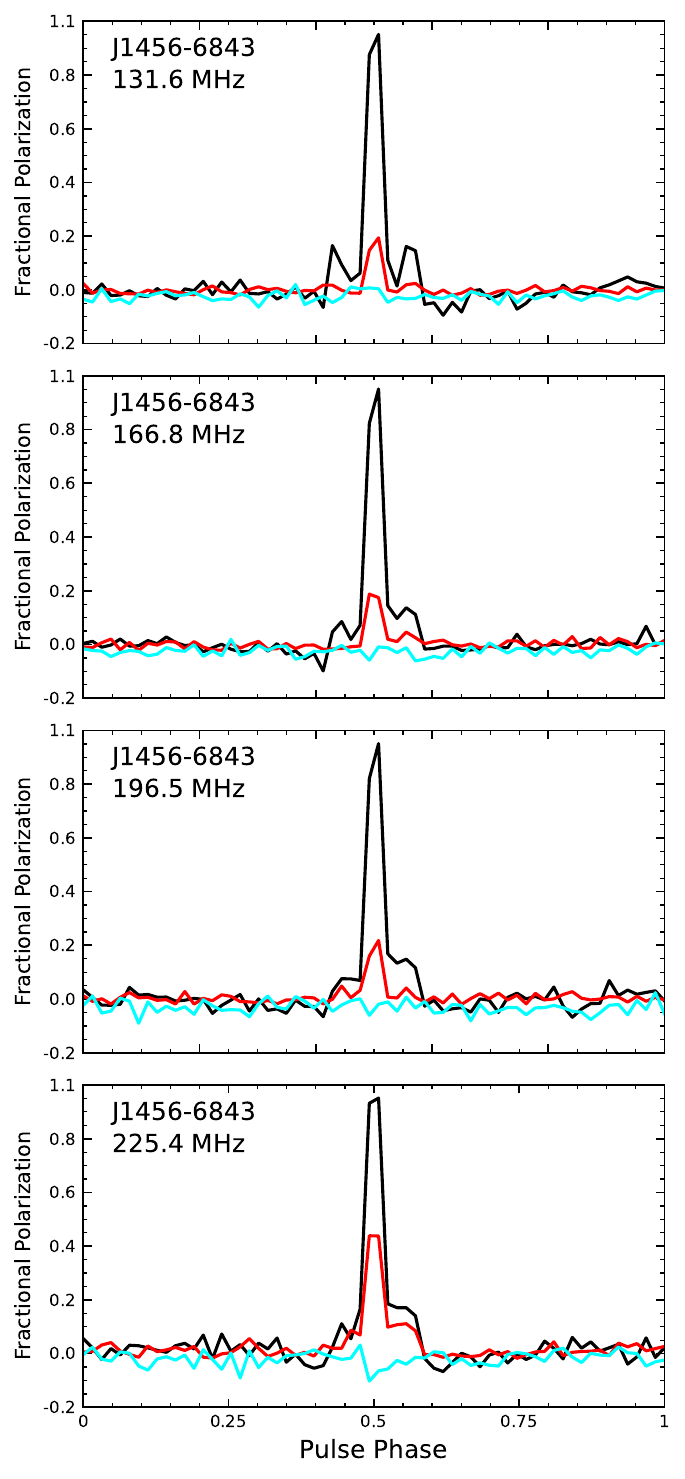}
\caption{Fig \ref{fig:multipol} Continued}
\end{figure*}
\begin{figure*}[htbp!]
\centering
\includegraphics[width=\textwidth,height=22cm]{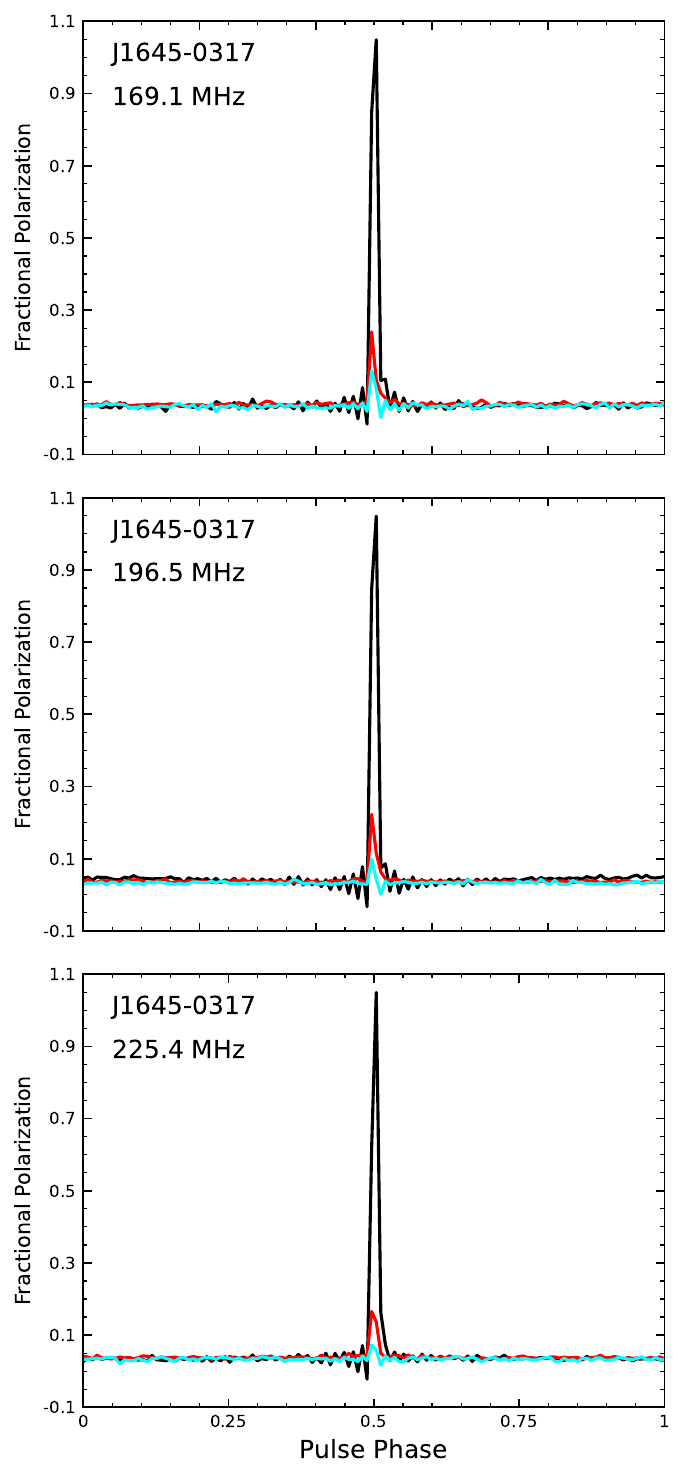}
\caption{Fig \ref{fig:multipol} Continued}
\end{figure*}
\begin{figure*}[htbp!]
\centering
\includegraphics[width=\textwidth,height=22cm]{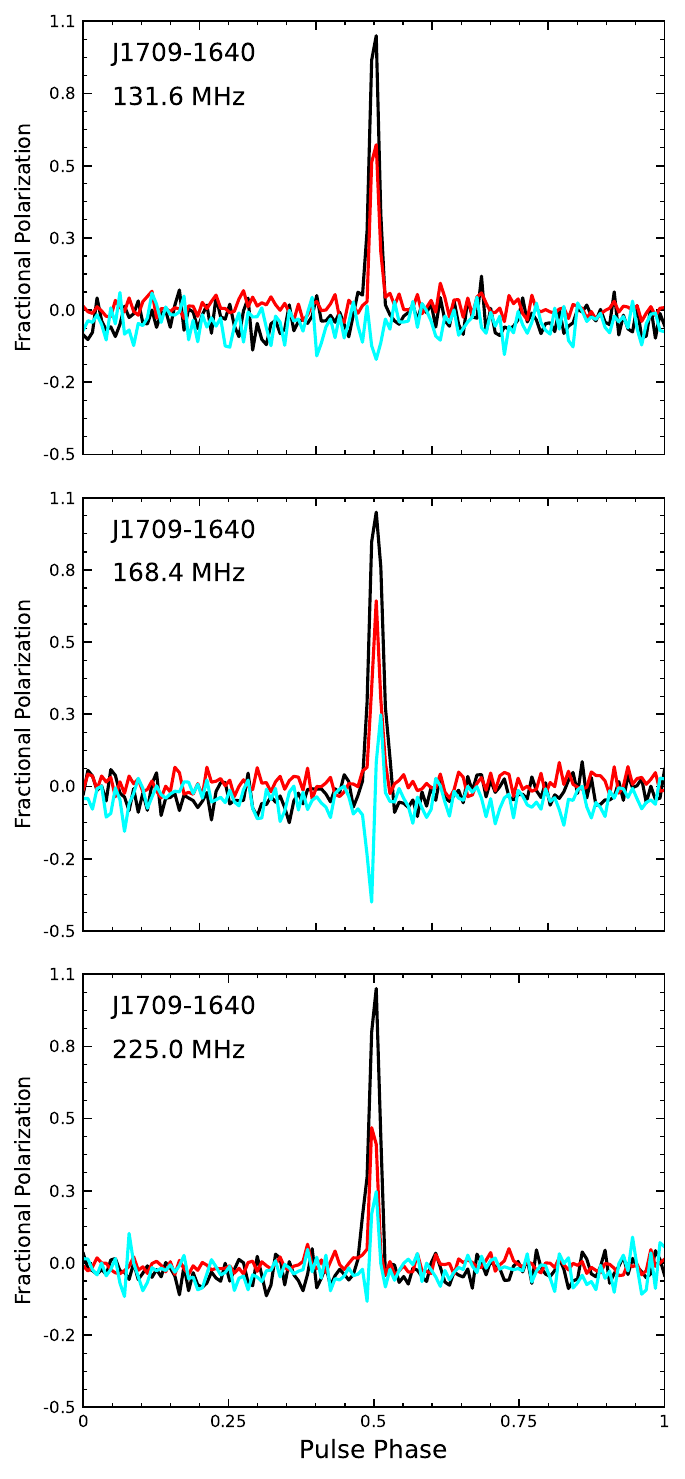}
\caption{Fig \ref{fig:multipol} Continued}
\end{figure*}
\begin{figure*}[htbp!]
\centering
\includegraphics[width=\textwidth,height=22cm]{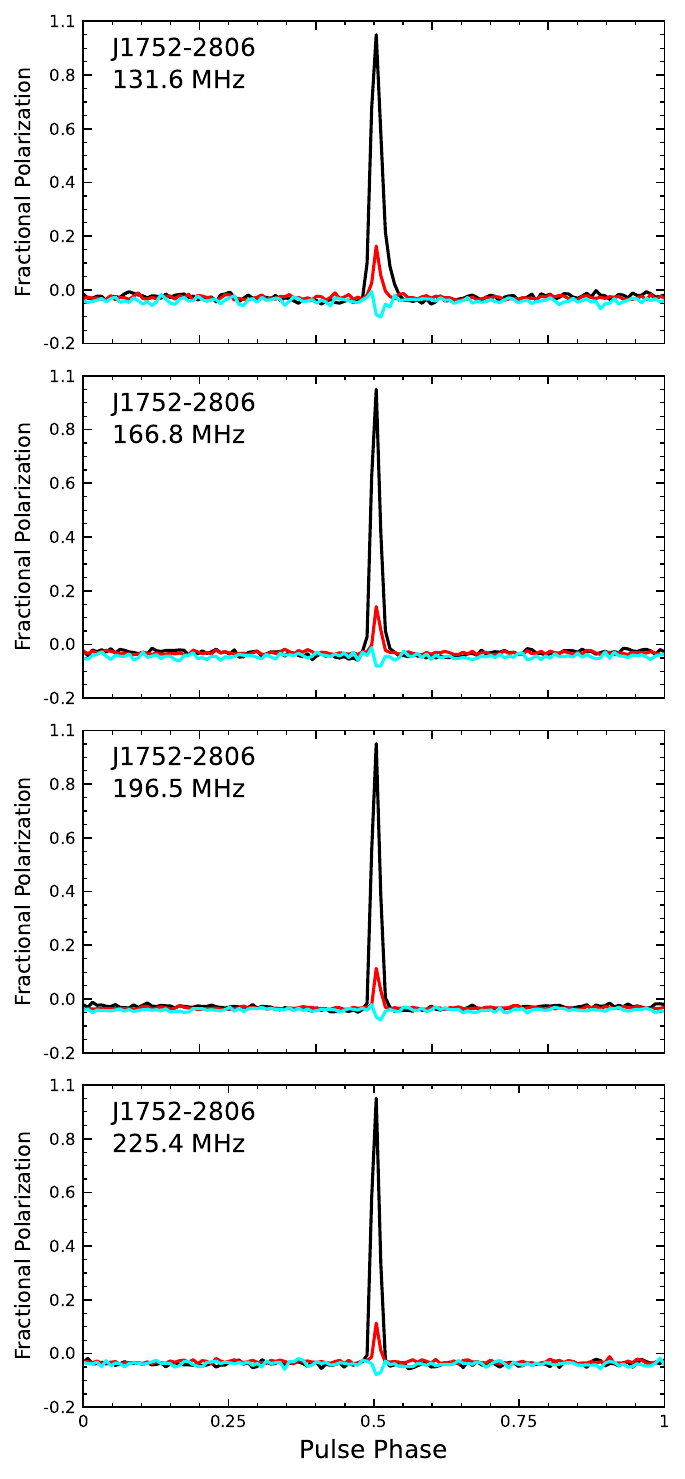}
\caption{Fig \ref{fig:multipol} Continued}
\end{figure*}
\begin{figure*}[htbp!]
\centering
\includegraphics[width=\textwidth,height=22cm]{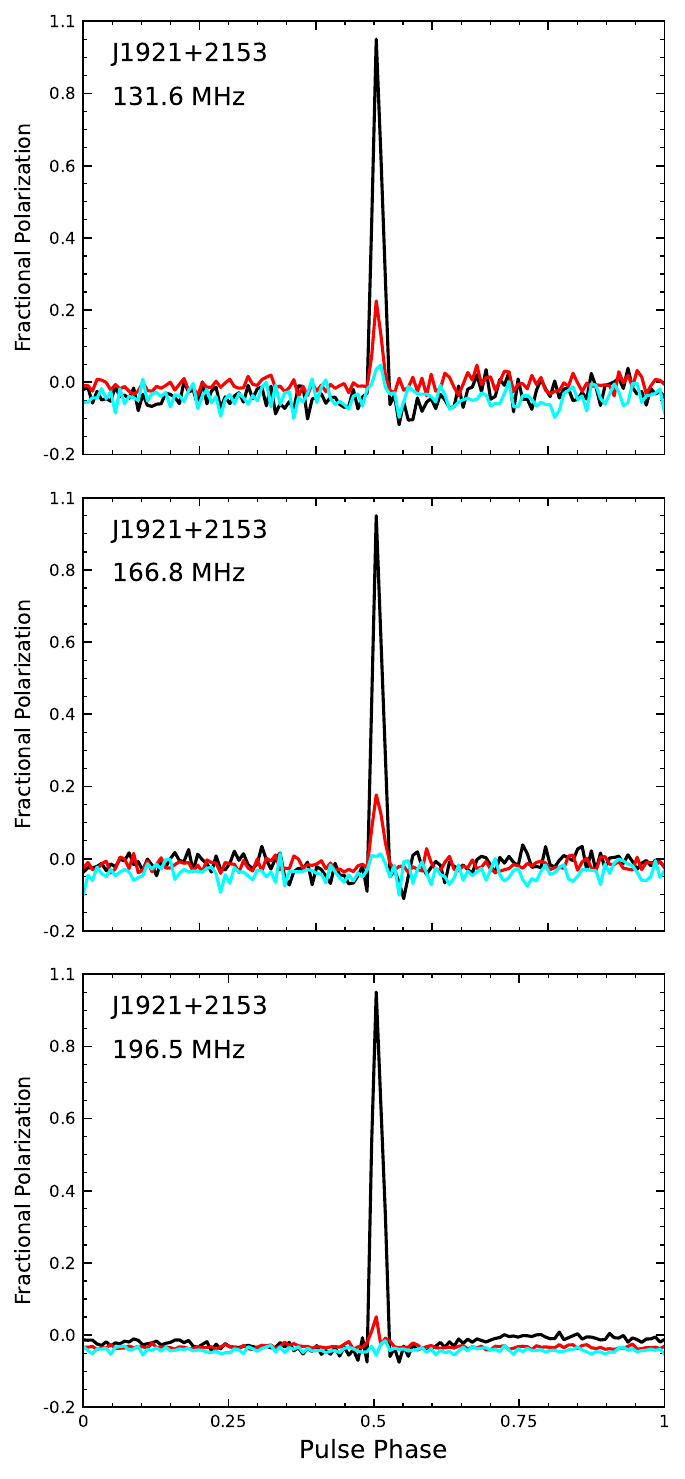}
\caption{Fig \ref{fig:multipol} Continued}
\end{figure*}
\begin{figure*}[htbp!]
\centering
\includegraphics[width=\textwidth,height=22cm]{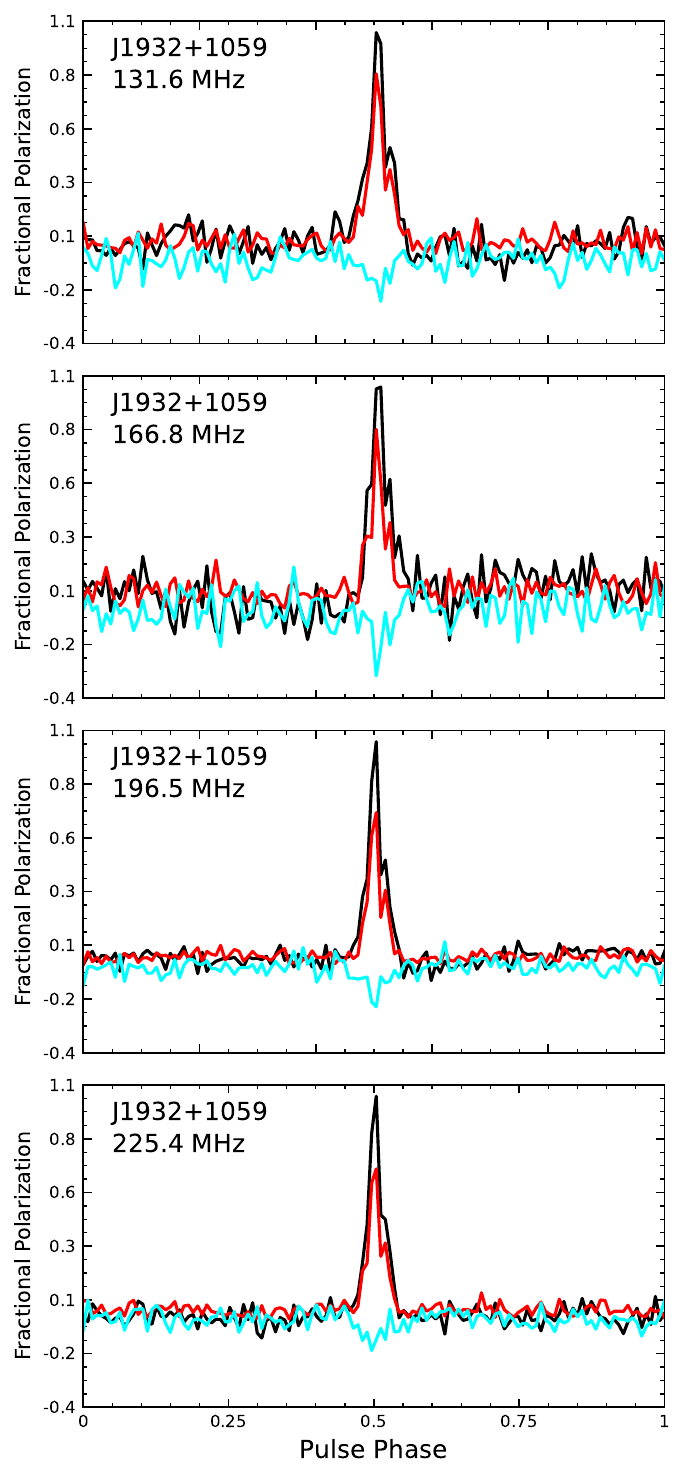}
\caption{Fig \ref{fig:multipol} Continued}
\end{figure*}
\begin{figure*}[htbp!]
\centering
\includegraphics[width=\textwidth,height=22cm]{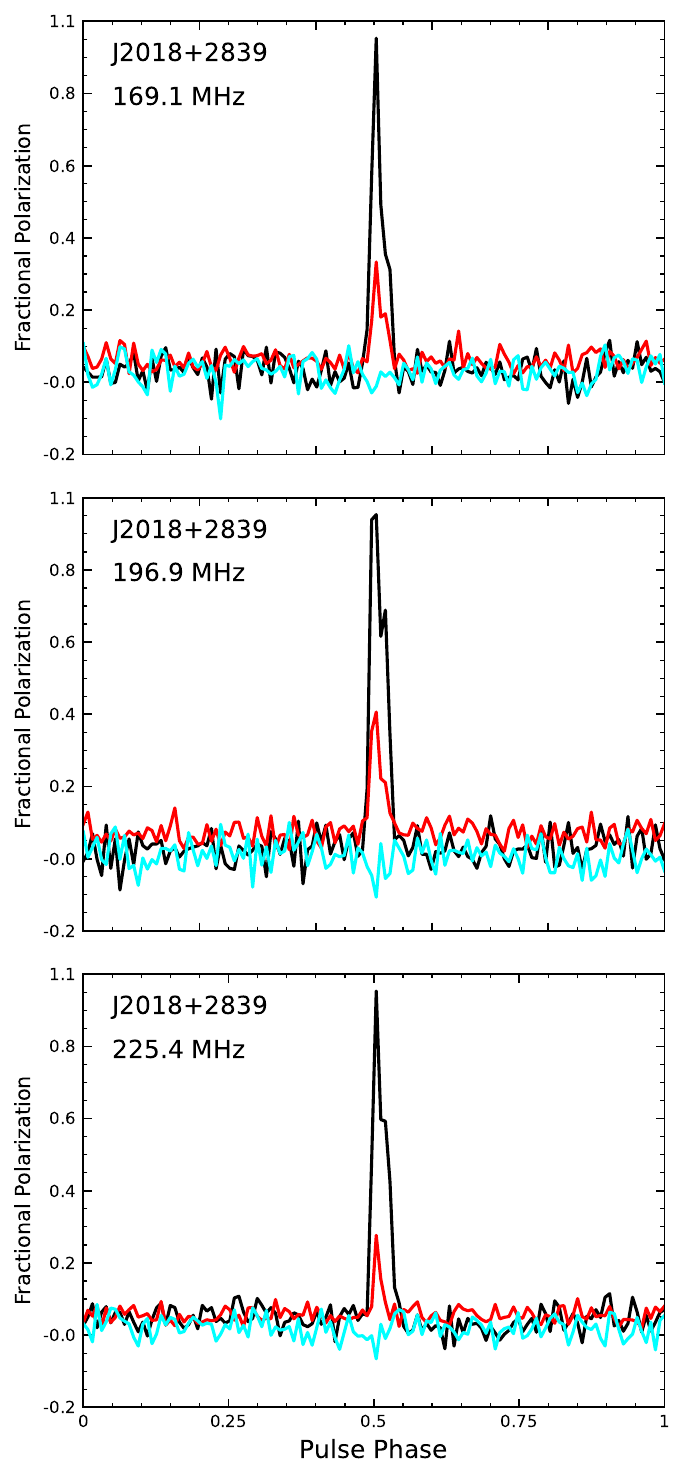}
\caption{Fig \ref{fig:multipol} Continued}
\end{figure*}
\begin{figure*}[htbp!]
\centering
\includegraphics[width=\textwidth,height=22cm]{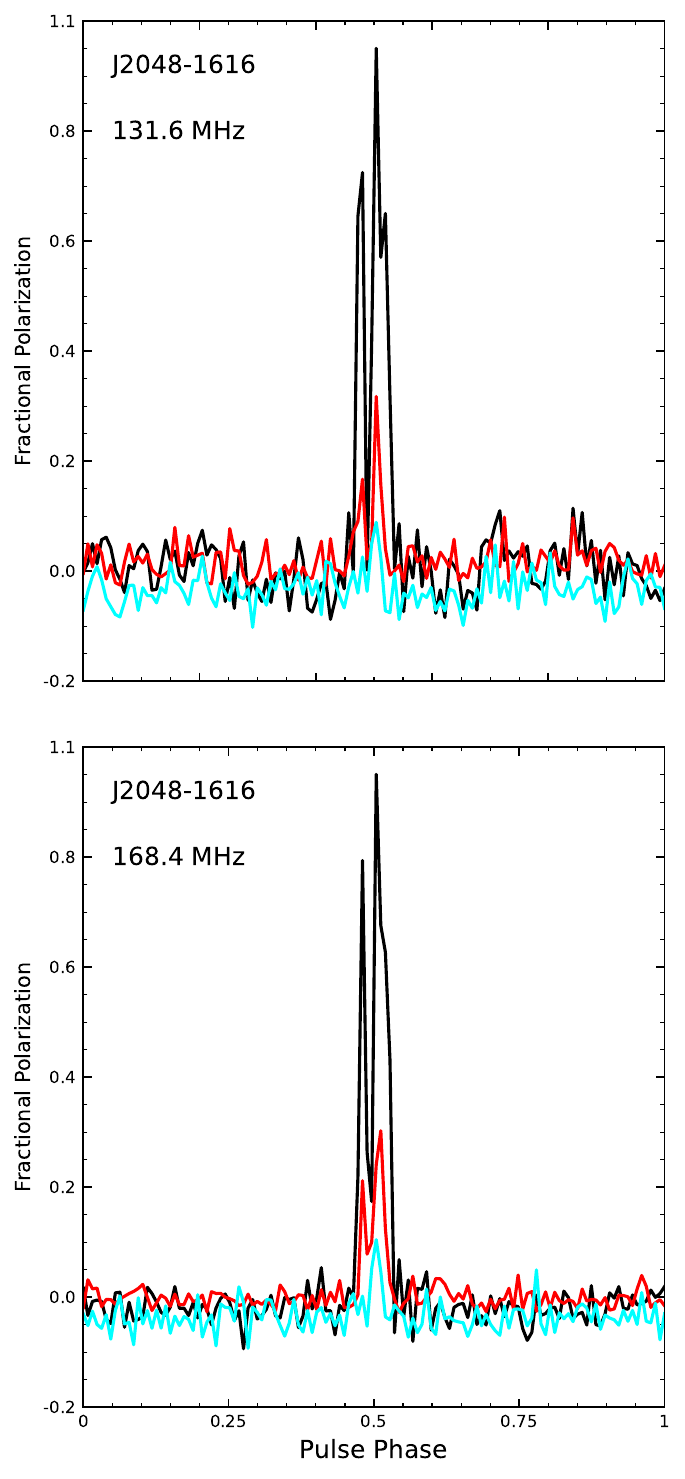}
\caption{Fig \ref{fig:multipol} Continued}
\end{figure*}
\begin{figure*}[htbp!]
\centering
\includegraphics[width=\textwidth,height=22cm]{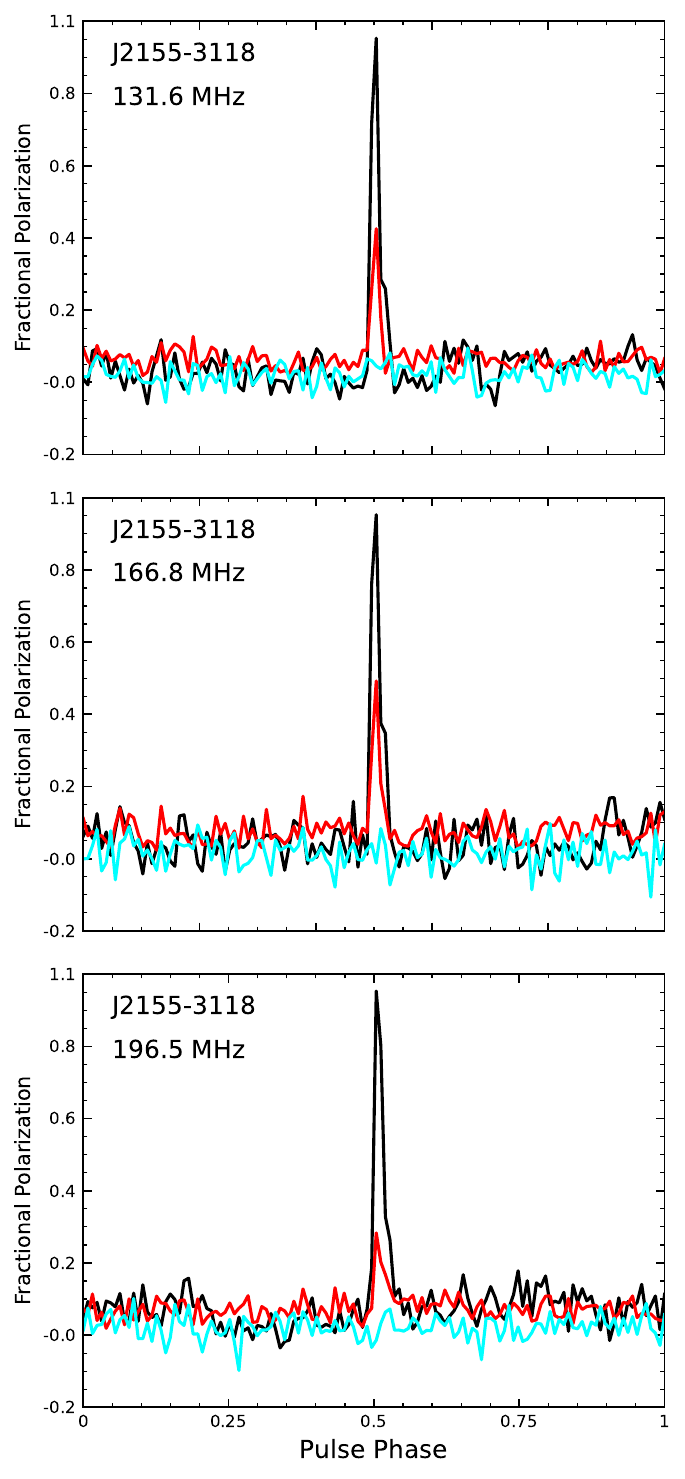}
\caption{Fig \ref{fig:multipol} Continued}
\end{figure*}
\begin{figure*}[htbp!]
\centering
\includegraphics[width=\textwidth,height=22cm]{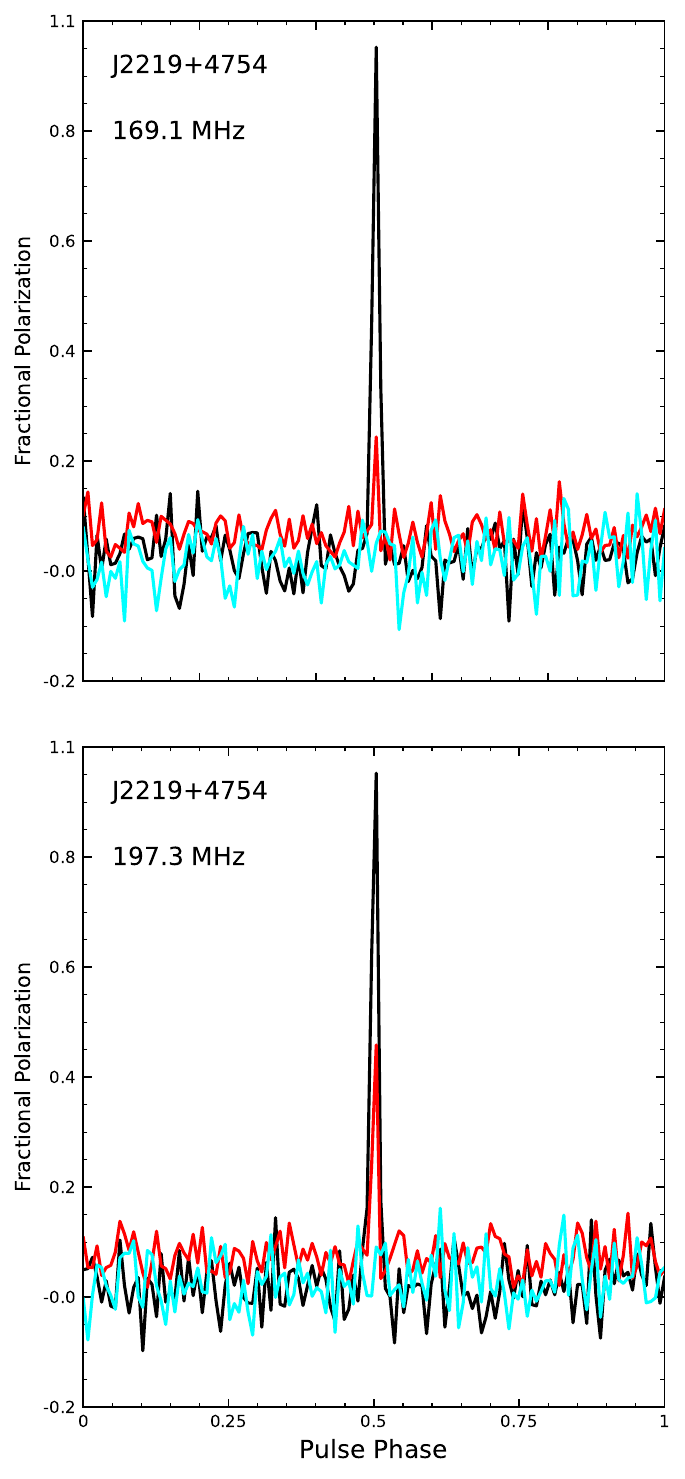}
\caption{Fig \ref{fig:multipol} Continued}
\end{figure*}
\end{document}

%% file: tables/dmtable.tex
\startlongtable
\centerwidetable
\begin{deluxetable*}{cccccccc}
\tablecolumns{8}
\tabletypesize{\small}
\tablewidth{0pt}
\tablecaption{Period and Dispersion measure of Pulsars detected by the EDA2 \label{tab:dmperiod}}
\tablehead{
    \colhead{Name} & \colhead{Period} & \colhead{DM$_{\mathrm{ATNF}}$} & \colhead{DM$_{\mathrm{Epoch}}^{\mathrm{ATNF}}$} & \colhead{DM$_{\mathrm{EDA2}}$} & \colhead{DM$_{\mathrm{Epoch}}^{\mathrm{EDA2}}$} & \colhead{S/N$_{pdmp}$ } & \colhead{DM$_{corr}$}\\
    \colhead{} & \colhead{(s) } & \colhead{(pc cm$^{-3}$)} & \colhead{(MJD) } & \colhead{(pc cm$^{-3}$)} & \colhead{(MJD) } & \colhead{ } & \colhead{(pc cm$^{-3}$)}
}
\startdata
        J0034-0534* & 0.00188 & 13.7652(0) & 55000 & 13.7652(0) & 60507 & --- & --- \\ \hline
        J0034-0721 & 0.94295 & 10.922(6) & 56843 & 10.852(87) & 60455 & 54.5 & -0.07 \\ \hline
        J0038-2501 & 0.25693 & 5.71(0) & 57474 & 5.59(8) & 60482 & 10.8 & -0.12 \\ \hline
        J0051+0423 & 0.35473 & 13.9(0) & 49800 & 13.9(1) & 60469 & 10.8 & -0.03 \\ \hline
        J0151-0635 & 1.46466 & 25.7(0) & 49347 & 25.9(1) & 60494 & 15.4 & 0.24 \\ \hline
        J0152-1637 & 0.83274 & 11.9258(0) & 56955 & 11.9758(805) & 60518 & 88.4 & 0.05 \\ \hline
        J0206-4028 & 0.63055 & 12.9(120) & 57600 & 12.98(9) & 60522 & 32.0 & 0.08 \\ \hline
        J0304+1932 & 1.38758 & 15.6568(3) & 56703 & 15.3768(13630) & 60503 & 8.2 & -0.28 \\
\enddata
\tablenotetext{}{S/N$_{pdmp}$ is the signal-to-noise ratio found by \textit{pdmp} for the optimal DM, DM$_{corr}$ is the correction in dispersion measure obtained using \textit{pdmp}}
\tablenotetext{}{Pulsar with * indicate MSPs.}
\end{deluxetable*}

%% file: tables/fluxtable.tex
\startlongtable
\centerwidetable
\begin{deluxetable*}{cccccccc}
\tablecolumns{8}
\tabletypesize{\small}
\tablewidth{0pt}
\tablecaption{Mean flux density (S$_{\nu}$) of pulsars detected in this census, along with pulse width measurements.\label{tab:fluxdensity}}
\tablehead{
    \colhead{Name} & \colhead{$\nu$} & \colhead{BW} & \colhead{W$_{50}$} & \colhead{W$_{10}$} & \colhead{T} & \colhead{S/N} & \colhead{S$_{\nu}$}\\
    \colhead{} & \colhead{(MHZ) } & \colhead{(MHz)} & \colhead{(phase)} & \colhead{(phase)} & \colhead{(s)} & \colhead{ } & \colhead{(mJy)}
}
\startdata
        J0034-0534 & 71.1 & 33.59 & 0.465 & 0.86 & 3578 & 579 & 7055(3527) \\ \hline
        J0034-0534 & 131.6 & 45.31 & 0.489 & 0.69 & 3119 & 248 & 892(446) \\ \hline
        J0034-0534 & 169.1 & 28.12 & 0.438 & 0.63 & 2790 & 182 & 751(375) \\ \hline
        J0034-0534 & 197.3 & 28.12 & 0.438 & 0.63 & 2639 & 89 & 497(248) \\ \hline
        J0034-0721 & 71.1 & 33.59 & 0.272 & 0.42 & 1743 & 147 & 1674(837) \\ \hline
        J0034-0721 & 131.2 & 44.53 & 0.199 & 0.32 & 1399 & 168 & 464(232) \\ \hline
        J0034-0721 & 168.4 & 28.12 & 0.162 & 0.28 & 1529 & 188 & 504(252) \\ \hline
        J0034-0721 & 196.5 & 28.12 & 0.162 & 0.27 & 1440 & 114 & 399(199) \\
\enddata
\tablenotetext{}{$\nu$,BW,T, and S/N are the center frequency, bandwidth, integration time and signal-to-noise ratio of the average pulse profile}
\tablenotetext{}{W$_{50}$ and W$_{10}$ are the fraction pulse width at FWHM and 10$\%$}
\tablenotetext{}{S$_{\nu}$ is the flux density calculted with W$_{50}$ and equation \ref{eq:2}}
\end{deluxetable*}

%% file: tables/spectratable.tex
\movetabledown=2cm
\begin{longrotatetable}
\begin{deluxetable*}{cccccccccc}
\tablecolumns{10}
\tabletypesize{\scriptsize}
\tablewidth{0pt}
\tablecaption{Best fitting model parameter for spectral fitting of pulsars detected with EDA2. The table is published in its entirety in a machine readable online format. \label{tab:spectable}}
\tablehead{
    \colhead{Name} & \colhead{$\nu_{range}^{a}$} & \colhead{$\nu_0^a$} & \colhead{a$_1^c$} & \colhead{a$_2^d$} & \colhead{a$_33^e$} & \colhead{a$_4^f$} & \colhead{P$^{b}_{best}$} & \colhead{Type$^b$} & \colhead{References}\\
    \colhead{} & \colhead{(MHz) } & \colhead{(MHz)} & \colhead{} & \colhead{(MHz)} & \colhead{} & \colhead{} & \colhead{} & \colhead{} &\colhead{}
}
\startdata
        J0034-0534 & 49-1660 & 287 & -4.69(28) & 84(0) & 16.59(1484) & 1.29(9) & 0.99 & DT & 5, 10, 11, 21, 39, 45, 46, 47, 58, 60, 65, 71, 72, 73, 77, 82 \\ \hline
        J0034-0721 & 20-1435 & 169 & -3.52(17) & 72(2) & 14.34(931) & 0.7(4) & 0.89 & DT & 4, 5, 10, 21, 25, 29, 31, 35, 45, 49, 51, 55, 60, 65, 69, 75, 82, 83, 86 \\ \hline
        J0038-2501 & 131-350 & 214 & -2.21(77) & --- & --- & --- & 0.92 & PL & 2, 5, 60 \\ \hline
        J0051+0423* & 25-1360 & 184 & -1.13(12) & --- & --- & --- & 0.85 & PL & 5, 10, 35, 45, 60, 67, 68, 83 \\ \hline
        J0151-0635 & 25-1408 & 187 & 0.52(32) & 733(157) & -2.75(111) & --- & 0.69 & BPL & 5, 21, 31, 35, 51, 55, 56, 60, 69, 83 \\ \hline
        J0152-1637 & 35-4750 & 408 & -2.92(24) & 20(8) & --- & 0.27(2) & 0.71 & PLL & 5, 10, 21, 29, 31, 35, 45, 51, 56, 60, 66, 69, 75, 86 \\ \hline
        J0206-4028 & 107-1440 & 392 & -1.25(45) & 1829(597) & --- & --- & 0.47 & PLH & 5, 31, 35, 56, 65, 66, 76, 86 \\ \hline
        J0304+1932 & 24-4850 & 346 & -7.89(9) & 179(11) & --- & 0.11(0) & 0.65 & PLL & 5, 8, 9, 21, 25, 28, 29, 35, 38, 44, 45, 51, 55, 56, 69, 75, 80 \\
\enddata
\end{deluxetable*}
\end{longrotatetable}

%% file: tables/rmtable.tex
\startlongtable
\centerwidetable
\begin{deluxetable*}{ccccccccc}
\tablecolumns{9}
\tabletypesize{\small}
\tablewidth{0pt}
\tablecaption{EDA2 RM measurements for 40 pulsars. For comparison we give the RM reported in ATNF pulsar catalog \citep{atnf} and with MeerKAT \citep{Posselt}, along with $\Delta$ RM between our measurements and these two catalog. Pulsar whose RM could be affected instrumental polarization near RM=0, are marked by asterisk. The table is published in its entirety in a machine readable online format. \label{tab:rmtable}}
\tablehead{
    \colhead{Name} & \colhead{RM$_{obs}$} & \colhead{RM$_{ion}$} & \colhead{RM$_{ISM}$} & \colhead{RM$^{ATNF}_{ISM}$} & \colhead{$\Delta$ RM$^{ATNF}_{Obs}$} & \colhead{RM$^{MeerKat}_{ISM}$} & \colhead{$\delta$ RM$^{MeerKat}_{Obs}$} & \colhead{B$_{\parallel}$}\\
    \colhead{} & \colhead{(rad $m^{-2}$) } & \colhead{rad $m^{-2}$} & \colhead{(rad $m^{-2}$)} & \colhead{(rad $m^{-2}$)} & \colhead{(rad $m^{-2}$)} & \colhead{(rad $m^{-2}$)} & \colhead{(rad $m^{-2}$)} & \colhead{($\mu$G)}
}
\startdata
        J0034-0721 & 8.31(5) & -1.65(10) & 9.96(11) & 9.89(7) & 0.07(13) & 9.9(29) & 0.06(290) & 1.12 \\ \hline
        J0437-4715* & -0.59(14) & -5.45(10) & 4.86(17) & 0.15(1) & 4.7(17) & --- & --- & 0.07 \\ \hline
        J0534+2200 & -44.53(8) & -1.16(8) & -43.36(11) & -45.44(8) & 2.07(14) & --- & --- & -0.99 \\ \hline
        J0630-2834 & 43.34(1) & -3.24(14) & 46.59(14) & 46.53(12) & 0.06(19) & 46.5(290) & 0.09(290) & 1.67 \\ \hline
        J0742-2822 & 147.3(7) & -3.58(10) & 150.89(13) & 149.94(5) & 0.94(14) & 149.9(290) & 0.99(290) & 2.50 \\ \hline
        J0820-1350* & 0.46(19) & -3.65(2) & 4.11(20) & -1.2(40) & 5.31(44) & -1.2(290) & 5.31(290) & -0.04 \\ \hline
        J0826+2637* & -0.04(21) & -2.28(5) & 2.24(22) & 5.38(6) & -3.13(22) & --- & --- & 0.34 \\ \hline
        J0835-4510 & 42.43(13) & -3.66(4) & 46.1(14) & 31.38(1) & 14.72(14) & 39.4(290) & 6.7(290) & 0.57 \\
\enddata
\tablenotetext{a}{$\Delta$ RM$^{ATNF}_{Obs}$ and $\delta$ RM$^{MeerKat}_{Obs}$ are difference in ionospheric corrected RM between ATNF and EDA2, and MeerKAT and EDA2 respectively.}
\tablenotetext{b}{B$_{\parallel}$ is calculated using equation \ref{eq:1} with RM$_{ISM}$ and DM$_{\mathrm{EDA2}}$ from Table \ref{tab:dmperiod}}
\end{deluxetable*}

%% file: census.bib
@ARTICLE{2022ApJ...935..167A,
       author = {{Astropy Collaboration} and {Price-Whelan}, Adrian M. and {Lim}, Pey Lian and {Earl}, Nicholas and {Starkman}, Nathaniel and {Bradley}, Larry and {Shupe}, David L. and {Patil}, Aarya A. and {Corrales}, Lia and {Brasseur}, C.~E. and {N{\"o}the}, Maximilian and {Donath}, Axel and {Tollerud}, Erik and {Morris}, Brett M. and {Ginsburg}, Adam and {Vaher}, Eero and {Weaver}, Benjamin A. and {Tocknell}, James and {Jamieson}, William and {van Kerkwijk}, Marten H. and {Robitaille}, Thomas P. and {Merry}, Bruce and {Bachetti}, Matteo and {G{\"u}nther}, H. Moritz and {Aldcroft}, Thomas L. and {Alvarado-Montes}, Jaime A. and {Archibald}, Anne M. and {B{\'o}di}, Attila and {Bapat}, Shreyas and {Barentsen}, Geert and {Baz{\'a}n}, Juanjo and {Biswas}, Manish and {Boquien}, M{\'e}d{\'e}ric and {Burke}, D.~J. and {Cara}, Daria and {Cara}, Mihai and {Conroy}, Kyle E. and {Conseil}, Simon and {Craig}, Matthew W. and {Cross}, Robert M. and {Cruz}, Kelle L. and {D'Eugenio}, Francesco and {Dencheva}, Nadia and {Devillepoix}, Hadrien A.~R. and {Dietrich}, J{\"o}rg P. and {Eigenbrot}, Arthur Davis and {Erben}, Thomas and {Ferreira}, Leonardo and {Foreman-Mackey}, Daniel and {Fox}, Ryan and {Freij}, Nabil and {Garg}, Suyog and {Geda}, Robel and {Glattly}, Lauren and {Gondhalekar}, Yash and {Gordon}, Karl D. and {Grant}, David and {Greenfield}, Perry and {Groener}, Austen M. and {Guest}, Steve and {Gurovich}, Sebastian and {Handberg}, Rasmus and {Hart}, Akeem and {Hatfield-Dodds}, Zac and {Homeier}, Derek and {Hosseinzadeh}, Griffin and {Jenness}, Tim and {Jones}, Craig K. and {Joseph}, Prajwel and {Kalmbach}, J. Bryce and {Karamehmetoglu}, Emir and {Ka{\l}uszy{\'n}ski}, Miko{\l}aj and {Kelley}, Michael S.~P. and {Kern}, Nicholas and {Kerzendorf}, Wolfgang E. and {Koch}, Eric W. and {Kulumani}, Shankar and {Lee}, Antony and {Ly}, Chun and {Ma}, Zhiyuan and {MacBride}, Conor and {Maljaars}, Jakob M. and {Muna}, Demitri and {Murphy}, N.~A. and {Norman}, Henrik and {O'Steen}, Richard and {Oman}, Kyle A. and {Pacifici}, Camilla and {Pascual}, Sergio and {Pascual-Granado}, J. and {Patil}, Rohit R. and {Perren}, Gabriel I. and {Pickering}, Timothy E. and {Rastogi}, Tanuj and {Roulston}, Benjamin R. and {Ryan}, Daniel F. and {Rykoff}, Eli S. and {Sabater}, Jose and {Sakurikar}, Parikshit and {Salgado}, Jes{\'u}s and {Sanghi}, Aniket and {Saunders}, Nicholas and {Savchenko}, Volodymyr and {Schwardt}, Ludwig and {Seifert-Eckert}, Michael and {Shih}, Albert Y. and {Jain}, Anany Shrey and {Shukla}, Gyanendra and {Sick}, Jonathan and {Simpson}, Chris and {Singanamalla}, Sudheesh and {Singer}, Leo P. and {Singhal}, Jaladh and {Sinha}, Manodeep and {Sip{\H{o}}cz}, Brigitta M. and {Spitler}, Lee R. and {Stansby}, David and {Streicher}, Ole and {{\v{S}}umak}, Jani and {Swinbank}, John D. and {Taranu}, Dan S. and {Tewary}, Nikita and {Tremblay}, Grant R. and {de Val-Borro}, Miguel and {Van Kooten}, Samuel J. and {Vasovi{\'c}}, Zlatan and {Verma}, Shresth and {de Miranda Cardoso}, Jos{\'e} Vin{\'\i}cius and {Williams}, Peter K.~G. and {Wilson}, Tom J. and {Winkel}, Benjamin and {Wood-Vasey}, W.~M. and {Xue}, Rui and {Yoachim}, Peter and {Zhang}, Chen and {Zonca}, Andrea and {Astropy Project Contributors}},
        title = "{The Astropy Project: Sustaining and Growing a Community-oriented Open-source Project and the Latest Major Release (v5.0) of the Core Package}",
      journal = {\apj},
         year = 2022,
        month = aug,
       volume = {935},
       number = {2},
          eid = {167},
        pages = {167},
          doi = {10.3847/1538-4357/ac7c74},
archivePrefix = {arXiv},
       eprint = {2206.14220},
 primaryClass = {astro-ph.IM},
       adsurl = {https://ui.adsabs.harvard.edu/abs/2022ApJ...935..167A}
}

@ARTICLE{2018AJ....156..123A,
       author = {{Astropy Collaboration} and {Price-Whelan}, A.~M. and {Sip{\H{o}}cz}, B.~M. and {G{\"u}nther}, H.~M. and {Lim}, P.~L. and {Crawford}, S.~M. and {Conseil}, S. and {Shupe}, D.~L. and {Craig}, M.~W. and {Dencheva}, N. and {Ginsburg}, A. and {VanderPlas}, J.~T. and {Bradley}, L.~D. and {P{\'e}rez-Su{\'a}rez}, D. and {de Val-Borro}, M. and {Aldcroft}, T.~L. and {Cruz}, K.~L. and {Robitaille}, T.~P. and {Tollerud}, E.~J. and {Ardelean}, C. and {Babej}, T. and {Bach}, Y.~P. and {Bachetti}, M. and {Bakanov}, A.~V. and {Bamford}, S.~P. and {Barentsen}, G. and {Barmby}, P. and {Baumbach}, A. and {Berry}, K.~L. and {Biscani}, F. and {Boquien}, M. and {Bostroem}, K.~A. and {Bouma}, L.~G. and {Brammer}, G.~B. and {Bray}, E.~M. and {Breytenbach}, H. and {Buddelmeijer}, H. and {Burke}, D.~J. and {Calderone}, G. and {Cano Rodr{\'\i}guez}, J.~L. and {Cara}, M. and {Cardoso}, J.~V.~M. and {Cheedella}, S. and {Copin}, Y. and {Corrales}, L. and {Crichton}, D. and {D'Avella}, D. and {Deil}, C. and {Depagne}, {\'E}. and {Dietrich}, J.~P. and {Donath}, A. and {Droettboom}, M. and {Earl}, N. and {Erben}, T. and {Fabbro}, S. and {Ferreira}, L.~A. and {Finethy}, T. and {Fox}, R.~T. and {Garrison}, L.~H. and {Gibbons}, S.~L.~J. and {Goldstein}, D.~A. and {Gommers}, R. and {Greco}, J.~P. and {Greenfield}, P. and {Groener}, A.~M. and {Grollier}, F. and {Hagen}, A. and {Hirst}, P. and {Homeier}, D. and {Horton}, A.~J. and {Hosseinzadeh}, G. and {Hu}, L. and {Hunkeler}, J.~S. and {Ivezi{\'c}}, {\v{Z}}. and {Jain}, A. and {Jenness}, T. and {Kanarek}, G. and {Kendrew}, S. and {Kern}, N.~S. and {Kerzendorf}, W.~E. and {Khvalko}, A. and {King}, J. and {Kirkby}, D. and {Kulkarni}, A.~M. and {Kumar}, A. and {Lee}, A. and {Lenz}, D. and {Littlefair}, S.~P. and {Ma}, Z. and {Macleod}, D.~M. and {Mastropietro}, M. and {McCully}, C. and {Montagnac}, S. and {Morris}, B.~M. and {Mueller}, M. and {Mumford}, S.~J. and {Muna}, D. and {Murphy}, N.~A. and {Nelson}, S. and {Nguyen}, G.~H. and {Ninan}, J.~P. and {N{\"o}the}, M. and {Ogaz}, S. and {Oh}, S. and {Parejko}, J.~K. and {Parley}, N. and {Pascual}, S. and {Patil}, R. and {Patil}, A.~A. and {Plunkett}, A.~L. and {Prochaska}, J.~X. and {Rastogi}, T. and {Reddy Janga}, V. and {Sabater}, J. and {Sakurikar}, P. and {Seifert}, M. and {Sherbert}, L.~E. and {Sherwood-Taylor}, H. and {Shih}, A.~Y. and {Sick}, J. and {Silbiger}, M.~T. and {Singanamalla}, S. and {Singer}, L.~P. and {Sladen}, P.~H. and {Sooley}, K.~A. and {Sornarajah}, S. and {Streicher}, O. and {Teuben}, P. and {Thomas}, S.~W. and {Tremblay}, G.~R. and {Turner}, J.~E.~H. and {Terr{\'o}n}, V. and {van Kerkwijk}, M.~H. and {de la Vega}, A. and {Watkins}, L.~L. and {Weaver}, B.~A. and {Whitmore}, J.~B. and {Woillez}, J. and {Zabalza}, V. and {Astropy Contributors}},
        title = "{The Astropy Project: Building an Open-science Project and Status of the v2.0 Core Package}",
      journal = {\aj},
         year = 2018,
        month = sep,
       volume = {156},
       number = {3},
          eid = {123},
        pages = {123},
          doi = {10.3847/1538-3881/aabc4f},
archivePrefix = {arXiv},
       eprint = {1801.02634},
 primaryClass = {astro-ph.IM},
       adsurl = {https://ui.adsabs.harvard.edu/abs/2018AJ....156..123A}
}

@ARTICLE{2013A&A...558A..33A,
       author = {{Astropy Collaboration} and {Robitaille}, Thomas P. and
         {Tollerud}, Erik J. and {Greenfield}, Perry and {Droettboom}, Michael and
         {Bray}, Erik and {Aldcroft}, Tom and {Davis}, Matt and
         {Ginsburg}, Adam and {Price-Whelan}, Adrian M. and
         {Kerzendorf}, Wolfgang E. and {Conley}, Alexander and {Crighton}, Neil and
         {Barbary}, Kyle and {Muna}, Demitri and {Ferguson}, Henry and
         {Grollier}, Fr{\'e}d{\'e}ric and {Parikh}, Madhura M. and
         {Nair}, Prasanth H. and {Unther}, Hans M. and {Deil}, Christoph and
         {Woillez}, Julien and {Conseil}, Simon and {Kramer}, Roban and
         {Turner}, James E.~H. and {Singer}, Leo and {Fox}, Ryan and
         {Weaver}, Benjamin A. and {Zabalza}, Victor and {Edwards}, Zachary I. and
         {Azalee Bostroem}, K. and {Burke}, D.~J. and {Casey}, Andrew R. and
         {Crawford}, Steven M. and {Dencheva}, Nadia and {Ely}, Justin and
         {Jenness}, Tim and {Labrie}, Kathleen and {Lim}, Pey Lian and
         {Pierfederici}, Francesco and {Pontzen}, Andrew and {Ptak}, Andy and
         {Refsdal}, Brian and {Servillat}, Mathieu and {Streicher}, Ole},
        title = "{Astropy: A community Python package for astronomy}",
      journal = {\aap},
         year = "2013",
        month = "Oct",
       volume = {558},
          eid = {A33},
        pages = {A33},
          doi = {10.1051/0004-6361/201322068},
archivePrefix = {arXiv},
       eprint = {1307.6212},
 primaryClass = {astro-ph.IM},
       adsurl = {https://ui.adsabs.harvard.edu/abs/2013A&A...558A..33A}
}

@ARTICLE{atnf,
       author = {{Manchester}, R.~N. and {Hobbs}, G.~B. and {Teoh}, A. and {Hobbs}, M.},
        title = "{The Australia Telescope National Facility Pulsar Catalogue}",
      journal = {\aj},
         year = 2005,
        month = apr,
       volume = {129},
       number = {4},
        pages = {1993-2006},
          doi = {10.1086/428488},
archivePrefix = {arXiv},
       eprint = {astro-ph/0412641},
 primaryClass = {astro-ph},
       adsurl = {https://ui.adsabs.harvard.edu/abs/2005AJ....129.1993M}
}

@ARTICLE{bilous2016,
       author = {{Bilous}, A.~V. and {Kondratiev}, V.~I. and {Kramer}, M. and {Keane}, E.~F. and {Hessels}, J.~W.~T. and {Stappers}, B.~W. and {Malofeev}, V.~M. and {Sobey}, C. and {Breton}, R.~P. and {Cooper}, S. and {Falcke}, H. and {Karastergiou}, A. and {Michilli}, D. and {Os{\l}owski}, S. and {Sanidas}, S. and {ter Veen}, S. and {van Leeuwen}, J. and {Verbiest}, J.~P.~W. and {Weltevrede}, P. and {Zarka}, P. and {Grie{\ss}meier}, J. -M. and {Serylak}, M. and {Bell}, M.~E. and {Broderick}, J.~W. and {Eisl{\"o}ffel}, J. and {Markoff}, S. and {Rowlinson}, A.},
        title = "{A LOFAR census of non-recycled pulsars: average profiles, dispersion measures, flux densities, and spectra}",
      journal = {\aap},
         year = 2016,
        month = jun,
       volume = {591},
          eid = {A134},
        pages = {A134},
          doi = {10.1051/0004-6361/201527702},
archivePrefix = {arXiv},
       eprint = {1511.01767},
 primaryClass = {astro-ph.SR},
       adsurl = {https://ui.adsabs.harvard.edu/abs/2016A&A...591A.134B}
}

@ARTICLE{malofeev,
       author = {{Malofeev}, V.~M. and {Malov}, O.~I. and {Shchegoleva}, N.~V.},
        title = "{Flux Densities of 235 Pulsars at 102.5 MHz}",
      journal = {Astronomy Reports},
         year = 2000,
        month = jul,
       volume = {44},
       number = {7},
        pages = {436-445},
          doi = {10.1134/1.163868},
       adsurl = {https://ui.adsabs.harvard.edu/abs/2000ARep...44..436M}
}

@ARTICLE{sanidaslotass,
       author = {{Sanidas}, S. and {Cooper}, S. and {Bassa}, C.~G. and {Hessels}, J.~W.~T. and {Kondratiev}, V.~I. and {Michilli}, D. and {Stappers}, B.~W. and {Tan}, C.~M. and {van Leeuwen}, J. and {Cerrigone}, L. and {Fallows}, R.~A. and {Iacobelli}, M. and {Orr{\'u}}, E. and {Pizzo}, R.~F. and {Shulevski}, A. and {Toribio}, M.~C. and {ter Veen}, S. and {Zucca}, P. and {Bondonneau}, L. and {Grie{\ss}meier}, J. -M. and {Karastergiou}, A. and {Kramer}, M. and {Sobey}, C.},
        title = "{The LOFAR Tied-Array All-Sky Survey (LOTAAS): Survey overview and initial pulsar discoveries}",
      journal = {\aap},
         year = 2019,
        month = jun,
       volume = {626},
          eid = {A104},
        pages = {A104},
          doi = {10.1051/0004-6361/201935609},
archivePrefix = {arXiv},
       eprint = {1905.04977},
 primaryClass = {astro-ph.HE},
       adsurl = {https://ui.adsabs.harvard.edu/abs/2019A&A...626A.104S}
}

@ARTICLE{jankowski441psr,
       author = {{Jankowski}, F. and {van Straten}, W. and {Keane}, E.~F. and {Bailes}, M. and {Barr}, E.~D. and {Johnston}, S. and {Kerr}, M.},
        title = "{Spectral properties of 441 radio pulsars}",
      journal = {\mnras},
         year = 2018,
        month = feb,
       volume = {473},
       number = {4},
        pages = {4436-4458},
          doi = {10.1093/mnras/stx2476},
archivePrefix = {arXiv},
       eprint = {1709.08864},
 primaryClass = {astro-ph.HE},
       adsurl = {https://ui.adsabs.harvard.edu/abs/2018MNRAS.473.4436J}
}

@ARTICLE{hewish,
       author = {{Hewish}, A. and {Bell}, S.~J. and {Pilkington}, J.~D.~H. and {Scott}, P.~F. and {Collins}, R.~A.},
        title = "{Observation of a Rapidly Pulsating Radio Source}",
      journal = {\nat},
         year = 1968,
        month = feb,
       volume = {217},
       number = {5130},
        pages = {709-713},
          doi = {10.1038/217709a0},
       adsurl = {https://ui.adsabs.harvard.edu/abs/1968Natur.217..709H}
}

@ARTICLE{bates,
       author = {{Bates}, S.~D. and {Lorimer}, D.~R. and {Verbiest}, J.~P.~W.},
        title = "{The pulsar spectral index distribution}",
      journal = {\mnras},
         year = 2013,
        month = may,
       volume = {431},
       number = {2},
        pages = {1352-1358},
          doi = {10.1093/mnras/stt257},
archivePrefix = {arXiv},
       eprint = {1302.2053},
 primaryClass = {astro-ph.SR},
       adsurl = {https://ui.adsabs.harvard.edu/abs/2013MNRAS.431.1352B}
}

@ARTICLE{Haslam,
       author = {{Haslam}, C.~G.~T. and {Salter}, C.~J. and {Stoffel}, H. and {Wilson}, W.~E.},
        title = "{A 408-MHZ All-Sky Continuum Survey. II. The Atlas of Contour Maps}",
      journal = {\aaps},
         year = 1982,
        month = jan,
       volume = {47},
        pages = {1},
       adsurl = {https://ui.adsabs.harvard.edu/abs/1982A&AS...47....1H}
}

@INPROCEEDINGS{zarka,
  author={Zarka, P. and Tagger, M. and Denis, L. and Girard, J. N. and Konovalenko, A. and Atemkeng, M. and Arnaud, M. and Azarian, S. and Barsuglia, M. and Bonafede, A. and Boone, F. and Bosma, A. and Boyer, R. and Branchesi, M. and Briand, C. and Cecconi, B. and Célestin, S. and Charrier, D. and Chassande-Mottin, E. and Coffre, A. and Cognard, I. and Combes, F. and Corbel, S. and Courte, C. and Dabbech, A. and Daiboo, S. and Dallier, R. and Dumez-Viou, C. and El Korso, M. N. and Falgarone, E. and Falkovych, I. and Ferrari, A. and Ferrari, C. and Ferrière, K. and Fevotte, C. and Fialkov, A. and Fullekrug, M. and Gérard, E. and Grießmeier, J.-M. and Guiderdoni, B. and Guillemot, L. and Hessels, J. and Koopmans, L. and Kondratiev, V. and Lamy, L. and Lanz, T. and Larzabal, P. and Lehnert, M. and Levrier, F. and Loh, A. and Macario, G. and Maintoux, J.-J. and Martin, L. and Mary, D. and Masson, S. and Miville-Deschenes, M.-A. and Oberoi, D. and Panchenko, M. and Pandey-Pommier, M. and Petiteau, A. and Pinçon, J.-L. and Revenu, B. and Rible, F. and Richard, C. and Rucker, H. O. and Salomé, P. and Semelin, B. and Serylak, M. and Smirnov, O. and Stappers, B. and Taffoureau, C. and Tasse, C. and Theureau, G. and Tokarsky, P. and Torchinsky, S. and Ulyanov, O. and van Driel, W. and Vasylieva, I. and Vaubaillon, J. and Vazza, F. and Vergani, S. and Was, M. and Weber, R. and Zakharenko, V.},
  booktitle={2015 International Conference on Antenna Theory and Techniques (ICATT)}, 
  title={NenUFAR: Instrument description and science case}, 
  year={2015},
  volume={},
  number={},
  pages={1-6},
  doi={10.1109/ICATT.2015.7136773}}

@ARTICLE{Tingay,
       author = {{Tingay}, S.~J. and {Goeke}, R. and {Bowman}, J.~D. and {Emrich}, D. and {Ord}, S.~M. and {Mitchell}, D.~A. and {Morales}, M.~F. and {Booler}, T. and {Crosse}, B. and {Wayth}, R.~B. and {Lonsdale}, C.~J. and {Tremblay}, S. and {Pallot}, D. and {Colegate}, T. and {Wicenec}, A. and {Kudryavtseva}, N. and {Arcus}, W. and {Barnes}, D. and {Bernardi}, G. and {Briggs}, F. and {Burns}, S. and {Bunton}, J.~D. and {Cappallo}, R.~J. and {Corey}, B.~E. and {Deshpande}, A. and {Desouza}, L. and {Gaensler}, B.~M. and {Greenhill}, L.~J. and {Hall}, P.~J. and {Hazelton}, B.~J. and {Herne}, D. and {Hewitt}, J.~N. and {Johnston-Hollitt}, M. and {Kaplan}, D.~L. and {Kasper}, J.~C. and {Kincaid}, B.~B. and {Koenig}, R. and {Kratzenberg}, E. and {Lynch}, M.~J. and {Mckinley}, B. and {Mcwhirter}, S.~R. and {Morgan}, E. and {Oberoi}, D. and {Pathikulangara}, J. and {Prabu}, T. and {Remillard}, R.~A. and {Rogers}, A.~E.~E. and {Roshi}, A. and {Salah}, J.~E. and {Sault}, R.~J. and {Udaya-Shankar}, N. and {Schlagenhaufer}, F. and {Srivani}, K.~S. and {Stevens}, J. and {Subrahmanyan}, R. and {Waterson}, M. and {Webster}, R.~L. and {Whitney}, A.~R. and {Williams}, A. and {Williams}, C.~L. and {Wyithe}, J.~S.~B.},
        title = "{The Murchison Widefield Array: The Square Kilometre Array Precursor at Low Radio Frequencies}",
      journal = {\pasa},
         year = 2013,
        month = jan,
       volume = {30},
          eid = {e007},
        pages = {e007},
          doi = {10.1017/pasa.2012.007},
archivePrefix = {arXiv},
       eprint = {1206.6945},
 primaryClass = {astro-ph.IM},
       adsurl = {https://ui.adsabs.harvard.edu/abs/2013PASA...30....7T}
}

@ARTICLE{Vanhaarlem,
       author = {{van Haarlem}, M.~P. and {Wise}, M.~W. and {Gunst}, A.~W. and {Heald}, G. and {McKean}, J.~P. and {Hessels}, J.~W.~T. and {de Bruyn}, A.~G. and {Nijboer}, R. and {Swinbank}, J. and {Fallows}, R. and {Brentjens}, M. and {Nelles}, A. and {Beck}, R. and {Falcke}, H. and {Fender}, R. and {H{\"o}randel}, J. and {Koopmans}, L.~V.~E. and {Mann}, G. and {Miley}, G. and {R{\"o}ttgering}, H. and {Stappers}, B.~W. and {Wijers}, R.~A.~M.~J. and {Zaroubi}, S. and {van den Akker}, M. and {Alexov}, A. and {Anderson}, J. and {Anderson}, K. and {van Ardenne}, A. and {Arts}, M. and {Asgekar}, A. and {Avruch}, I.~M. and {Batejat}, F. and {B{\"a}hren}, L. and {Bell}, M.~E. and {Bell}, M.~R. and {van Bemmel}, I. and {Bennema}, P. and {Bentum}, M.~J. and {Bernardi}, G. and {Best}, P. and {B{\^\i}rzan}, L. and {Bonafede}, A. and {Boonstra}, A. -J. and {Braun}, R. and {Bregman}, J. and {Breitling}, F. and {van de Brink}, R.~H. and {Broderick}, J. and {Broekema}, P.~C. and {Brouw}, W.~N. and {Br{\"u}ggen}, M. and {Butcher}, H.~R. and {van Cappellen}, W. and {Ciardi}, B. and {Coenen}, T. and {Conway}, J. and {Coolen}, A. and {Corstanje}, A. and {Damstra}, S. and {Davies}, O. and {Deller}, A.~T. and {Dettmar}, R. -J. and {van Diepen}, G. and {Dijkstra}, K. and {Donker}, P. and {Doorduin}, A. and {Dromer}, J. and {Drost}, M. and {van Duin}, A. and {Eisl{\"o}ffel}, J. and {van Enst}, J. and {Ferrari}, C. and {Frieswijk}, W. and {Gankema}, H. and {Garrett}, M.~A. and {de Gasperin}, F. and {Gerbers}, M. and {de Geus}, E. and {Grie{\ss}meier}, J. -M. and {Grit}, T. and {Gruppen}, P. and {Hamaker}, J.~P. and {Hassall}, T. and {Hoeft}, M. and {Holties}, H.~A. and {Horneffer}, A. and {van der Horst}, A. and {van Houwelingen}, A. and {Huijgen}, A. and {Iacobelli}, M. and {Intema}, H. and {Jackson}, N. and {Jelic}, V. and {de Jong}, A. and {Juette}, E. and {Kant}, D. and {Karastergiou}, A. and {Koers}, A. and {Kollen}, H. and {Kondratiev}, V.~I. and {Kooistra}, E. and {Koopman}, Y. and {Koster}, A. and {Kuniyoshi}, M. and {Kramer}, M. and {Kuper}, G. and {Lambropoulos}, P. and {Law}, C. and {van Leeuwen}, J. and {Lemaitre}, J. and {Loose}, M. and {Maat}, P. and {Macario}, G. and {Markoff}, S. and {Masters}, J. and {McFadden}, R.~A. and {McKay-Bukowski}, D. and {Meijering}, H. and {Meulman}, H. and {Mevius}, M. and {Middelberg}, E. and {Millenaar}, R. and {Miller-Jones}, J.~C.~A. and {Mohan}, R.~N. and {Mol}, J.~D. and {Morawietz}, J. and {Morganti}, R. and {Mulcahy}, D.~D. and {Mulder}, E. and {Munk}, H. and {Nieuwenhuis}, L. and {van Nieuwpoort}, R. and {Noordam}, J.~E. and {Norden}, M. and {Noutsos}, A. and {Offringa}, A.~R. and {Olofsson}, H. and {Omar}, A. and {Orr{\'u}}, E. and {Overeem}, R. and {Paas}, H. and {Pandey-Pommier}, M. and {Pandey}, V.~N. and {Pizzo}, R. and {Polatidis}, A. and {Rafferty}, D. and {Rawlings}, S. and {Reich}, W. and {de Reijer}, J. -P. and {Reitsma}, J. and {Renting}, G.~A. and {Riemers}, P. and {Rol}, E. and {Romein}, J.~W. and {Roosjen}, J. and {Ruiter}, M. and {Scaife}, A. and {van der Schaaf}, K. and {Scheers}, B. and {Schellart}, P. and {Schoenmakers}, A. and {Schoonderbeek}, G. and {Serylak}, M. and {Shulevski}, A. and {Sluman}, J. and {Smirnov}, O. and {Sobey}, C. and {Spreeuw}, H. and {Steinmetz}, M. and {Sterks}, C.~G.~M. and {Stiepel}, H. -J. and {Stuurwold}, K. and {Tagger}, M. and {Tang}, Y. and {Tasse}, C. and {Thomas}, I. and {Thoudam}, S. and {Toribio}, M.~C. and {van der Tol}, B. and {Usov}, O. and {van Veelen}, M. and {van der Veen}, A. -J. and {ter Veen}, S. and {Verbiest}, J.~P.~W. and {Vermeulen}, R. and {Vermaas}, N. and {Vocks}, C. and {Vogt}, C. and {de Vos}, M. and {van der Wal}, E. and {van Weeren}, R. and {Weggemans}, H. and {Weltevrede}, P. and {White}, S. and {Wijnholds}, S.~J. and {Wilhelmsson}, T. and {Wucknitz}, O. and {Yatawatta}, S. and {Zarka}, P. and {Zensus}, A. and {van Zwieten}, J.},
        title = "{LOFAR: The LOw-Frequency ARray}",
      journal = {\aap},
         year = 2013,
        month = aug,
       volume = {556},
          eid = {A2},
        pages = {A2},
          doi = {10.1051/0004-6361/201220873},
archivePrefix = {arXiv},
       eprint = {1305.3550},
 primaryClass = {astro-ph.IM},
       adsurl = {https://ui.adsabs.harvard.edu/abs/2013A&A...556A...2V}
}

@ARTICLE{gtaylor,
       author = {{Taylor}, G.~B. and {Ellingson}, S.~W. and {Kassim}, N.~E. and {Craig}, J. and {Dowell}, J. and {Wolfe}, C.~N. and {Hartman}, J. and {Bernardi}, G. and {Clarke}, T. and {Cohen}, A. and {Dalal}, N.~P. and {Erickson}, W.~C. and {Hicks}, B. and {Greenhill}, L.~J. and {Jacoby}, B. and {Lane}, W. and {Lazio}, J. and {Mitchell}, D. and {Navarro}, R. and {Ord}, S.~M. and {Pihlstr{\"o}m}, Y. and {Polisensky}, E. and {Ray}, P.~S. and {Rickard}, L.~J. and {Schinzel}, F.~K. and {Schmitt}, H. and {Sigman}, E. and {Soriano}, M. and {Stewart}, K.~P. and {Stovall}, K. and {Tremblay}, S. and {Wang}, D. and {Weiler}, K.~W. and {White}, S. and {Wood}, D.~L.},
        title = "{First Light for the First Station of the Long Wavelength Array}",
      journal = {Journal of Astronomical Instrumentation},
         year = 2012,
        month = dec,
       volume = {1},
       number = {1},
          eid = {1250004-284},
        pages = {1250004-284},
          doi = {10.1142/S2251171712500043},
archivePrefix = {arXiv},
       eprint = {1206.6733},
 primaryClass = {astro-ph.IM},
       adsurl = {https://ui.adsabs.harvard.edu/abs/2012JAI.....150004T}
}

@BOOK{handbook,
       author = {{Lorimer}, D.~R. and {Kramer}, M.},
        title = "{Handbook of Pulsar Astronomy}",
         year = 2004,
       volume = {4},
       adsurl = {https://ui.adsabs.harvard.edu/abs/2004hpa..book.....L},
      publisher={Cambridge University Press},
       isbn={9780521828239},
       lccn={2004057044},
       series={Cambridge Observing Handbooks for Research Astronomers},
        url={https://books.google.com/books?id=OZ8tdN6qJcsC},
}

@ARTICLE{Bansal,
       author = {{Bansal}, K. and {Taylor}, G.~B. and {Stovall}, Kevin and {Dowell}, Jayce},
        title = "{Scattering Study of Pulsars below 100 MHz Using LWA1}",
      journal = {\apj},
         year = 2019,
        month = apr,
       volume = {875},
       number = {2},
          eid = {146},
        pages = {146},
          doi = {10.3847/1538-4357/ab0d8f},
archivePrefix = {arXiv},
       eprint = {1903.03457},
 primaryClass = {astro-ph.HE},
       adsurl = {https://ui.adsabs.harvard.edu/abs/2019ApJ...875..146B}
}

@ARTICLE{Donner,
       author = {{Donner}, J.~Y. and {Verbiest}, J.~P.~W. and {Tiburzi}, C. and {Os{\l}owski}, S. and {Michilli}, D. and {Serylak}, M. and {Anderson}, J.~M. and {Horneffer}, A. and {Kramer}, M. and {Grie{\ss}meier}, J. -M. and {K{\"u}nsem{\"o}ller}, J. and {Hessels}, J.~W.~T. and {Hoeft}, M. and {Miskolczi}, A.},
        title = "{First detection of frequency-dependent, time-variable dispersion measures}",
      journal = {\aap},
         year = 2019,
        month = apr,
       volume = {624},
          eid = {A22},
        pages = {A22},
          doi = {10.1051/0004-6361/201834059},
archivePrefix = {arXiv},
       eprint = {1902.03814},
 primaryClass = {astro-ph.GA},
       adsurl = {https://ui.adsabs.harvard.edu/abs/2019A&A...624A..22D}
}

@ARTICLE{Tiburzi,
       author = {{Tiburzi}, C. and {Shaifullah}, G.~M. and {Bassa}, C.~G. and {Zucca}, P. and {Verbiest}, J.~P.~W. and {Porayko}, N.~K. and {van der Wateren}, E. and {Fallows}, R.~A. and {Main}, R.~A. and {Janssen}, G.~H. and {Anderson}, J.~M. and {Bak Nielsen}, A. -S. and {Donner}, J.~Y. and {Keane}, E.~F. and {K{\"u}nsem{\"o}ller}, J. and {Os{\l}owski}, S. and {Grie{\ss}meier}, J. -M. and {Serylak}, M. and {Br{\"u}ggen}, M. and {Ciardi}, B. and {Dettmar}, R. -J. and {Hoeft}, M. and {Kramer}, M. and {Mann}, G. and {Vocks}, C.},
        title = "{The impact of solar wind variability on pulsar timing}",
      journal = {\aap},
         year = 2021,
        month = mar,
       volume = {647},
          eid = {A84},
        pages = {A84},
          doi = {10.1051/0004-6361/202039846},
archivePrefix = {arXiv},
       eprint = {2012.11726},
 primaryClass = {astro-ph.HE},
       adsurl = {https://ui.adsabs.harvard.edu/abs/2021A&A...647A..84T}
}

@ARTICLE{Kumar,
       author = {{Kumar}, P. and {White}, S.~M. and {Stovall}, K. and {Dowell}, J. and {Taylor}, G.~B.},
        title = "{Pulsar observations at low frequencies: applications to pulsar timing and solar wind models}",
      journal = {\mnras},
         year = 2022,
        month = apr,
       volume = {511},
       number = {3},
        pages = {3937-3950},
          doi = {10.1093/mnras/stac316},
archivePrefix = {arXiv},
       eprint = {2202.01251},
 primaryClass = {astro-ph.HE},
       adsurl = {https://ui.adsabs.harvard.edu/abs/2022MNRAS.511.3937K}
}

@ARTICLE{Sobey,
       author = {{Sobey}, C. and {Bilous}, A.~V. and {Grie{\ss}meier}, J. -M. and {Hessels}, J.~W.~T. and {Karastergiou}, A. and {Keane}, E.~F. and {Kondratiev}, V.~I. and {Kramer}, M. and {Michilli}, D. and {Noutsos}, A. and {Pilia}, M. and {Polzin}, E.~J. and {Stappers}, B.~W. and {Tan}, C.~M. and {van Leeuwen}, J. and {Verbiest}, J.~P.~W. and {Weltevrede}, P. and {Heald}, G. and {Alves}, M.~I.~R. and {Carretti}, E. and {En{\ss}lin}, T. and {Haverkorn}, M. and {Iacobelli}, M. and {Reich}, W. and {Van Eck}, C.},
        title = "{Low-frequency Faraday rotation measures towards pulsars using LOFAR: probing the 3D Galactic halo magnetic field}",
      journal = {\mnras},
         year = 2019,
        month = apr,
       volume = {484},
       number = {3},
        pages = {3646-3664},
          doi = {10.1093/mnras/stz214},
archivePrefix = {arXiv},
       eprint = {1901.07738},
 primaryClass = {astro-ph.GA},
       adsurl = {https://ui.adsabs.harvard.edu/abs/2019MNRAS.484.3646S}
}

@ARTICLE{vonhoensbroech,
       author = {{von Hoensbroech}, A. and {Lesch}, H. and {Kunzl}, T.},
        title = "{Natural polarization modes in pulsar magnetospheres}",
      journal = {\aap},
         year = 1998,
        month = aug,
       volume = {336},
        pages = {209-219},
          doi = {10.48550/arXiv.astro-ph/9804318},
archivePrefix = {arXiv},
       eprint = {astro-ph/9804318},
 primaryClass = {astro-ph},
       adsurl = {https://ui.adsabs.harvard.edu/abs/1998A&A...336..209V}
}

@ARTICLE{Noutsos,
       author = {{Noutsos}, A. and {Sobey}, C. and {Kondratiev}, V.~I. and {Weltevrede}, P. and {Verbiest}, J.~P.~W. and {Karastergiou}, A. and {Kramer}, M. and {Kuniyoshi}, M. and {Alexov}, A. and {Breton}, R.~P. and {Bilous}, A.~V. and {Cooper}, S. and {Falcke}, H. and {Grie{\ss}meier}, J. -M. and {Hassall}, T.~E. and {Hessels}, J.~W.~T. and {Keane}, E.~F. and {Os{\l}owski}, S. and {Pilia}, M. and {Serylak}, M. and {Stappers}, B.~W. and {ter Veen}, S. and {van Leeuwen}, J. and {Zagkouris}, K. and {Anderson}, K. and {B{\"a}hren}, L. and {Bell}, M. and {Broderick}, J. and {Carbone}, D. and {Cendes}, Y. and {Coenen}, T. and {Corbel}, S. and {Eisl{\"o}ffel}, J. and {Fender}, R. and {Garsden}, H. and {Jonker}, P. and {Law}, C. and {Markoff}, S. and {Masters}, J. and {Miller-Jones}, J. and {Molenaar}, G. and {Osten}, R. and {Pietka}, M. and {Rol}, E. and {Rowlinson}, A. and {Scheers}, B. and {Spreeuw}, H. and {Staley}, T. and {Stewart}, A. and {Swinbank}, J. and {Wijers}, R. and {Wijnands}, R. and {Wise}, M. and {Zarka}, P. and {van der Horst}, A.},
        title = "{Pulsar polarisation below 200 MHz: Average profiles and propagation effects}",
      journal = {\aap},
         year = 2015,
        month = apr,
       volume = {576},
          eid = {A62},
        pages = {A62},
          doi = {10.1051/0004-6361/201425186},
archivePrefix = {arXiv},
       eprint = {1501.03312},
 primaryClass = {astro-ph.GA},
       adsurl = {https://ui.adsabs.harvard.edu/abs/2015A&A...576A..62N}
}

@ARTICLE{Sieber,
       author = {{Sieber}, W.},
        title = "{Pulsar Spectra}",
      journal = {\aap},
         year = 1973,
        month = oct,
       volume = {28},
        pages = {237},
       adsurl = {https://ui.adsabs.harvard.edu/abs/1973A&A....28..237S}
}

@ARTICLE{Cordes,
       author = {{Cordes}, J.~M.},
        title = "{Observational limits on the location of pulsar emission regions.}",
      journal = {\apj},
         year = 1978,
        month = jun,
       volume = {222},
        pages = {1006-1011},
          doi = {10.1086/156218},
       adsurl = {https://ui.adsabs.harvard.edu/abs/1978ApJ...222.1006C}
}

@ARTICLE{Rankin,
       author = {{Rankin}, J.~M.},
        title = "{Toward an empirical theory of pulsar emission. II. On the spectral behavior of component width.}",
      journal = {\apj},
         year = 1983,
        month = nov,
       volume = {274},
        pages = {359-368},
          doi = {10.1086/161451},
       adsurl = {https://ui.adsabs.harvard.edu/abs/1983ApJ...274..359R}
}

@ARTICLE{Kijak,
       author = {{Kijak}, J. and {Gupta}, Y. and {Krzeszowski}, K.},
        title = "{Turn-over in pulsar spectra around 1 GHz}",
      journal = {\aap},
         year = 2007,
        month = feb,
       volume = {462},
       number = {2},
        pages = {699-702},
          doi = {10.1051/0004-6361:20066125},
archivePrefix = {arXiv},
       eprint = {astro-ph/0611445},
 primaryClass = {astro-ph},
       adsurl = {https://ui.adsabs.harvard.edu/abs/2007A&A...462..699K}
}

@ARTICLE{Kuzmin,
       author = {{Kuzmin}, A.~D. and {Losovsky}, B. Ya.},
        title = "{No low-frequency turn-over in the spectra of millisecond pulsars}",
      journal = {\aap},
         year = 2001,
        month = mar,
       volume = {368},
        pages = {230-238},
          doi = {10.1051/0004-6361:20000507},
       adsurl = {https://ui.adsabs.harvard.edu/abs/2001A&A...368..230K}
}

@ARTICLE{vanstraten,
       author = {{van Straten}, Willem and {Demorest}, Paul and {Oslowski}, Stefan},
        title = "{Pulsar Data Analysis with PSRCHIVE}",
      journal = {Astronomical Research and Technology},
         year = 2012,
        month = jul,
       volume = {9},
       number = {3},
        pages = {237-256},
archivePrefix = {arXiv},
       eprint = {1205.6276},
 primaryClass = {astro-ph.IM},
       adsurl = {https://ui.adsabs.harvard.edu/abs/2012AR&T....9..237V}
}

@article{Alken2021,
  doi = {10.1186/s40623-020-01288-x},
  url = {https://doi.org/10.1186/s40623-020-01288-x},
  year = {2021},
  month = feb,
  publisher = {Springer Science and Business Media {LLC}},
  volume = {73},
  number = {1},
  author = {P. Alken and E. Th{\'{e}}bault and C. D. Beggan and H. Amit and J. Aubert and J. Baerenzung and T. N. Bondar and W. J. Brown and S. Califf and A. Chambodut and A. Chulliat and G. A. Cox and C. C. Finlay and A. Fournier and N. Gillet and A. Grayver and M. D. Hammer and M. Holschneider and L. Huder and G. Hulot and T. Jager and C. Kloss and M. Korte and W. Kuang and A. Kuvshinov and B. Langlais and J.-M. L{\'{e}}ger and V. Lesur and P. W. Livermore and F. J. Lowes and S. Macmillan and W. Magnes and M. Mandea and S. Marsal and J. Matzka and M. C. Metman and T. Minami and A. Morschhauser and J. E. Mound and M. Nair and S. Nakano and N. Olsen and F. J. Pav{\'{o}}n-Carrasco and V. G. Petrov and G. Ropp and M. Rother and T. J. Sabaka and S. Sanchez and D. Saturnino and N. R. Schnepf and X. Shen and C. Stolle and A. Tangborn and L. T{\o}ffner-Clausen and H. Toh and J. M. Torta and J. Varner and F. Vervelidou and P. Vigneron and I. Wardinski and J. Wicht and A. Woods and Y. Yang and Z. Zeren and B. Zhou},
  title = {International Geomagnetic Reference Field: the thirteenth generation},
  journal = {Earth,  Planets and Space}
}

@article{bell,
    author = {Bell, M. E. and Murphy, Tara and Johnston, S. and Kaplan, D. L. and Croft, S. and Hancock, P. and Callingham, J. R. and Zic, A. and Dobie, D. and Swiggum, J. K. and Rowlinson, A. and Hurley-Walker, N. and Offringa, A. R. and Bernardi, G. and Bowman, J. D. and Briggs, F. and Cappallo, R. J. and Deshpande, A. A. and Gaensler, B. M. and Greenhill, L. J. and Hazelton, B. J. and Johnston-Hollitt, M. and Lonsdale, C. J. and McWhirter, S. R. and Mitchell, D. A. and Morales, M. F. and Morgan, E. and Oberoi, D. and Ord, S. M. and Prabu, T. and Shankar, N. Udaya and Srivani, K. S. and Subrahmanyan, R. and Tingay, S. J. and Wayth, R. B. and Webster, R. L. and Williams, A. and Williams, C. L.},
    title = "{Time-domain and spectral properties of pulsars at 154 MHz}",
    journal = {Monthly Notices of the Royal Astronomical Society},
    volume = {461},
    number = {1},
    pages = {908-921},
    year = {2016},
    month = {05},
    issn = {0035-8711},
    doi = {10.1093/mnras/stw1293},
    url = {https://doi.org/10.1093/mnras/stw1293},
    eprint = {https://academic.oup.com/mnras/article-pdf/461/1/908/13484764/stw1293.pdf},
}

@ARTICLE{Lorimer1995,
       author = {{Lorimer}, D.~R. and {Yates}, J.~A. and {Lyne}, A.~G. and {Gould}, D.~M.},
        title = "{Multifrequency flux density measurements of 280 pulsars}",
      journal = {\mnras},
         year = 1995,
        month = mar,
       volume = {273},
       number = {2},
        pages = {411-421},
          doi = {10.1093/mnras/273.2.411},
       adsurl = {https://ui.adsabs.harvard.edu/abs/1995MNRAS.273..411L}
}

@article{Lyne1988,
    author = {Lyne, A. G. and Manchester, R. N.},
    title = "{The shape of pulsar radio beams}",
    journal = {Monthly Notices of the Royal Astronomical Society},
    volume = {234},
    number = {3},
    pages = {477-508},
    year = {1988},
    month = {10},
    issn = {0035-8711},
    doi = {10.1093/mnras/234.3.477},
    url = {https://doi.org/10.1093/mnras/234.3.477},
    eprint = {https://academic.oup.com/mnras/article-pdf/234/3/477/18522547/mnras234-0477.pdf},
}

@ARTICLE{Izvekova,
       author = {{Izvekova}, V.~A. and {Kuzmin}, A.~D. and {Malofeev}, V.~M. and {Shitov}, Iu. P.},
        title = "{Radio Spectra of Pulsars - Part One - Observations of Flux Densities at Meter Wavelengths and Analysis of the Spectra}",
      journal = {\apss},
         year = 1981,
        number = {1},
        pages = {45-72},
          doi = {10.1007/BF00654022},
       adsurl = {https://ui.adsabs.harvard.edu/abs/1981Ap&SS..78...45I}
}

@ARTICLE{Malov,
       author = {{Malov}, I.~F.},
        title = "{On the nature of the low-frequency dropoff in pulsar spectra}",
      journal = {\sovast},
         year = 1979,
        month = apr,
       volume = {23},
        pages = {205},
       adsurl = {https://ui.adsabs.harvard.edu/abs/1979SvA....23..205M}
}

@ARTICLE{Lammodel,
       author = {{Lam}, M.~T. and {Cordes}, J.~M. and {Chatterjee}, S. and {Jones}, M.~L. and {McLaughlin}, M.~A. and {Armstrong}, J.~W.},
        title = "{Systematic and Stochastic Variations in Pulsar Dispersion Measures}",
      journal = {\apj},
         year = 2016,
        month = apr,
       volume = {821},
       number = {1},
          eid = {66},
        pages = {66},
        doi = {10.3847/0004-637X/821/1/66},
archivePrefix = {arXiv},
       eprint = {1512.02203},
 primaryClass = {astro-ph.HE},
       adsurl = {https://ui.adsabs.harvard.edu/abs/2016ApJ...821...66L}
}

@ARTICLE{Ocker,
       author = {{Ocker}, Stella Koch and {Cordes}, James M. and {Chatterjee}, Shami and {Dolch}, Timothy},
        title = "{An In Situ Study of Turbulence near Stellar Bow Shocks}",
      journal = {\apj},
         year = 2021,
        month = dec,
       volume = {922},
       number = {2},
          eid = {233},
        pages = {233},
          doi = {10.3847/1538-4357/ac2b28},
archivePrefix = {arXiv},
       eprint = {2107.10371},
 primaryClass = {astro-ph.GA},
       adsurl = {https://ui.adsabs.harvard.edu/abs/2021ApJ...922..233O}
}

@ARTICLE{Armstrong,
    author = {{Armstrong}, J.~W. and {Rickett}, B.~J. and {Spangler}, S.~R.},
    title = "{Electron Density Power Spectrum in the Local Interstellar Medium}",
    journal = {\apj},
    year = 1995,
    month = apr,
    volume = {443},
    pages = {209},
    doi = {10.1086/175515},
    adsurl = {https://ui.adsabs.harvard.edu/abs/1995ApJ...443..209A}
}

@ARTICLE{Hobbs,
       author = {{Hobbs}, G. and {Lyne}, A.~G. and {Kramer}, M. and {Martin}, C.~E. and {Jordan}, C.},
        title = "{Long-term timing observations of 374 pulsars}",
      journal = {\mnras},
     keywords = {methods: data analysis, astrometry, pulsars: general},
         year = 2004,
        month = oct,
       volume = {353},
       number = {4},
        pages = {1311-1344},
          \doi = {10.1111/j.1365-2966.2004.08157.x},
       adsurl = {https://ui.adsabs.harvard.edu/abs/2004MNRAS.353.1311H},
      adsnote = {Provided by the SAO/NASA Astrophysics Data System}
}

@ARTICLE{Backer,
       author = {{Backer}, D.~C. and {Hama}, S. and {van Hook}, S. and {Foster}, R.~S.},
        title = "{Temporal Variations of Pulsar Dispersion Measures}",
      journal = {\apj},
         year = 1993,
        month = feb,
       volume = {404},
        pages = {636},
          doi = {10.1086/172317},
       adsurl = {https://ui.adsabs.harvard.edu/abs/1993ApJ...404..636B}
}

@ARTICLE{Hobbsvel,
       author = {{Hobbs}, G. and {Lorimer}, D.~R. and {Lyne}, A.~G. and {Kramer}, M.},
        title = "{A statistical study of 233 pulsar proper motions}",
      journal = {\mnras},
         year = 2005,
        month = jul,
       volume = {360},
       number = {3},
        pages = {974-992},
          doi = {10.1111/j.1365-2966.2005.09087.x},
archivePrefix = {arXiv},
       eprint = {astro-ph/0504584},
 primaryClass = {astro-ph},
       adsurl = {https://ui.adsabs.harvard.edu/abs/2005MNRAS.360..974H}
}

@INPROCEEDINGS{janssen,
       author = {{Janssen}, G. and {Hobbs}, G. and {McLaughlin}, M. and {Bassa}, C. and {Deller}, A. and {Kramer}, M. and {Lee}, K. and {Mingarelli}, C. and {Rosado}, P. and {Sanidas}, S. and {Sesana}, A. and {Shao}, L. and {Stairs}, I. and {Stappers}, B. and {Verbiest}, J.~P.~W.},
        title = "{Gravitational Wave Astronomy with the SKA}",
    booktitle = {Advancing Astrophysics with the Square Kilometre Array (AASKA14)},
         year = 2015,
        month = apr,
          eid = {37},
        pages = {37},
          doi = {10.22323/1.215.0037},
archivePrefix = {arXiv},
       eprint = {1501.00127},
 primaryClass = {astro-ph.IM},
       adsurl = {https://ui.adsabs.harvard.edu/abs/2015aska.confE..37J}
}

@ARTICLE{kerr,
       author = {{Johnston}, Simon and {Kerr}, Matthew},
        title = "{Polarimetry of 600 pulsars from observations at 1.4 GHz with the Parkes radio telescope}",
      journal = {\mnras},
         year = 2018,
        month = mar,
       volume = {474},
       number = {4},
        pages = {4629-4636},
          doi = {10.1093/mnras/stx3095},
archivePrefix = {arXiv},
       eprint = {1711.10092},
 primaryClass = {astro-ph.HE},
       adsurl = {https://ui.adsabs.harvard.edu/abs/2018MNRAS.474.4629J}
}

@ARTICLE{Wang,
       author = {{Wang}, P.~F. and {Wang}, C. and {Han}, J.~L.},
        title = "{On the frequency dependence of pulsar linear polarization}",
      journal = {\mnras},
         year = 2015,
        month = mar,
       volume = {448},
       number = {1},
        pages = {771-780},
          doi = {10.1093/mnras/stu2765},
archivePrefix = {arXiv},
       eprint = {1501.00066},
 primaryClass = {astro-ph.HE},
       adsurl = {https://ui.adsabs.harvard.edu/abs/2015MNRAS.448..771W}
}

@ARTICLE{olszanskimitra,
       author = {{Olszanski}, Timothy E.~E. and {Mitra}, Dipanjan and {Rankin}, Joanna M.},
        title = "{Arecibo 4.5/1.4/0.33-GHz polarimetric single-pulse emission survey}",
      journal = {\mnras},
         year = 2019,
        month = oct,
       volume = {489},
       number = {2},
        pages = {1543-1555},
          doi = {10.1093/mnras/stz2172},
archivePrefix = {arXiv},
       eprint = {1909.09685},
 primaryClass = {astro-ph.HE},
       adsurl = {https://ui.adsabs.harvard.edu/abs/2019MNRAS.489.1543O}
}

@ARTICLE{skrzypczak,
       author = {{Skrzypczak}, Anna and {Basu}, Rahul and {Mitra}, Dipanjan and {Melikidze}, George I. and {Maciesiak}, Krzysztof and {Koralewska}, Olga and {Filothodoros}, Alexandros},
        title = "{Meterwavelength Single-pulse Polarimetric Emission Survey. IV. The Period Dependence of Component Widths of Pulsars}",
      journal = {\apj},
         year = 2018,
        month = feb,
       volume = {854},
       number = {2},
          eid = {162},
        pages = {162},
          doi = {10.3847/1538-4357/aaa758},
archivePrefix = {arXiv},
       eprint = {1801.02720},
 primaryClass = {astro-ph.HE},
       adsurl = {https://ui.adsabs.harvard.edu/abs/2018ApJ...854..162S}
}

@ARTICLE{basumitrageorge,
       author = {{Basu}, Rahul and {Mitra}, Dipanjan and {Melikidze}, George I.},
        title = "{Meterwavelength Single-pulse Polarimetric Emission Survey. V. Flux Density, Component Spectral Variation, and Emission States}",
      journal = {\apj},
         year = 2021,
        month = aug,
       volume = {917},
       number = {1},
          eid = {48},
        pages = {48},
          doi = {10.3847/1538-4357/ac0828},
archivePrefix = {arXiv},
       eprint = {2106.01402},
 primaryClass = {astro-ph.HE},
       adsurl = {https://ui.adsabs.harvard.edu/abs/2021ApJ...917...48B}
}

@ARTICLE{rahulsharan,
       author = {{Sharan}, Rahul and {Bhattacharyya}, Bhaswati and {Kumari}, Sangita and {Roy}, Jayanta and {Ghosh}, Ankita},
        title = "{Flux Density Stability and Temporal Changes in the Spectra of Millisecond Pulsars Using the GMRT}",
      journal = {\apj},
         year = 2024,
        month = aug,
       volume = {971},
       number = {2},
          eid = {174},
        pages = {174},
          doi = {10.3847/1538-4357/ad55c8},
archivePrefix = {arXiv},
       eprint = {2406.03939},
 primaryClass = {astro-ph.HE},
       adsurl = {https://ui.adsabs.harvard.edu/abs/2024ApJ...971..174S}
}

@ARTICLE{bhat2004,
       author = {{Bhat}, N.~D. Ramesh and {Cordes}, James M. and {Camilo}, Fernando and {Nice}, David J. and {Lorimer}, Duncan R.},
        title = "{Multifrequency Observations of Radio Pulse Broadening and Constraints on Interstellar Electron Density Microstructure}",
      journal = {\apj},
         year = 2004,
        month = apr,
       volume = {605},
       number = {2},
        pages = {759-783},
          doi = {10.1086/382680},
archivePrefix = {arXiv},
       eprint = {astro-ph/0401067},
 primaryClass = {astro-ph},
       adsurl = {https://ui.adsabs.harvard.edu/abs/2004ApJ...605..759B}
}

@ARTICLE{Hemberger2008,
       author = {{Hemberger}, Daniel A. and {Stinebring}, Daniel R.},
        title = "{Time Variability of Interstellar Scattering and Improvements to Pulsar Timing}",
      journal = {\apjl},
         year = 2008,
        month = feb,
       volume = {674},
       number = {1},
        pages = {L37},
          doi = {10.1086/528985},
       adsurl = {https://ui.adsabs.harvard.edu/abs/2008ApJ...674L..37H}
}

@ARTICLE{Kirsten2019,
       author = {{Kirsten}, F. and {Bhat}, N.~D.~R. and {Meyers}, B.~W. and {Macquart}, J. -P. and {Tremblay}, S.~E. and {Ord}, S.~M.},
        title = "{Probing Pulsar Scattering between 120 and 280 MHz with the MWA}",
      journal = {\apj},
         year = 2019,
        month = apr,
       volume = {874},
       number = {2},
          eid = {179},
        pages = {179},
          doi = {10.3847/1538-4357/ab0c05},
archivePrefix = {arXiv},
       eprint = {1903.02087},
 primaryClass = {astro-ph.HE},
       adsurl = {https://ui.adsabs.harvard.edu/abs/2019ApJ...874..179K}
}

@ARTICLE{Kumar2025,
       author = {{Kumar}, P. and {Taylor}, G.~B. and {Stovall}, K. and {Dowell}, J. and {White}, S.~M.},
        title = "{A Multifrequency Census of 100 Pulsars below 100 MHz with LWA: A Systematic Study of Flux Density, Spectra, Timing, Dispersion, Polarization, and Its Variation from a Decade of Observations}",
      journal = {\apj},
         year = 2025,
        month = apr,
       volume = {982},
       number = {2},
          eid = {132},
        pages = {132},
          doi = {10.3847/1538-4357/adb97a},
archivePrefix = {arXiv},
       eprint = {2501.11862},
 primaryClass = {astro-ph.HE},
       adsurl = {https://ui.adsabs.harvard.edu/abs/2025ApJ...982..132K}
}

@ARTICLE{Wayth2022,
       author = {{Wayth}, Randall and {Sokolowski}, Marcin and {Broderick}, Jess and {Tingay}, Steven J. and {Bhushan}, Raunaq and {Booler}, Tom and {Chiello}, Riccardo and {Davidson}, David B. and {Emrich}, David and {Juswardy}, Budi and {Kenney}, David and {Macario}, Giulia and {Magro}, Alessio and {Mattana}, Andrea and {Minchin}, David and {Monari}, Jader and {McPhail}, Andrew and {Perini}, Federico and {Pupillo}, Giuseppe and {Schiaffino}, Marco and {Subrahmanyan}, Ravi and {van Es}, Andre and {Walker}, Mia and {Waterson}, Mark},
        title = "{Engineering Development Array 2: design, performance, and lessons from an SKA-Low prototype station}",
      journal = {Journal of Astronomical Telescopes, Instruments, and Systems},
         year = 2022,
        month = jan,
       volume = {8},
          eid = {011010},
        pages = {011010},
          doi = {10.1117/1.JATIS.8.1.011010},
archivePrefix = {arXiv},
       eprint = {2112.00908},
 primaryClass = {astro-ph.IM},
       adsurl = {https://ui.adsabs.harvard.edu/abs/2022JATIS...8a1010W}
}

@ARTICLE{naldi2017,
       author = {{Naldi}, Giovanni and {Mattana}, Andrea and {Pastore}, Sandro and {Alderighi}, Monica and {Zarb Adami}, Kristian and {Schillir{\`o}}, Francesco and {Aminaei}, Amin and {Baker}, Jeremy and {Belli}, Carolina and {Comoretto}, Gianni and {Chiarucci}, Simone and {Chiello}, Riccardo and {D'Angelo}, Sergio and {Dalle Mura}, Gabriele and {De Marco}, Andrea and {Halsall}, Rob and {Magro}, Alessio and {Monari}, Jader and {Roberts}, Matt and {Perini}, Federico and {Poloni}, Marco and {Pupillo}, Giuseppe and {Rusticelli}, Simone and {Schiaffino}, Marco and {Zaccaro}, Emanuele},
        title = "{The Digital Signal Processing Platform for the Low Frequency Aperture Array: Preliminary Results on the Data Acquisition Unit}",
      journal = {Journal of Astronomical Instrumentation},
         year = 2017,
        month = mar,
       volume = {6},
       number = {1},
          eid = {1641014},
        pages = {1641014},
          doi = {10.1142/S2251171716410142},
       adsurl = {https://ui.adsabs.harvard.edu/abs/2017JAI.....641014N}
}

@ARTICLE{lee2022,
       author = {{Lee}, C.~P. and {Bhat}, N.~D.~R. and {Sokolowski}, M. and {Swainston}, N.~A. and {Ung}, D. and {Magro}, A. and {Chiello}, R.},
        title = "{Spectral analysis of 22 radio pulsars using SKA-Low precursor stations}",
      journal = {\pasa},
         year = 2022,
        month = sep,
       volume = {39},
          eid = {e042},
        pages = {e042},
          doi = {10.1017/pasa.2022.40},
archivePrefix = {arXiv},
       eprint = {2208.07182},
 primaryClass = {astro-ph.HE},
       adsurl = {https://ui.adsabs.harvard.edu/abs/2022PASA...39...42L}
}

@ARTICLE{tremblay2015,
       author = {{Tremblay}, S.~E. and {Ord}, S.~M. and {Bhat}, N.~D.~R. and {Tingay}, S.~J. and {Crosse}, B. and {Pallot}, D. and {Oronsaye}, S.~I. and {Bernardi}, G. and {Bowman}, J.~D. and {Briggs}, F. and {Cappallo}, R.~J. and {Corey}, B.~E. and {Deshpande}, A.~A. and {Emrich}, D. and {Goeke}, R. and {Greenhill}, L.~J. and {Hazelton}, B.~J. and {Johnston-Hollitt}, M. and {Kaplan}, D.~L. and {Kasper}, J.~C. and {Kratzenberg}, E. and {Lonsdale}, C.~J. and {Lynch}, M.~J. and {McWhirter}, S.~R. and {Mitchell}, D.~A. and {Morales}, M.~F. and {Morgan}, E. and {Oberoi}, D. and {Prabu}, T. and {Rogers}, A.~E.~E. and {Roshi}, A. and {Udaya Shankar}, N. and {Srivani}, K.~S. and {Subrahmanyan}, R. and {Waterson}, M. and {Wayth}, R.~B. and {Webster}, R.~L. and {Whitney}, A.~R. and {Williams}, A. and {Williams}, C.~L.},
        title = "{The High Time and Frequency Resolution Capabilities of the Murchison Widefield Array}",
      journal = {\pasa},
         year = 2015,
        month = feb,
       volume = {32},
          eid = {e005},
        pages = {e005},
          doi = {10.1017/pasa.2015.6},
archivePrefix = {arXiv},
       eprint = {1501.05723},
 primaryClass = {astro-ph.IM},
       adsurl = {https://ui.adsabs.harvard.edu/abs/2015PASA...32....5T}
}

@ARTICLE{swainston2022,
       author = {{Swainston}, N.~A. and {Lee}, C.~P. and {McSweeney}, S.~J. and {Bhat}, N.~D.~R.},
        title = "{pulsar\_spectra: A pulsar flux density catalogue and spectrum fitting repository}",
      journal = {\pasa},
         year = 2022,
        month = nov,
       volume = {39},
          eid = {e056},
        pages = {e056},
          doi = {10.1017/pasa.2022.52},
archivePrefix = {arXiv},
       eprint = {2209.13324},
 primaryClass = {astro-ph.HE},
       adsurl = {https://ui.adsabs.harvard.edu/abs/2022PASA...39...56S}
}

@ARTICLE{sokolowski2022,
       author = {{Sokolowski}, M. and {Tingay}, S.~J. and {Davidson}, D.~B. and {Wayth}, R.~B. and {Ung}, D. and {Broderick}, J. and {Juswardy}, B. and {Kovaleva}, M. and {Macario}, G. and {Pupillo}, G. and {Sutinjo}, A.},
        title = "{What is the SKA-Low sensitivity for your favourite radio source?}",
      journal = {\pasa},
         year = 2022,
        month = apr,
       volume = {39},
          eid = {e015},
        pages = {e015},
          doi = {10.1017/pasa.2021.63},
archivePrefix = {arXiv},
       eprint = {2204.05873},
 primaryClass = {astro-ph.IM},
       adsurl = {https://ui.adsabs.harvard.edu/abs/2022PASA...39...15S}
}

@ARTICLE{dylan,
       author = {{Grigg}, Dylan and {Tingay}, Steven and {Prabu}, Steve and {Sokolowski}, Marcin and {Indermuehle}, Balthasar},
        title = "{Enhanced detection and identification of satellites using an all-sky multi-frequency survey with prototype SKA-Low stations}",
      journal = {\pasa},
         year = 2025,
        month = jan,
       volume = {42},
          eid = {e015},
        pages = {e015},
          doi = {10.1017/pasa.2024.136},
archivePrefix = {arXiv},
       eprint = {2412.14483},
 primaryClass = {astro-ph.EP},
       adsurl = {https://ui.adsabs.harvard.edu/abs/2025PASA...42...15G}
}

@ARTICLE{dspsr,
       author = {{van Straten}, W. and {Bailes}, M.},
        title = "{DSPSR: Digital Signal Processing Software for Pulsar Astronomy}",
      journal = {\pasa},
         year = 2011,
        month = jan,
       volume = {28},
       number = {1},
        pages = {1-14},
          doi = {10.1071/AS10021},
archivePrefix = {arXiv},
       eprint = {1008.3973},
 primaryClass = {astro-ph.IM},
       adsurl = {https://ui.adsabs.harvard.edu/abs/2011PASA...28....1V}
}

@ARTICLE{Sutinjo,
       author = {{Sutinjo}, A.~T. and {Sokolowski}, M. and {Kovaleva}, M. and {Ung}, D.~C.~X. and {Broderick}, J.~W. and {Wayth}, R.~B. and {Davidson}, D.~B. and {Tingay}, S.~J.},
        title = "{Sensitivity of a low-frequency polarimetric radio interferometer}",
      journal = {\aap},
         year = 2021,
        month = feb,
       volume = {646},
          eid = {A143},
        pages = {A143},
          doi = {10.1051/0004-6361/202039445},
archivePrefix = {arXiv},
       eprint = {2012.08075},
 primaryClass = {astro-ph.IM},
       adsurl = {https://ui.adsabs.harvard.edu/abs/2021A&A...646A.143S}
}

@ARTICLE{BIC,
       author = {{Liddle}, Andrew R.},
        title = "{Information criteria for astrophysical model selection}",
      journal = {\mnras},
         year = 2007,
        month = may,
       volume = {377},
       number = {1},
        pages = {L74-L78},
          doi = {10.1111/j.1745-3933.2007.00306.x},
archivePrefix = {arXiv},
       eprint = {astro-ph/0701113},
 primaryClass = {astro-ph},
       adsurl = {https://ui.adsabs.harvard.edu/abs/2007MNRAS.377L..74L}
}

@ARTICLE{Burns1966,
       author = {{Burn}, B.~J.},
        title = "{On the depolarization of discrete radio sources by Faraday dispersion}",
      journal = {\mnras},
         year = 1966,
        month = jan,
       volume = {133},
        pages = {67},
          doi = {10.1093/mnras/133.1.67},
       adsurl = {https://ui.adsabs.harvard.edu/abs/1966MNRAS.133...67B}
}

@ARTICLE{Brentjens,
       author = {{Brentjens}, M.~A. and {de Bruyn}, A.~G.},
        title = "{Faraday rotation measure synthesis}",
      journal = {\aap},
         year = 2005,
        month = oct,
       volume = {441},
       number = {3},
        pages = {1217-1228},
          doi = {10.1051/0004-6361:20052990},
archivePrefix = {arXiv},
       eprint = {astro-ph/0507349},
 primaryClass = {astro-ph},
       adsurl = {https://ui.adsabs.harvard.edu/abs/2005A&A...441.1217B}
}

@ARTICLE{Wanghan,
       author = {{Wang}, Chen and {Han}, J.~L. and {Lai}, Dong},
        title = "{The Faraday rotation in the pulsar magnetosphere}",
      journal = {\mnras},
         year = 2011,
        month = oct,
       volume = {417},
       number = {2},
        pages = {1183-1191},
          doi = {10.1111/j.1365-2966.2011.19333.x},
archivePrefix = {arXiv},
       eprint = {1105.2602},
 primaryClass = {astro-ph.SR},
       adsurl = {https://ui.adsabs.harvard.edu/abs/2011MNRAS.417.1183W}
}

@ARTICLE{Noutsos2009,
       author = {{Noutsos}, A. and {Karastergiou}, A. and {Kramer}, M. and {Johnston}, S. and {Stappers}, B.~W.},
        title = "{Phase-resolved Faraday rotation in pulsars}",
      journal = {\mnras},
         year = 2009,
        month = jul,
       volume = {396},
       number = {3},
        pages = {1559-1572},
          doi = {10.1111/j.1365-2966.2009.14806.x},
archivePrefix = {arXiv},
       eprint = {0903.5511},
 primaryClass = {astro-ph.GA},
       adsurl = {https://ui.adsabs.harvard.edu/abs/2009MNRAS.396.1559N}
}

@ARTICLE{Heald,
       author = {{Heald}, G. and {Braun}, R. and {Edmonds}, R.},
        title = "{The Westerbork SINGS survey. II Polarization, Faraday rotation, and magnetic fields}",
      journal = {\aap},
         year = 2009,
        month = aug,
       volume = {503},
       number = {2},
        pages = {409-435},
          doi = {10.1051/0004-6361/200912240},
archivePrefix = {arXiv},
       eprint = {0905.3995},
 primaryClass = {astro-ph.GA},
       adsurl = {https://ui.adsabs.harvard.edu/abs/2009A&A...503..409H}
}

@ARTICLE{Schnitzeler,
       author = {{Schnitzeler}, D.~H.~F.~M. and {Katgert}, P. and {de Bruyn}, A.~G.},
        title = "{WSRT Faraday tomography of the Galactic ISM at {\ensuremath{\lambda}} \raisebox{-0.5ex}\textasciitilde 0.86 m. I. The GEMINI data set at (l, b) = (181{\textdegree}, 20{\textdegree})}",
      journal = {\aap},
         year = 2009,
        month = feb,
       volume = {494},
       number = {2},
        pages = {611-622},
          doi = {10.1051/0004-6361:20078912},
archivePrefix = {arXiv},
       eprint = {0810.4211},
 primaryClass = {astro-ph},
       adsurl = {https://ui.adsabs.harvard.edu/abs/2009A&A...494..611S}
}

@ARTICLE{Sotomayor,
       author = {{Sotomayor-Beltran}, C. and {Sobey}, C. and {Hessels}, J.~W.~T. and {de Bruyn}, G. and {Noutsos}, A. and {Alexov}, A. and {Anderson}, J. and {Asgekar}, A. and {Avruch}, I.~M. and {Beck}, R. and {Bell}, M.~E. and {Bell}, M.~R. and {Bentum}, M.~J. and {Bernardi}, G. and {Best}, P. and {Birzan}, L. and {Bonafede}, A. and {Breitling}, F. and {Broderick}, J. and {Brouw}, W.~N. and {Br{\"u}ggen}, M. and {Ciardi}, B. and {de Gasperin}, F. and {Dettmar}, R. -J. and {van Duin}, A. and {Duscha}, S. and {Eisl{\"o}ffel}, J. and {Falcke}, H. and {Fallows}, R.~A. and {Fender}, R. and {Ferrari}, C. and {Frieswijk}, W. and {Garrett}, M.~A. and {Grie{\ss}meier}, J. and {Grit}, T. and {Gunst}, A.~W. and {Hassall}, T.~E. and {Heald}, G. and {Hoeft}, M. and {Horneffer}, A. and {Iacobelli}, M. and {Juette}, E. and {Karastergiou}, A. and {Keane}, E. and {Kohler}, J. and {Kramer}, M. and {Kondratiev}, V.~I. and {Koopmans}, L.~V.~E. and {Kuniyoshi}, M. and {Kuper}, G. and {van Leeuwen}, J. and {Maat}, P. and {Macario}, G. and {Markoff}, S. and {McKean}, J.~P. and {Mulcahy}, D.~D. and {Munk}, H. and {Orru}, E. and {Paas}, H. and {Pandey-Pommier}, M. and {Pilia}, M. and {Pizzo}, R. and {Polatidis}, A.~G. and {Reich}, W. and {R{\"o}ttgering}, H. and {Serylak}, M. and {Sluman}, J. and {Stappers}, B.~W. and {Tagger}, M. and {Tang}, Y. and {Tasse}, C. and {ter Veen}, S. and {Vermeulen}, R. and {van Weeren}, R.~J. and {Wijers}, R.~A.~M.~J. and {Wijnholds}, S.~J. and {Wise}, M.~W. and {Wucknitz}, O. and {Yatawatta}, S. and {Zarka}, P.},
        title = "{Calibrating high-precision Faraday rotation measurements for LOFAR and the next generation of low-frequency radio telescopes}",
      journal = {\aap},
         year = 2013,
        month = apr,
       volume = {552},
          eid = {A58},
        pages = {A58},
          doi = {10.1051/0004-6361/201220728},
archivePrefix = {arXiv},
       eprint = {1303.6230},
 primaryClass = {astro-ph.IM},
       adsurl = {https://ui.adsabs.harvard.edu/abs/2013A&A...552A..58S}
}

@ARTICLE{Posselt,
       author = {{Posselt}, B. and {Karastergiou}, A. and {Johnston}, S. and {Parthasarathy}, A. and {Oswald}, L.~S. and {Main}, R.~A. and {Basu}, A. and {Keith}, M.~J. and {Song}, X. and {Weltevrede}, P. and {Tiburzi}, C. and {Bailes}, M. and {Buchner}, S. and {Geyer}, M. and {Kramer}, M. and {Spiewak}, R. and {Krishnan}, V. Venkatraman},
        title = "{The Thousand Pulsar Array program on MeerKAT - IX. The time-averaged properties of the observed pulsar population}",
      journal = {\mnras},
         year = 2023,
        month = apr,
       volume = {520},
       number = {3},
        pages = {4582-4600},
          doi = {10.1093/mnras/stac3383},
archivePrefix = {arXiv},
       eprint = {2211.11849},
 primaryClass = {astro-ph.HE},
       adsurl = {https://ui.adsabs.harvard.edu/abs/2023MNRAS.520.4582P}
}

@ARTICLE{Xue_2017,
       author = {{Xue}, Mengyao and {Bhat}, N.~D.~R. and {Tremblay}, S.~E. and {Ord}, S.~M. and {Sobey}, C. and {Swainston}, N.~A. and {Kaplan}, D.~L. and {Johnston}, Simon and {Meyers}, B.~W. and {McSweeney}, S.~J.},
        title = "{A Census of Southern Pulsars at 185 MHz}",
      journal = {\pasa},
         year = 2017,
        month = dec,
       volume = {34},
          eid = {e070},
        pages = {e070},
          doi = {10.1017/pasa.2017.66},
archivePrefix = {arXiv},
       eprint = {1711.08933},
 primaryClass = {astro-ph.HE},
       adsurl = {https://ui.adsabs.harvard.edu/abs/2017PASA...34...70X}
}

@ARTICLE{Bhat_2023,
       author = {{Bhat}, N.~D.~R. and {Swainston}, N.~A. and {McSweeney}, S.~J. and {Xue}, M. and {Meyers}, B.~W. and {Kudale}, S. and {Dai}, S. and {Tremblay}, S.~E. and {van Straten}, W. and {Shannon}, R.~M. and {Smith}, K.~R. and {Sokolowski}, M. and {Ord}, S.~M. and {Sleap}, G. and {Williams}, A. and {Hancock}, P.~J. and {Lange}, R. and {Tocknell}, J. and {Johnston-Hollitt}, M. and {Kaplan}, D.~L. and {Tingay}, S.~J. and {Walker}, M.},
        title = "{The Southern-sky MWA Rapid Two-metre (SMART) pulsar survey{\textemdash}II. Survey status, pulsar census, and first pulsar discoveries}",
      journal = {\pasa},
         year = 2023,
        month = may,
       volume = {40},
          eid = {e020},
        pages = {e020},
          doi = {10.1017/pasa.2023.18},
archivePrefix = {arXiv},
       eprint = {2302.11920},
 primaryClass = {astro-ph.HE},
       adsurl = {https://ui.adsabs.harvard.edu/abs/2023PASA...40...20B}
}

@ARTICLE{Gould_1998,
       author = {{Gould}, D.~M. and {Lyne}, A.~G.},
        title = "{Multifrequency polarimetry of 300 radio pulsars}",
      journal = {\mnras},
         year = 1998,
        month = nov,
       volume = {301},
       number = {1},
        pages = {235-260},
          doi = {10.1046/j.1365-8711.1998.02018.x},
       adsurl = {https://ui.adsabs.harvard.edu/abs/1998MNRAS.301..235G}
}

@ARTICLE{Coles2015,
       author = {{Coles}, W.~A. and {Kerr}, M. and {Shannon}, R.~M. and {Hobbs}, G.~B. and {Manchester}, R.~N. and {You}, X. -P. and {Bailes}, M. and {Bhat}, N.~D.~R. and {Burke-Spolaor}, S. and {Dai}, S. and {Keith}, M.~J. and {Levin}, Y. and {Os{\l}owski}, S. and {Ravi}, V. and {Reardon}, D. and {Toomey}, L. and {van Straten}, W. and {Wang}, J.~B. and {Wen}, L. and {Zhu}, X.~J.},
        title = "{Pulsar Observations of Extreme Scattering Events}",
      journal = {\apj},
         year = 2015,
        month = aug,
       volume = {808},
       number = {2},
          eid = {113},
        pages = {113},
          doi = {10.1088/0004-637X/808/2/113},
archivePrefix = {arXiv},
       eprint = {1506.07948},
 primaryClass = {astro-ph.SR},
       adsurl = {https://ui.adsabs.harvard.edu/abs/2015ApJ...808..113C}
}

@ARTICLE{Yaoscat,
       author = {{Yao}, J.~M. and {Manchester}, R.~N. and {Wang}, N.},
        title = "{A New Electron-density Model for Estimation of Pulsar and FRB Distances}",
      journal = {\apj},
         year = 2017,
        month = jan,
       volume = {835},
       number = {1},
          eid = {29},
        pages = {29},
          doi = {10.3847/1538-4357/835/1/29},
archivePrefix = {arXiv},
       eprint = {1610.09448},
 primaryClass = {astro-ph.GA},
       adsurl = {https://ui.adsabs.harvard.edu/abs/2017ApJ...835...29Y}
}

@ARTICLE{Pygedm,
       author = {{Price}, D.~C. and {Flynn}, C. and {Deller}, A.},
        title = "{A comparison of Galactic electron density models using PyGEDM}",
      journal = {\pasa},
         year = 2021,
        month = aug,
       volume = {38},
          eid = {e038},
        pages = {e038},
          doi = {10.1017/pasa.2021.33},
archivePrefix = {arXiv},
       eprint = {2106.15816},
 primaryClass = {astro-ph.GA},
       adsurl = {https://ui.adsabs.harvard.edu/abs/2021PASA...38...38P}
}

@ARTICLE{velaXue,
       author = {{Xue}, Mengyao and {Ord}, S.~M. and {Tremblay}, S.~E. and {Bhat}, N.~D.~R. and {Sobey}, C. and {Meyers}, B.~W. and {McSweeney}, S.~J. and {Swainston}, N.~A.},
        title = "{MWA tied-array processing II: Polarimetric verification and analysis of two bright southern pulsars}",
      journal = {\pasa},
         year = 2019,
        month = jul,
       volume = {36},
          eid = {e025},
        pages = {e025},
          doi = {10.1017/pasa.2019.19},
archivePrefix = {arXiv},
       eprint = {1905.00598},
 primaryClass = {astro-ph.HE},
       adsurl = {https://ui.adsabs.harvard.edu/abs/2019PASA...36...25X}
}

@ARTICLE{crabgiant,
       author = {{Sokolowski}, M. and {Kumar}, P. and {Dhavali}, S. and {Meyers}, B.~W. and {Bhat}, N.~D.~R. and {Bera}, A. and {McSweeney}, S.},
        title = "{100,000 Crab giant pulses at 215 MHz detected with an SKA-Low prototype station}",
      journal = {arXiv e-prints},
         year = 2025,
        month = jun,
          eid = {arXiv:2506.07422},
        pages = {arXiv:2506.07422},
          doi = {10.48550/arXiv.2506.07422},
archivePrefix = {arXiv},
       eprint = {2506.07422},
 primaryClass = {astro-ph.HE},
       adsurl = {https://ui.adsabs.harvard.edu/abs/2025arXiv250607422S}
}

@INPROCEEDINGS{marcintransient,
       author = {{Sokolowski}, Marcin and {Price}, Danny C. and {Wayth}, B., Randall},
        title = "{A High Time Resolution All-Sky Monitor for Fast Radio Bursts and Technosignatures}",
    booktitle = {2022 3rd URSI Atlantic and Asia Pacific Radio Science Meeting (AT-AP-RASC},
         year = 2022,
        month = jul,
          eid = {1},
        pages = {1},
          doi = {10.23919/AT-AP-RASC54737.2022.9814380},
       adsurl = {https://ui.adsabs.harvard.edu/abs/2022aapr.confE...1S}
}

@techreport{caiazzo2017ska,
  title={SKA phase 1 system requirements specification v11},
  author={Caiazzo, M and others},
  year={2017},
  institution={Tech. Rep. SKA-TEL-SKO-0000008 (SKA Organisation)}
}

@Article{numpy,
 title         = {Array programming with {NumPy}},
 author        = {Charles R. Harris and K. Jarrod Millman and St{\'{e}}fan J.
                 van der Walt and Ralf Gommers and Pauli Virtanen and David
                 Cournapeau and Eric Wieser and Julian Taylor and Sebastian
                 Berg and Nathaniel J. Smith and Robert Kern and Matti Picus
                 and Stephan Hoyer and Marten H. van Kerkwijk and Matthew
                 Brett and Allan Haldane and Jaime Fern{\'{a}}ndez del
                 R{\'{i}}o and Mark Wiebe and Pearu Peterson and Pierre
                 G{\'{e}}rard-Marchant and Kevin Sheppard and Tyler Reddy and
                 Warren Weckesser and Hameer Abbasi and Christoph Gohlke and
                 Travis E. Oliphant},
 year          = {2020},
 month         = sep,
 journal       = {Nature},
 volume        = {585},
 number        = {7825},
 pages         = {357--362},
 doi           = {10.1038/s41586-020-2649-2},
 publisher     = {Springer Science and Business Media {LLC}},
 url           = {https://doi.org/10.1038/s41586-020-2649-2}
}

@Article{Hunter:2007,
  Author    = {Hunter, J. D.},
  Title     = {Matplotlib: A 2D graphics environment},
  Journal   = {Computing in Science \& Engineering},
  Volume    = {9},
  Number    = {3},
  Pages     = {90--95},
  publisher = {IEEE COMPUTER SOC},
  doi       = {10.1109/MCSE.2007.55},
  year      = 2007
}

@ARTICLE{2020SciPy-NMeth,
  author  = {Virtanen, Pauli and Gommers, Ralf and Oliphant, Travis E. and
            Haberland, Matt and Reddy, Tyler and Cournapeau, David and
            Burovski, Evgeni and Peterson, Pearu and Weckesser, Warren and
            Bright, Jonathan and {van der Walt}, St{\'e}fan J. and
            Brett, Matthew and Wilson, Joshua and Millman, K. Jarrod and
            Mayorov, Nikolay and Nelson, Andrew R. J. and Jones, Eric and
            Kern, Robert and Larson, Eric and Carey, C J and
            Polat, {\.I}lhan and Feng, Yu and Moore, Eric W. and
            {VanderPlas}, Jake and Laxalde, Denis and Perktold, Josef and
            Cimrman, Robert and Henriksen, Ian and Quintero, E. A. and
            Harris, Charles R. and Archibald, Anne M. and
            Ribeiro, Ant{\^o}nio H. and Pedregosa, Fabian and
            {van Mulbregt}, Paul and {SciPy 1.0 Contributors}},
  title   = {{{SciPy} 1.0: Fundamental Algorithms for Scientific
            Computing in Python}},
  journal = {Nature Methods},
  year    = {2020},
  volume  = {17},
  pages   = {261--272},
  adsurl  = {https://rdcu.be/b08Wh},
  doi     = {10.1038/s41592-019-0686-2},
}

@ARTICLE{2004PASA...21..302H,
       author = {{Hotan}, A.~W. and {van Straten}, W. and {Manchester}, R.~N.},
        title = "{PSRCHIVE and PSRFITS: An Open Approach to Radio Pulsar Data Storage and Analysis}",
      journal = {\pasa},
         year = 2004,
        month = jan,
       volume = {21},
       number = {3},
        pages = {302-309},
          doi = {10.1071/AS04022},
archivePrefix = {arXiv},
       eprint = {astro-ph/0404549},
 primaryClass = {astro-ph},
       adsurl = {https://ui.adsabs.harvard.edu/abs/2004PASA...21..302H}
}
